\documentclass[]{aastex631}
\usepackage{amsmath}
\usepackage{graphicx}
\usepackage{threeparttable}
\usepackage{txfonts}
\usepackage{soul}
\usepackage{bm} 
\usepackage{tabularx} 
\usepackage{booktabs}
\usepackage{varwidth}
\usepackage{graphicx}

\shorttitle{Calibrating $\rm{DM_{IGM}} - \mathit{\mathit{z}}$ relation using host galaxies of FRBs}
\shortauthors{Li et al.}

\begin{document}
\title{Calibrating $\rm{DM_{IGM}}- \mathit{\mathit{z}}$ Relation using Host Galaxy Properties of Fast Radio Bursts}

\correspondingauthor{Fa-Yin Wang}
\email{fayinwang@nju.edu.cn}

\author[0009-0007-3326-7827]{Rui-Nan Li}
\affiliation{School of Astronomy and Space Science, Nanjing University Nanjing 210023, People's Republic of China}

\author[0000-0002-8046-984X]{Ke Xu}
\affiliation{School of Astronomy and Space Science, Nanjing University Nanjing 210023, People's Republic of China}

\author[0000-0001-7176-8170]{Dao-Hong Gao}
\affiliation{School of Astronomy and Space Science, Nanjing University Nanjing 210023, People's Republic of China}

\author[0000-0001-6021-5933]{Qin Wu}
\affiliation{School of Astronomy and Space Science, Nanjing University Nanjing 210023, People's Republic of China}

\author[0000-0003-0672-5646]{Shuang-Xi Yi}
\affiliation{School of Physics and Physical Engineering, Qufu Normal University, Qufu 273165, People's Republic of China}

\author[0000-0003-4157-7714]{Fa-Yin Wang}
\affiliation{School of Astronomy and Space Science, Nanjing University Nanjing 210023, People's Republic of China}
\affiliation{Key Laboratory of Modern Astronomy and Astrophysics (Nanjing University) Ministry of Education, People's Republic of China}

\begin{abstract}

Fast radio bursts (FRBs) are extragalactic radio transients that offer valuable insight of intergalactic medium (IGM). However, the dispersion measure (DM) contributed by IGM ($\rm{DM_{IGM}}$) is degenerated with that from the host galaxy ($\rm{DM_{host}}$), necessitating calibration of the $\rm{DM_{IGM}}$$-z$ relation for cosmological applications. As $\rm{DM_{host}}$ is expected to correlate with host galaxy properties, it is feasible to estimate $\rm{DM_{host}}$ from observable host characteristics. In this study, we conduct spectral energy distribution (SED) and Sérsic model fittings to derive the parameters of FRB host galaxies. Then, we examine the correlations between the excess dispersion measure ($\rm{DM_{exc}}$) and host galaxy parameters, including star formation rate (SFR), stellar mass, specific star formation rate (sSFR), inclination angle, and projected area. A tight correlation between $\rm{DM_{exc}}$ and sSFR is found. This correlation is utilized to estimate the $\rm{DM_{host}}$ of FRBs, providing a method to calibrate the DM$_{\rm IGM}-z$ relation. This approach leads to a notable improvement in calibration performance.

\end{abstract}

\keywords{Radio bursts (1339) --- Interstellar medium (847) --- Radio transient sources (2008)}

\section{Introduction} \label{sec:intro}
Fast radio bursts (FRBs) are intense bursts of radio waves lasting a few milliseconds, originating from sources located at cosmological distances \citep{Lorimer2007, Xiao2021, Zhang2023Review}. To date, nearly 1000 FRBs are detected, among which over 110 FRBs are localized to their host galaxies (e.g. \cite{Bhardwaj2024, Connor2024,Amiri2025}). FRBs have been observed with redshifts reaching up to $\mathit{\mathit{z}} \sim 1.354$ \citep{Connor2024}, placing them at distances where the intergalactic medium (IGM) significantly influences their dispersion measures (DMs). The radio pulses of FRBs are dispersed when traveling through the ionized mediums. The DM is defined as the integral of the electron column density along the line of sight (LOS): $\mathrm{DM}= \int_0^L n_e(l) /(1+z)\mathrm{d} l $. The total DM of an FRB at redshift $\mathit{\mathit{z}}$ can be split up into the sum of the contributions of its line-of-sight components 
\begin{equation} \label{equ:1}
\rm{}DM_{total}(\mathit{\mathit{z}}) = DM_{MW,ISM} + DM_{MW,halo} + DM_{IGM}(\mathit{\mathit{z}}) + {DM_{extra}} + \frac{DM_{host}}{1+\mathit{\mathit{z}}},
\end{equation}
where $\rm{}DM_{MW,ISM}$ and $\rm{}DM_{MW,halo}$ represent the contribution from interstellar medium (ISM) and halo in the Milky Way, respectively, $\rm{DM_{IGM}(\mathit{\mathit{z}})}$ represents the contribution from the IGM which has been found to be a function of redshift $\mathit{\mathit{z}}$ \citep{Macquart2020}, also known as the Macquart relation. $\rm DM_{extra}$ is defined as the combined contribution from the haloes of the intervening galaxies, and the local environment surrounding the FRB source. $\rm{}{DM_{host}}$ is the contribution from the ISM in the host galaxy in the rest frame and the factor
$1+z$ accounts for the cosmic dilation. $\rm{}{DM_{exc}} \equiv \rm DM_{total} - \rm{}{DM_{MW}}$, representing the extragalactic contribution to dispersion measure.

Consequently, once the $\rm{DM_{IGM}}$ is accurately determined, FRBs become invaluable tools for cosmological studies \citep{Wu2024}. They enable precise measurements of the Hubble constant \citep{Wu2022, Hagstotz2022, James2022, Wei2023,Gao2024,Kalita2024,Zhang2025,Gao2025}, and the cosmic baryon content \citep{McQuinn2014, Macquart2020,li2020b,Beniamini2021, YangKB2022, Lee2022,Lin2023, Connor2024, Acharya2025,ZhangZ2025}. 

Previous studies often relied on assumed values for the components of DM from $\rm{DM_{IGM}}$ and $\rm{DM_{host}}$ due to challenges in decoupling their respective contributions in $\rm{DM_{exc}}$. The probability density distributions for $\rm{DM_{IGM}}$ and $\rm{DM_{host}}$ have been utilized to mitigate the ``degeneracy problem'' \citep{Macquart2020, Zhang2020,ZhangZJ2021}. Observationally, $\rm{DM_{host}}$ can be estimated by the extinction-corrected  $\rm{H\alpha}$ emission line \citep{Cardelli1989} and its emission measure \citep{Cordes2016, Tendulkar2017, Ocker2022}. Scattering measurements of FRB pulses can also serve as an estimator for $\rm{DM_{host}}$, as the intergalactic medium contributes minimally to the scattering time \citep{Cordes2022, Mo2025}. 

Recent studies demonstrate that the types of host galaxies for FRBs are diverse, spanning elliptical, transiting, spiral, irregular, and merging galaxies in morphology classification \citep{Bhardwaj2024, Amiri2025, Eftekhari2025}. The early-type galaxies would contribute less average DM than late-type galaxies, such as spirals \citep{XuHan2015,Chawla2022}. In classification based on galaxy luminosity, both star-forming and quiescent galaxies are found in the host galaxy samples \citep{Acharya2025}. In classification based on stellar mass, most of the hosts are high-mass galaxies, but three active repeaters coincide with persistent radio sources (e.g. FRB 20121102A, FRB 20190520B, FRB 20240114A) are localized to low-mass galaxies \citep{Tendulkar2017, Niu2022, Hewitt2024}. The properties of FRB host galaxies are expected to influence the observed values of $\rm{DM_{host}}$, with galaxies exhibiting higher star formation rates (SFR) generally showing larger electron densities in their interstellar medium \citep{Herrera2016, Wada2007, Li2025SFR, Luo2018, LiZX2019}. Thus, $\rm{DM_{host}}$ is expected to be proportional to the square-root of the SFR of host galaxy \citep{Luo2018}. \cite{Kaasinen2017} suggested that $n_{\rm{e}}$ in the ISM of high-$\mathit{\mathit{z}}$ star-forming galaxies increases with their higher SFRs. Additionally, FRBs located at larger projected offsets from the centers of their host galaxies are expected to exhibit lower $\rm{DM_{host}}$, owing to the decreasing electron density with galactocentric distance \citep{Wada2007, Leonel2010, Ocker2020}. Furthermore, the inclination angles of host galaxies introduce a selection bias in the observation of FRBs, as more significant scattering effects for FRB sources located in edge-on galaxies \citep{Bhardwaj2024}. Although the proposed correlations between $\rm{DM_{host}}$ and SFR or projected offset are motivated by theoretical models and studies of general galaxy populations, they have not yet been systematically tested using the growing sample of localized FRBs. The correlation between galaxy inclination and $\rm{DM_{exc}}$ has already been investigated using 23 FRBs by \citet{Bhardwaj2024}, we expand that analysis to a significantly larger sample of 73 localized FRBs in this work. Analogous to Type Ia supernovae, establishing such empirical correlations is crucial for calibrating the $\rm{DM_{IGM}}$–$z$ relation, which underpins cosmological applications of FRBs.

In this work, we investigate the correlations between host galaxy properties and the $\rm{DM_{exc}}$ for a sample of localized FRBs. By utilizing the tight correlations identified, we calibrate the $\rm{DM_{IGM}} - \mathit{z}$ relation. Details of the FRB sample and host galaxy fitting methods are provided in Section~\ref{subsec:data} and Section~\ref{subsec:fitting methods}, respectively. The correlation analysis is presented in Section~\ref{sec:correlation analysis}, followed by the application of the results to calibrate $\rm{DM_{IGM}} - \mathit{z}$ relation in Section~\ref{sec:application}. Discussion is given in Section~\ref{sec:dis}. The summary is shown in Section~\ref{sec:con}.

\section{Methodology} \label{sec:method}
\subsection{Data Collection and FRB Sample}\label{subsec:data}
In recent years, thanks to the development of various projects designed for rapid localization of FRBs, the number of localized FRB has increased rapidly, making it possible to conduct statistical studies.

A total of 42 FRBs localized by the DSA-110, a radio interferometer designed to maximize the detection and localization rate of FRBs, is included in our sample \citep{Ravi2019, Ravi2022, Ravi2023, Connor2024, Sharma2024, Law2024}.
The Commensal Real-time ASKAP Fast Transient (CRAFT) Survey Science Project, which continuously searches for fast radio transients such as FRBs, contributes 31 localized FRBs that are included in our sample \citep{Mahony2018, Prochaska2019, Bannister2019, Bhandari2020, Bhandari2022, Chittidi2021, Heintz2020, Ryder2023, Shannon2024}.
From the first catalog of FRB host galaxies localized via the combination of CHIME baseband and CHIME-KKO VLBI, 19 FRBs are incorporated into our sample \citep{Amiri2025}, along with 10 additional CHIME baseband-localized FRBs that are also included in this study \citep{Bhardwaj2021, Bhardwaj2024, Ibik2024, Michilli2023}.
Moreover, 7 FRBs localized by the MeerKAT 64-dish interferometer are included in our sample \citep{Rajwade2022, Caleb2023, Driessen2024, Gordon2023, Rajwade2024}.
In addition, 6 FRBs localized by the Very Large Array (VLA) \citep{Tendulkar2017, Law2020, Niu2022, LiYe2025, Anna-Thomas2025} and 2 FRBs localized via very-long-baseline interferometry (VLBI) \citep{Marcote2020, Cassanelli2024} are also incorporated. 
The repeating burst FRB 20200120E is excluded from the analysis due to its proximity and non-cosmological redshift \citep{Bhardwaj2021b}.
In this work, we collected a total of 117 FRBs with localized host galaxies with redshift and DM measurements. We perform spectral energy distribution (SED) fitting for galaxies with photometric data that meet the selection criteria described in Section \ref{subsec:fitting methods}, and Sérsic profile fitting for those with available imaging data from DESI or Pan-STARRS. The datasets used in this work are summarized in Table \ref{tab:data1}.

\subsection{SED and Sérsic Model Fitting} \label{subsec:fitting methods}
The imaging and photometry data utilized in this analysis include the Panoramic Survey Telescope and Rapid Response System (Pan-STARRS) survey \citep{Panstarr}, the Dark Energy Spectroscopic Instrument (DESI) Legacy Imaging Surveys \citep{DESI,Dey2019}, Sloan Digital Sky Survey (SDSS) \citep{SDSS}, Two Micron All Sky Survey (2MASS) \citep{2MASS}, Wide-field Infrared Survey Explorer (WISE) \citep{WISE} and Galaxy Evolution Explorer (GALEX) \citep{GALEX}. 

Sérsic profile fitting is primarily performed using DESI imaging data due to its greater depth ($\rm{}m_{AB,DESI} \lesssim 23.5~\rm{}mag$) compared to Pan-STARRS ($\rm{}m_{AB,PAN} \lesssim22~\rm{}mag$). Pan-STARRS data are used only when the FRB host galaxy lies outside the survey footprint of DESI. The 73 host galaxies used in this work are summarized in Table \ref{tab:data1}. The best-fitted Sérsic model is determined by Python package \textsc{PetroFit}~\citep{Geda2022}. In each run, we set the galaxy center position, effective radius, Sérsic index, axis ratio, postion angle and flux amplitude as free parameters. Later, the axis ratio ($b/a$) is coverted into the cosine of the galaxy inclination ($\cos i$) using the formula~\citep{Hubble1926,Bhardwaj2024b}:
\begin{equation}
    \cos i=\sqrt{\frac{(b/a)^2-q_0^2}{1-q_0^2}},\label{equ:incl}
\end{equation}
where $q_0$ is the intrinsic axial ratio of edge-on galaxies, assumed to be 0.2~\citep{Tully2000}. The area of galaxies ($A$) is further calculated from the effective radius ($r_{\rm e}$) as
\begin{equation}
    A=r_{\rm e}^2/\cos i.\label{equ:area}
\end{equation}

Aside from SED fitting, there are other methods to estimate the SFR of host galaxies. The $H_\alpha$ luminosity, $L_{\rm{}H_\alpha}$, is commonly used to calculate the SFR using the correlation: $\rm{}SFR(\rm{}H_\alpha) = 7.9 \times 10^{-42}~M_{\odot}~yr^{-1}\times(\mathit{L}_{\rm{}H_\alpha}/erg~s^{-1})$ \citep{Kennicutt1994}. Additionally, the SFR can be derived from the [OII] luminosity ($L_{\rm{}OII}$), UV luminosity ($L_{\rm{}UV}$), and infrared luminosity ($L_{\rm{}FIR}$) \citep{Kennicutt1998}. However, these methods have their limitations. $L_{\rm{}UV}$ is highly affected by dust absorption, which reduces its reliability in estimating the total SFR, especially in dust-rich environments. $L_{\rm{}H_\alpha}$ primarily traces recent star formation in HII regions but is also limited by dust extinction and typically reflects a short timescale of star formation. $L_{\rm{}FIR}$, on the other hand, cannot directly differentiate between emission from stars and emission from dust-heated material. Furthermore, different software used for SED fitting can result in varying SFR estimates. 

Hence, we used \textsc{BAGPIPES} \citep{Carnall2018} to fit the SED for all galaxies in our sample with matched photometries including GALEX (FUV and NUV matched from MAST\footnote{https://mast.stsci.edu/portal/Mashup/Clients/Mast/Portal.html}), SDSS ($u$, $g$, $r$, $i$, $z$ band composite model magnitudes), DESI Legacy Imaging Survey ($g$, $r$, $i$, $z$ band model fluxes\footnote{https://www.legacysurvey.org/dr10/catalogs/\#galactic-extinction-coefficients}), Pan-STARRS ($g$, $r$, $i$, $z$, $y$ band Kron fluxes), 2MASS ($J$, $H$, $K_{\rm s}$ band extrapolation-fit magnitudes) and WISE ($W1$,$W2$,$W3$ band instrumental profile-fit photometry magnitudes). WISE $W4$ is not used because its low resolution could lead to the contamination from nearby objects. 
The photometry data from SDSS, DESI and Pan-STARRS are all used in the SED fitting. For galaxies with multiple observations in the same band, the contribution of each observation to the fitting result are determined by their error weighting.
In SED fitting, we adopted Stellar Population Synthesis model from 2016 version of \cite{Bruzual2003} (age $\in \rm [0.1,\;15]~Gyr$ and metallicity $Z/Z_\odot \in [0.01,\;2.5]$), the delayed exponentially declining star-formation history (timescale $\tau \in \rm[0.1,\;10]~Gyr$, the dust attenuation curve in the form of \cite{Calzetti2020} for young and old stellar population separately (separated at 0.01~Gyr; $V$ band attenuation $A_V \in [0,\;5]$), and the dust emission model with varied $q_{\rm PAH}$ ($\in [0.1,\;4.58]$) while other parameters are fixed at default value as \cite{Sharma2024}, and the nebular emission constructed following \cite{Byler2017} with \textsc{Cloudy}~\citep{Ferland2017} (ionization parameter $\log U \in \rm [-2,\;-5]$).
We evaluated the SED fitting results visually and classified them with five quality levels. Quality = 0 represents good SED fitting results. Quality = 1 indicates that the galaxies have no near-infrared observations but are well-fitted in the optical bands. Quality = 2 corresponds to insufficient photometry ($<4$) in SED fitting. Quality = 3 means that the Kron fluxes from Pan-STARRS are not consistent to the model fluxes from DESI Legacy Imaging Survey, either due to the pollution of nearby objects or the extended galaxy morphology, but the SED fitting results are good when excluding Pan-STARRS photometry. Quality = 4 represents the missing data and bad SED fitting cases. 
The bad SED fitting cases are those galaxies with inconsistent photometry among SDSS, DESI and Pan-STARRS or those with more than three photometric points deviating $>3\sigma$ from the best-fitted SED.
In this work, galaxies with Quality = 0, 1 and 3 were included in the correlation analysis. 

\section{Correlation Analysis} \label{sec:correlation analysis}

The value of $\rm{}DM_{host}$ depends on the electron density ($n_{\rm{e}}$) in the host galaxy ISM and the specific path that the FRB traverses through it. Since these factors are inherently linked to the physical properties of the host galaxy, $\rm DM_{host}$ is expected to exhibit correlations with global galaxy parameters.  However, direct and precise measurements of $\rm DM_{host}$ remain unfeasible at present. To explore the potential connections, we investigate the correlations between $\rm DM_{exc}$ and a set of host galaxy parameters, including SFR, stellar mass, specific star formation rate (sSFR; star formation rate divided by stellar mass), projected offset of FRB, inclination angle ($i$), and galaxy area. Here, $\rm DM_{exc}$ is derived by subtracting the Milky Way contribution ($\rm DM_{MW,ISM}$) estimated using the NE2001 model \citep{ne2001}. In cases where specific uncertainties are unavailable, we adopt a fiducial uncertainty of $10\%$ for $\rm DM_{exc}$, inclination angle, offset, area and $\tau_{\rm exc}$.  Varying the error fraction, either higher or lower, does not affect the final results significantly. 

\begin{figure} 
\centering
\includegraphics[width=200 mm]{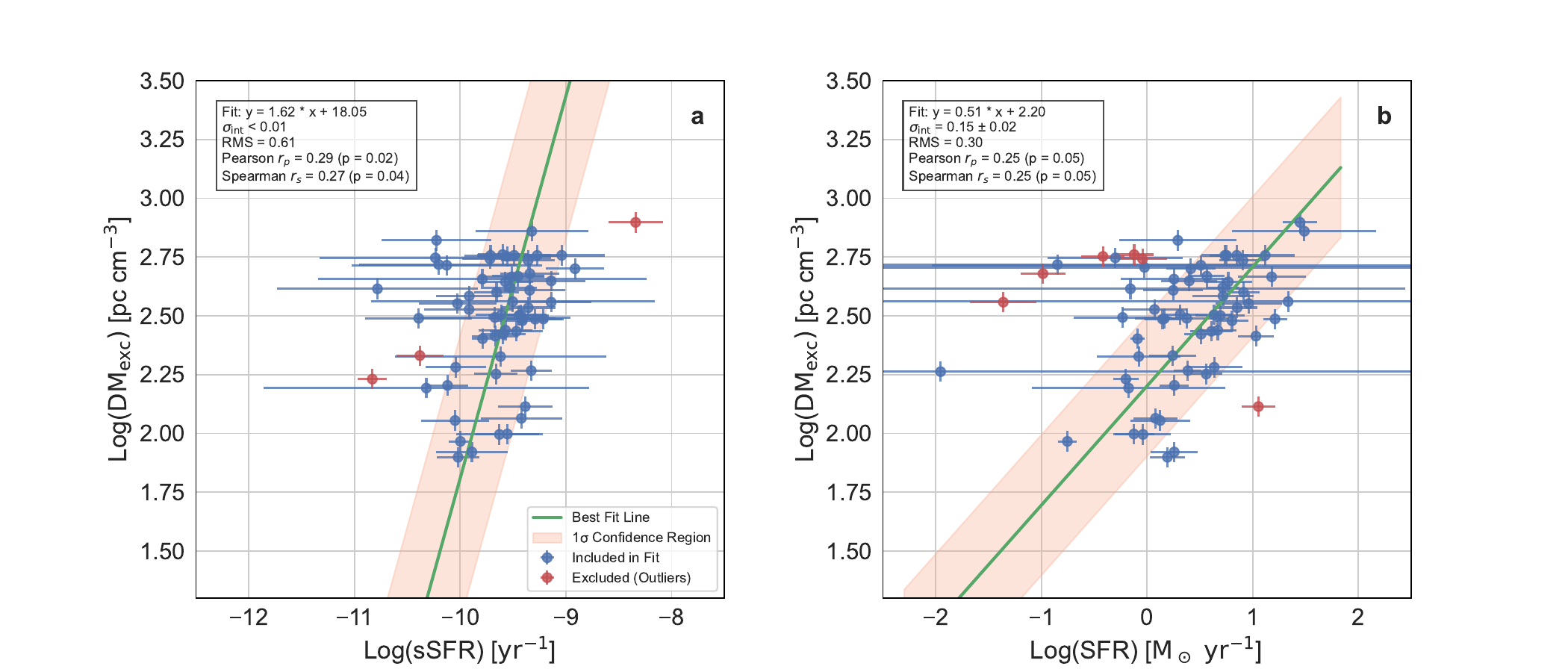}
\caption{The fitting results for the correlations between sSFR and SFR of host galaxies and $\rm{}DM_{exc}$. The blue circles with error bars in both the $x$ and $y$ axes represent the data points included in the fitting, while the red circles with error bars indicate outliers that significantly deviate from the main trends. The green solid line represents the linear fitting results. The shaded region corresponds to the $1\sigma$ confidence interval, calculated from the root mean square (RMS) deviation. The fitted parameters are summarized in Table \ref{Tab:fit}. 
Panel (a): the relation between the sSFR and $\rm{}DM_{exc}$ for 58 FRBs and 3 FRBs are selected as outliers. The sSFR is truncated at $\log(\mathrm{sSFR}) = -11$, as galaxies with $\log(\mathrm{sSFR}) < -11$ are generally classified as quiescent (or ``red'') galaxies, for which SFR measurements are considered unreliable.
Panel (b): the relation between the SFR and $\rm{}DM_{exc}$ for 57 FRBs and 6 FRBs are selected as outliers. FRBs with $\rm{}SFR\le0.01~M_{\odot}~yr^{-1}$ are excluded in fitting due to the large uncertainties.  
}
\label{Fig1}
\end{figure}

\begin{figure} 
\centering
\includegraphics[width=200 mm]{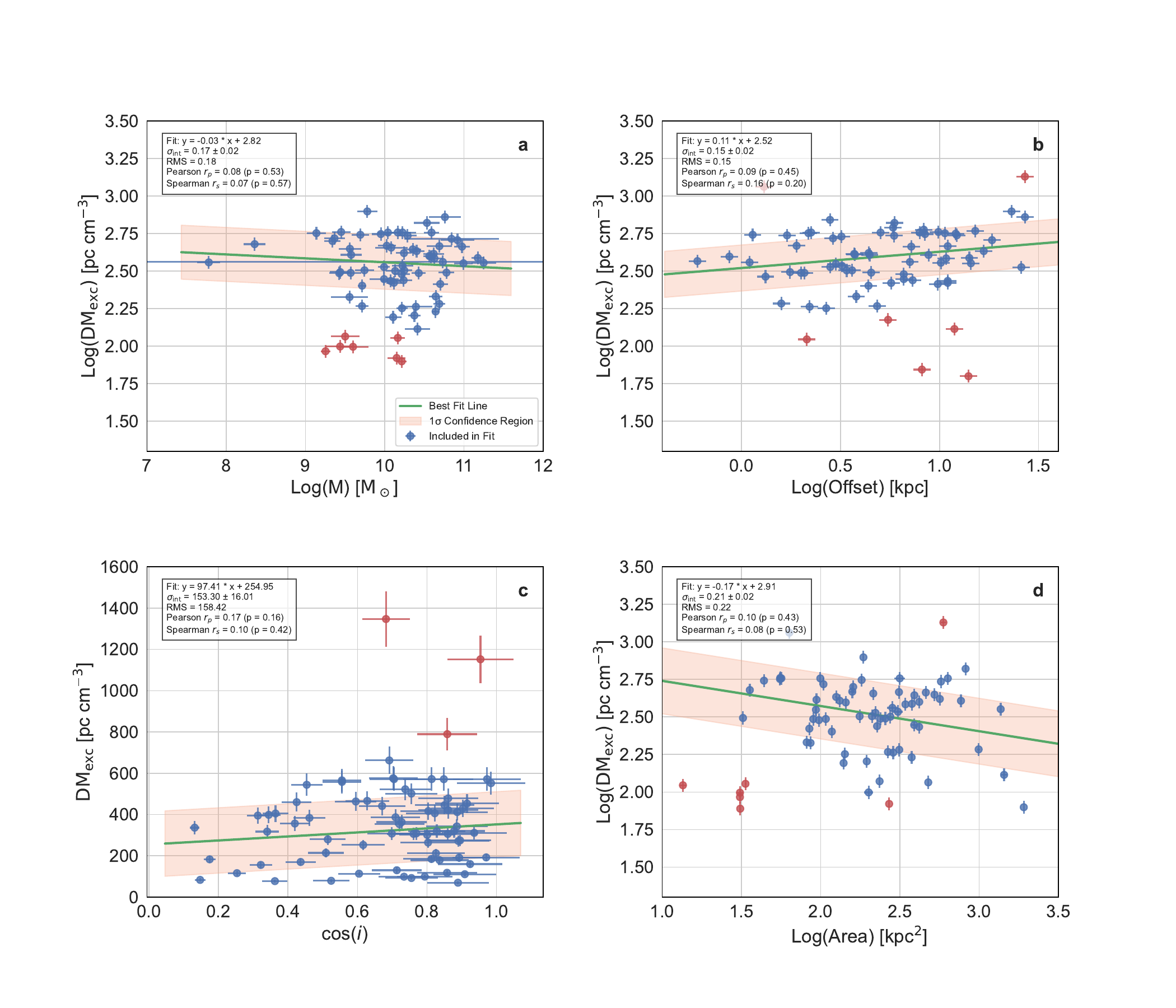}
\caption{The fitting results for the correlations between stellar mass, offset, inclination angle and projected area of host galaxies and $\rm{}DM_{exc}$. The blue circles with error bars in both the $x$ and $y$ axes represent the data points included in the fitting, while the red circles with error bars indicate outliers that significantly deviate from the main trends. The green solid line represents the linear fitting results. The shaded region corresponds to the $1\sigma$ confidence interval, calculated from the root mean square (RMS) deviation. The fitted parameters are summarized in Table \ref{Tab:fit}. 
Panel (a): the relation between the stellar mass and $\rm{}DM_{exc}$ for 62 FRBs and 7 FRBs are selected as outliers. 
Panel (b): the relation between the offset and $\rm{}DM_{exc}$ for 63 FRBs and 7 FRBs are selected as outliers. 
Panel (c): the relation between the cosine of the inclination angles and $\rm{}DM_{exc}$ for 70 FRBs and 3 FRBs are selected as outliers. 
Panel (d): the relation between the area and $\rm{}DM_{exc}$ for 64 FRBs and 7 FRBs are selected as outliers. 
}
\label{Fig2}
\end{figure}

The LtsFit package \citep{Cappellari2014} is employed for robust linear regression, optimized for hyperplane fitting in $N$-dimensional space while accounting for measurement uncertainties and intrinsic scatter in all variables. It implements the Least Trimmed Squares (LTS) method as described in \citet[][Section 3.2]{Cappellari2013}, iteratively excluding outliers \citep{Isobe2023}.
The best-fitting linear relation of the form $y = a(x - x_0) + b$ is determined by minimizing the following chi-square statistic \citep{Cappellari2013}
\begin{equation}
\chi^2=\sum_{j=1}^N \frac{\left[b+a\left(x_j-x_0\right)-y_j\right]^2}{\left(a \Delta x_j\right)^2+\left(\Delta y_j\right)^2+\sigma_{int}^2},
\end{equation}
where $\sigma_{\rm{}int}$ represents the intrinsic scatter in the $y$-direction around the fitted relation. Data points deviating by more than $2.6\sigma$ from the best fit are classified as outliers and excluded from the fitting process. However, the Pearson ($r_p$) and Spearman ($r_s$) correlation coefficients are computed using the full dataset. 
The fitting results are presented in Figure \ref{Fig1}, Figure \ref{Fig2} and Table \ref{Tab:fit}.

In Figures \ref{Fig1} and \ref{Fig2}, we present the correlations between $\rm DM_{exc}$ and various host galaxy properties. 
Among all tested parameters, only the sSFR and the SFR in Figure \ref{Fig1} show a statistically significant and physically meaningful correlation with $\rm DM_{exc}$. 
In panel (a) of Figure \ref{Fig1}, the best-fit slope is $1.62 \pm 0.36$, and the intrinsic scatter is close to zero ($\sigma_{\text{int}} < 0.01$), indicating a remarkably tight relation with minimal underlying variability. Both the Pearson ($r_p = 0.29$, $p = 0.02$) and Spearman ($r_s = 0.27$, $p = 0.04$) coefficients confirm the statistical significance. This suggests that FRBs occurring in galaxies with higher sSFR tend to exhibit larger $\rm DM_{exc}$ values, likely reflecting denser ionized environments. Since sSFR encapsulates both star formation activity and stellar mass, it is more sensitive to gas-rich, actively star-forming systems. This is consistent with theoretical models and simulations that link high sSFR to enhanced electron densities in the ISM \citep{Herrera2016, Wada2007, Li2025SFR}. The result also supports the idea that at least some FRBs trace recent star formation and may originate from core-collapse supernovae via magnetar formation \citep{Zhao2021,James2022b, Gordon2023, Bhardwaj2024, Sharma2024, Mo2025}, though some studies have reported a potential bias \citep{Zhang2019, Zhang2022b, Qiang2022, Chen2024, Champati2025}.

In comparison, panel (b) of Figure \ref{Fig1} shows a moderate positive trend between the SFR and $\rm DM_{exc}$, with a fitted slope of $0.51 \pm 0.12$ and a relatively small intrinsic scatter of $0.15 \pm 0.02$. Compared to panel (a), the correlation is also slightly weaker (Pearson $r_p = 0.25$, $p = 0.05$; Spearman $r_s = 0.25$, $p = 0.05$). Furthermore, after including observational uncertainties, a larger fraction of data points fall outside the 1$\sigma$ confidence interval than in panel (a), indicating a poorer fit. While this result is broadly consistent with predictions in \cite{Luo2018}, the weaker statistical confidence and higher scatter compared to sSFR indicate that SFR alone is a less reliable predictor of $\rm DM_{exc}$.

The results for other parameters shown in Figure \ref{Fig2} exhibit no statistical significance. 
In panel (a), the stellar mass shows virtually no correlation with $\rm DM_{exc}$. The best-fit slope is nearly flat ($-0.03 \pm 0.12$), and the Pearson and Spearman coefficients are close to zero ($r_p = 0.08$, $p = 0.53$; $r_s = 0.07$, $p = 0.57$). The intrinsic scatter is relatively high ($\sigma_{\text{int}} = 0.17 \pm 0.02$), further confirming that stellar mass does not contribute meaningfully to the variation in $\rm DM_{exc}$.
For the projected offset between the FRB and the host galaxy center in panel (b), we find a weakly positive trend (slope $= 0.11 \pm 0.12$), with non-significant correlations ($r_p = 0.09$, $p = 0.45$; $r_s = 0.16$, $p = 0.20$) and moderate intrinsic scatter ($\sigma_{\text{int}} = 0.15 \pm 0.02$). Theoretically, a negative correlation might be expected, as electron density typically decreases with galactocentric radius \citep{Wada2007,Ocker2020}. The discrepancy may be due to the small sample size or the influence of extreme outliers.

In panel (c), the inclination angle exhibits a very weak positive slope ($97.41 \pm 82.04$), but neither the Pearson ($r_p = 0.17$, $p = 0.16$) nor Spearman ($r_s = 0.10$, $p = 0.42$) coefficients are statistically significant. Moreover, the intrinsic scatter is extremely large ($\sigma_{\text{int}} = 153.27 \pm 16.06$), indicating high variability and suggesting that inclination has no reliable effect on $\rm DM_{exc}$. Further discussion is provided in Section \ref{sec:dis}.
Lastly, in panel (d), the projected area of the host galaxy shows a slight negative slope ($-0.17 \pm 0.02$). But, both the correlation coefficients are insignificant ($r_p = -0.10$, $p = 0.43$; $r_s = 0.08$, $p = 0.53$), and the intrinsic scatter is moderate ($\sigma_{\text{int}} = 0.21 \pm 0.02$). The inconsistency between the slope and correlation signs may stem from strong outliers. Additionally, galaxy morphology - an important factor influencing the area - was not explicitly modeled in our Sérsic fits, which may introduce further uncertainty.

Taken together, and considering both the statistical significance and intrinsic scatter of each correlation, only the correlation between sSFR and $\rm DM_{exc}$ stands out as robust. Since $\rm DM_{IGM}$ and $\rm DM_{extra}$ are largely uncorrelated with host galaxy properties, this correlation may reflects an intrinsic connection between sSFR and $\rm DM_{host}$. Therefore, we propose that sSFR may serve as a useful empirical proxy for estimating $\rm DM_{host}$ in FRB studies.

\section{Calibration the $\rm{DM_{IGM}} - \mathit{z}$ Relation} \label{sec:application}
The correlation between the $\rm{}\log(sSFR)$ and $\rm{}\log(DM_{exc})$ can be described by the following linear relation in logarithmic space
\begin{equation} \label{equ:2}
\rm{}\log(DM_{exc}) = \alpha\rm{} \log(sSFR) + \beta,
\end{equation}
which corresponds to the following power-law form in linear space
\begin{equation} \label{equ:3}
\rm{} DM_{exc} = 10^{\beta}~sSFR^{\alpha}.
\end{equation}
As shown in Equation (\ref{equ:1}), $\rm{}DM_{exc}$ consists of $\rm{}DM_{host}$, $\rm{}DM_{IGM}$ and $\rm{}DM_{extra}$. Since both $\rm{}DM_{IGM}$ and $\rm{}DM_{extra}$ are expected to be uncorrelated with sSFR, the correlation in Equation (\ref{equ:3}) should originate from $\rm{}DM_{host}$. Therefore, the $\rm{}DM_{host}$ in the rest frame can be reasonably modeled as a function of sSFR:
\begin{equation} \label{equ:4}
\rm{} DM_{host} = 10^{\beta}~sSFR^{\alpha} - constant,
\end{equation}
where the constant represents the combined contribution from $\rm DM_{IGM}$ and $\rm DM_{extra}$, which is independent of sSFR.  
Theoretically, the average value of $\rm{}DM_{IGM}$ is given by \citep{Deng2014,Gao2024}
\begin{equation}\label{equ:5}
\left\langle\mathrm{DM}_{\mathrm{IGM}}\right\rangle=\frac{21 c \Omega_b H_0}{64 \pi  G m_p} \times \int_0^\mathit{z} \frac{f_{\mathrm{IGM}}(1+\mathit{z}) d \mathit{z}}{\left[\Omega_m(1+\mathit{z})^3+1-\Omega_m\right]^{1 / 2}},
\end{equation}
where $f_{\mathrm{IGM}}$ represents the fraction of baryons in the IGM, $m_p$ is the proton mass, $\Omega_m$ is the cosmological matter density parameter assuming a flat universe, with $\Omega_\Lambda = 1 - \Omega_m$, and $\Omega_b$ is the baryon density parameter, determined from the Planck results based on Big Bang nucleosynthesis constraints and primordial deuterium abundance measurements \citep{Cooke2018}. $H_0$ is the Hubble constant, $G$ is the gravitational constant, and $c$ is the speed of light. The hydrogen and helium fractions are normalized to 1. Apart from $f_{\mathrm{IGM}}(1+\mathit{z})$ and $\mathit{z}$, other constant terms do not affect the relation with $\mathit{z}$, and $f_{\mathrm{IGM}}$ can be considered as a constant. At low redshifts ($z<1$), the integration $\int_0^\mathit{z} (1+z)/(\Omega_m (1+\mathit{z})^3 + 1 - \Omega_m)^{1/2}dz \approx z$. Therefore, Equation (\ref{equ:5}) can be rewritten as
\begin{equation}\label{equ:6}
\left\langle\mathrm{DM}_{\mathrm{IGM}}\right\rangle(z) \approx 771\times z,
\end{equation}
with cosmological parameters from \cite{Planck2020} and $f_{\mathrm{IGM}}=0.84$ \citep{Wu2022}. Based on Equation (\ref{equ:6}) and Equation (\ref{equ:4}), $\rm{DM_{exc}}$ can be simply expressed as
\begin{equation}\label{equ:7}
\begin{gathered}
\mathrm{DM}_{\mathrm{exc}} = \frac{\mathrm{DM}_{\mathrm{host}}}{1+z} + \mathrm{DM}_{\mathrm{IGM}}(z) + \mathrm{DM}_{\mathrm{extra}} + \mathrm{DM}_{\mathrm{MW,halo}} \approx \frac{\rm{}10^{\beta}~sSFR^{\alpha}}{1+z} + \mathit{Az} + B,
\end{gathered}
\end{equation}
where $B$ is considered as a constant to account for the sum of the average contribution of $\rm{DM_{extra}}$ and $\rm{DM_{MW,halo}}$. Equation (\ref{equ:7}) is employed to fit the observed $\rm DM_{exc}$, as shown in Figure~\ref{Fig2}, yielding best-fit parameters of $\alpha = 0.71\pm0.01$, $\beta = 8.53\pm0.03$, $A = 782.27\pm1.01$, and $\rm B = 156.99\pm0.51$. The value of $\alpha$ is lower than that derived in panel (a) of Figure~\ref{Fig1}, which may result from the inclusion of a $z$-dependent IGM term in the model of Equation~(\ref{equ:7}). For sources where $\rm DM_{exc}$ is primarily contributed by the IGM and the host galaxy contribution is minimal, the model tends to overpredict the observed DM values. This overprediction may arise from either the intrinsically low host DM due to large FRB–host offsets, or from potential overestimation of SFRs. The latter could be caused by a steeply declining star formation history or contamination from active galactic nuclei. However, the large uncertainties in FRB localization and the limited photometric coverage currently prevent a robust test of these interpretations. The fitted $A = 782.27\pm1.01$ is slightly below the values obtained from cosmological simulations \citep{ZhangZJ2021, Batten2021, Walker2024}. Notably, it agrees with the theoretical estimate of $A = 771$ in Equation~(\ref{equ:6}), based on the cosmological parameters from \citet{Planck2020}.

\begin{figure}[htbp]
\centering
\begin{minipage}{0.49\textwidth}
    \centering
    \includegraphics[width=\textwidth]{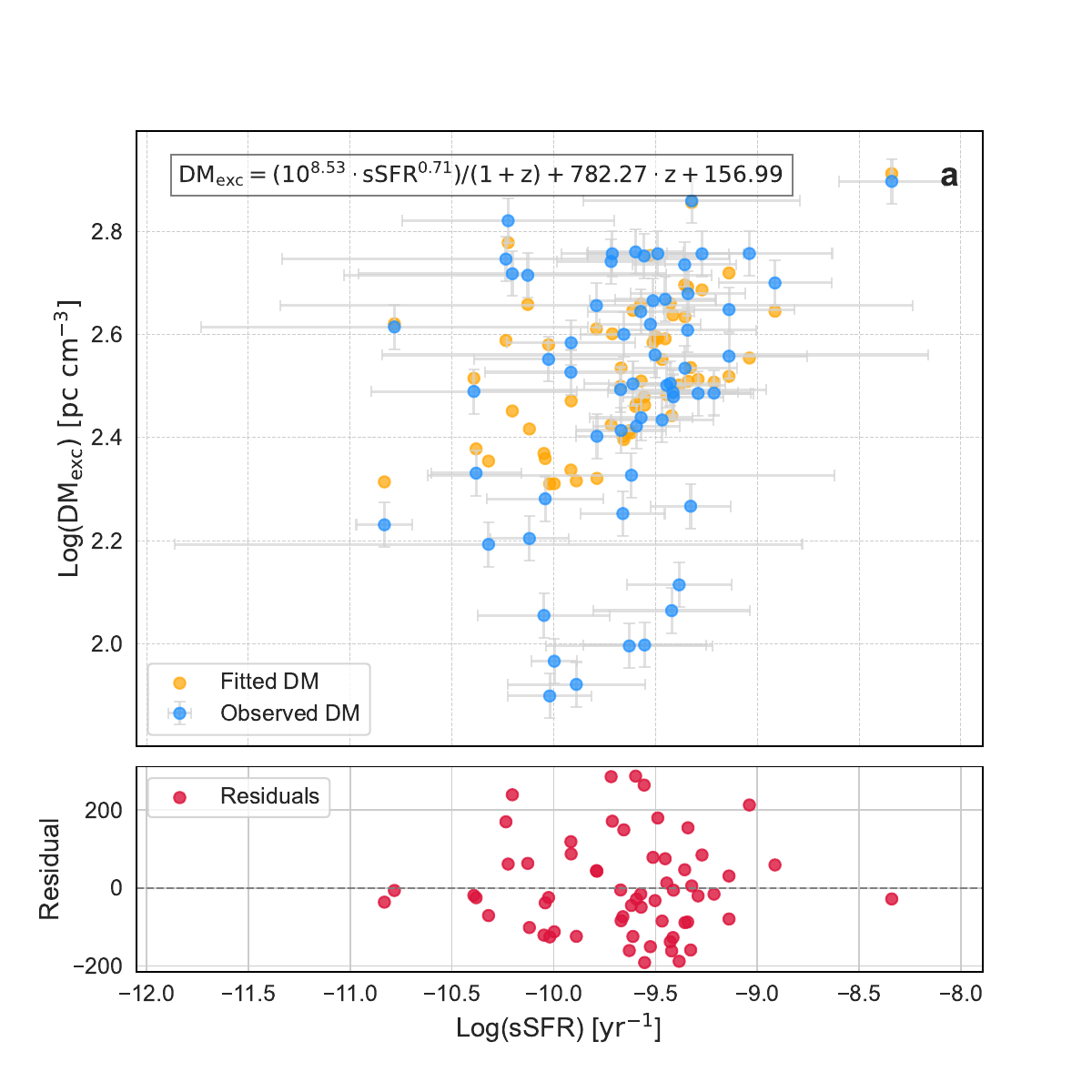}
\end{minipage}
\begin{minipage}{0.49\textwidth}
    \centering
    \includegraphics[width=\textwidth]{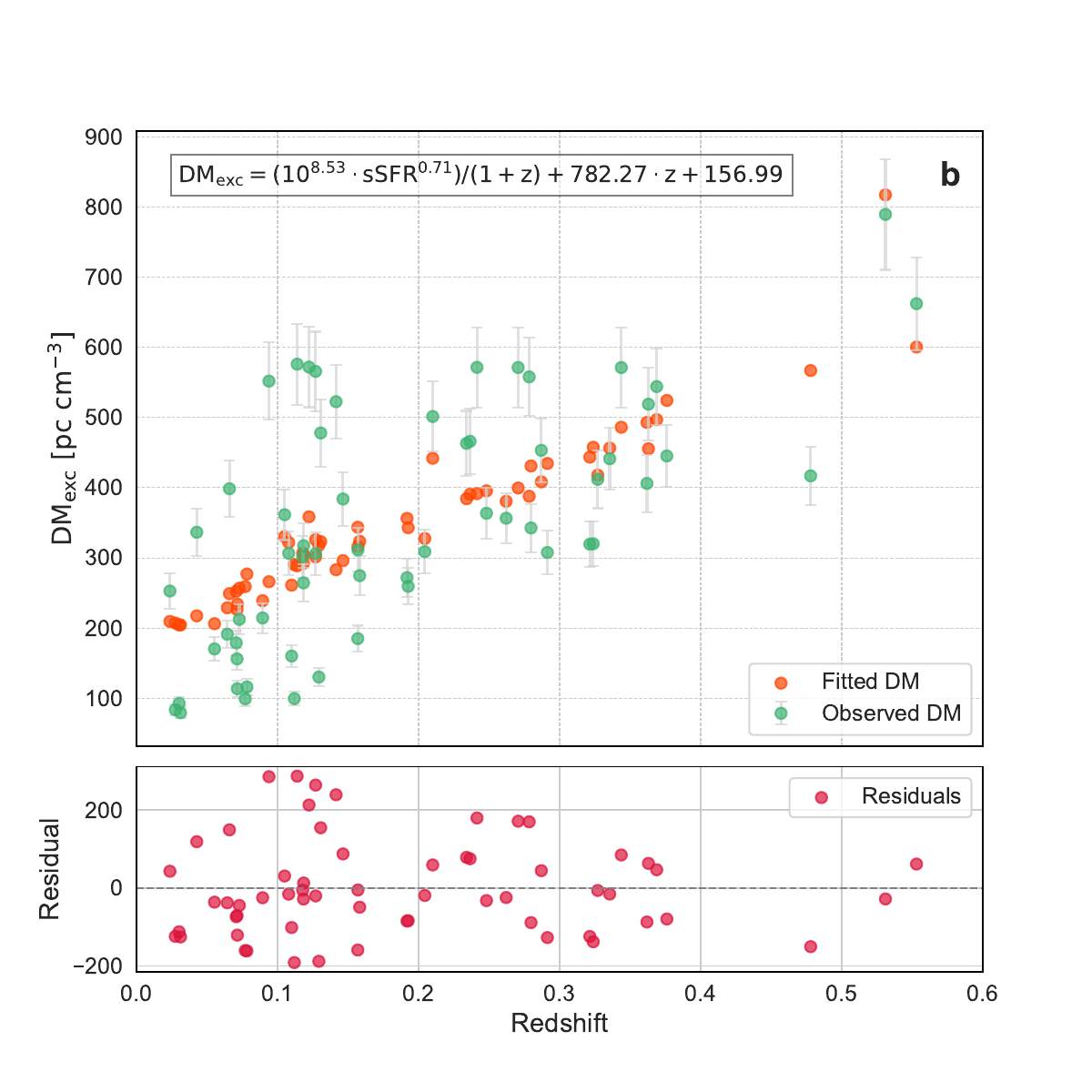}
\end{minipage}
\caption{Panel (a): The relation between the $\rm{}\log(sSFR)$ and $\rm{}\log(DM_{exc})$ with residuals. The blue circles represent the observed data and the orange circles indicate the fitted results. The fitting parameters are $\alpha = 0.71$, $\beta = 8.53$, $\rm{}A = 782.27$ and $\rm{}B = 156.99$. The root mean square error (RMS) is 120.25 and the coefficient of determination ($\rm{}R^2$) is 0.51. 
Panel (b): The fitted relation between the redshift and $\rm{}DM_{exc}$ with residuals. The green circles represent the observed data and the amber circles indicate the fitted results.}
\label{Fig3}
\end{figure}

\begin{figure}[htbp]
\centering
\begin{minipage}{0.48\textwidth}
    \centering
    \includegraphics[width=\textwidth]{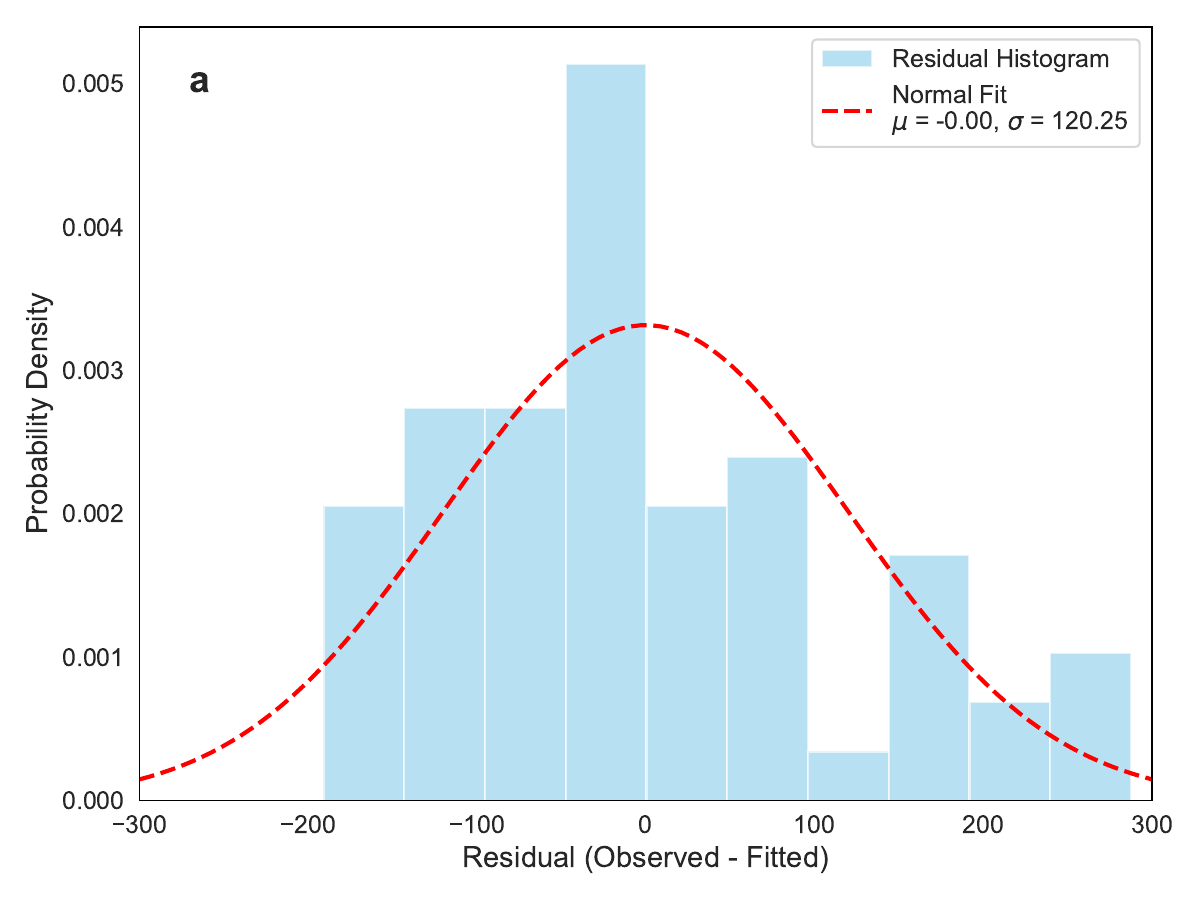}
\end{minipage}
\begin{minipage}{0.48\textwidth}
    \centering
    \includegraphics[width=\textwidth]{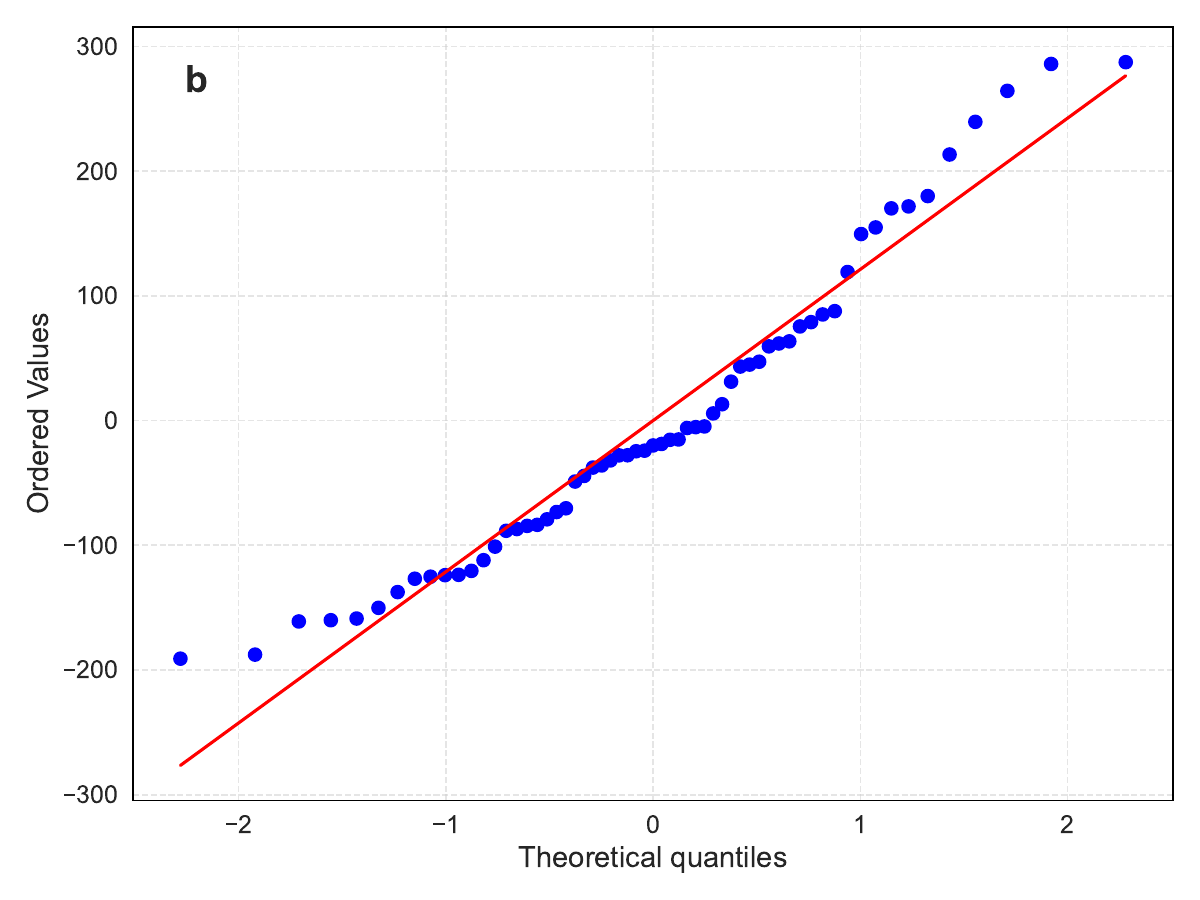}
\end{minipage}
\caption{Panel (a): Residual distribution with Gaussian Fit. The histogram shows the distribution of residuals with a Gaussian fit overlaid (red dashed line). The fitted Gaussian parameters, including the mean ($\mu = 0.00$) and standard deviation ($\sigma = 120.25$), are displayed. The plot suggests that the residuals follow a near-normal distribution with slight deviations in the tail.
Panel (b): Q-Q Plot of residuals. The quantile-quantile (Q-Q) plot compares the residuals against a normal distribution. The points lie near the diagonal, indicating that the distribution of residuals is approximately normal. Deviations from the line suggest some mild skewness or heavy tail in the residuals.}
\label{Fig4}
\end{figure}

\begin{table}[ht]
\centering
\caption{Fitting Results of Figure \ref{Fig1} and \ref{Fig2}}
\label{Tab:fit}
\begin{tabular}{lcccccc}
\hline
Parameters & sSFR  & SFR & $M$ & offset & cos($i$) & area \\
\hline
$a$ & $1.62 \pm 0.35$ & $0.51 \pm 0.06$ & $-0.03 \pm 0.04$  & $0.11 \pm 0.05$ & $97.41 \pm 85.01$ & $-0.17 \pm 0.07$ \\
$b$ & $18.05 \pm 3.36$ & $2.20 \pm 0.04$ & $2.82 \pm 0.03$  &  $2.52 \pm 0.05$ & $254.94 \pm 67.22$ & $2.91 \pm 0.17$ \\
$r_{p}$ & $0.29 (\rm{p} = 0.02)$ & $0.25 (\rm{p} = 0.05)$ & $0.08 (\rm{p} = 0.53)$ & $0.09 (\rm{p} = 0.45)$ & $0.17 (\rm{p} = 0.16)$ & $0.10 (\rm{p} = 0.43)$ \\
$r_{s}$ & $0.27 (\rm{p} = 0.04)$ & $0.25 (\rm{p} = 0.05)$ & $0.07 (\rm{p} = 0.57)$ & $0.16 (\rm{p} = 0.20)$ & $0.10 (\rm{p} = 0.42)$ & $0.08 (\rm{p} = 0.53)$ \\
$\sigma_{int}$ & $< 0.01$ & $0.15\pm0.02$ & $0.17\pm0.02$ & $0.15\pm0.02$ & $153.27\pm16.08$ & $0.21\pm0.02$  \\
\hline
\end{tabular}
\vspace{10pt}  
\begin{minipage}{0.8\textwidth}  
    \textit{Note:} 
    SFR, $M$, sSFR, offset, and area are fitted in logarithmic scale, while $\cos(i)$ is fitted in linear scale. The linear relation $y = ax + b$ is used.
\end{minipage}
\end{table}

To assess the goodness of fit, we performed a residual analysis by examining both the distribution of residuals and their conformity to a normal distribution. We first constructed a histogram of the residuals and fitted a Gaussian function to evaluate whether the deviations between the observed and modeled values are symmetrically distributed around zero. Panel (a) of Figure \ref{Fig4} illustrates that the resulting distribution exhibits approximate symmetry, suggesting that the residuals are largely consistent with Gaussian noise. 
Additionally, we generated a quantile–quantile (Q–Q) plot comparing the empirical quantiles of the residuals with those expected from a theoretical normal distribution. Panel (b) of Figure \ref{Fig4} demonstrates that the residuals lie close to the 1:1 reference line, particularly near the center, indicating that they approximately follow the normal distribution. Minor deviations in the tails suggest slight heavy-tailed behavior, but the overall agreement supports the statistical validity of the model within the assumed error structure.

By combining Equation (\ref{equ:6}) with the fitting results shown in Figure \ref{Fig3}, $\rm{DM_{host}}$ can be expressed as a function of $\rm{sSFR}$ and $z$,
\begin{equation} \label{equ:9}
\rm{} DM_{host} +  \rm{} DM_{extra} + \mathrm{DM}_{\mathrm{MW,halo}} = \frac{10^{8.53} ~ sSFR^{0.71}}{1+z} + 156.99.
\end{equation}
This equation is employed to infer $\rm{DM_{IGM}}$. $\rm{DM_{IGM}}$ can be derived by subtracting $\rm{DM_{host}}$, $\rm{DM_{extra}}$ and $\rm{DM_{MW,halo}}$. This allows that the $\rm{sSFR-DM_{exc}}$ relation is applied for calibrating the $\rm{DM_{IGM}} - \mathit{z}$ relation. In cases when the $\rm{DM_{IGM}}$ from Equation (\ref{equ:9}) results in unreasonable values, we impose a criteria $\rm{DM_{exc} - (DM_{host} + DM_{extra} + DM_{MW,halo})} \gtrsim 30~\text{pc}~\text{cm}^{-3}$  as estimated from Equation (\ref{equ:6}) for this FRB sample. 

\begin{figure}[htbp]
\centering
\begin{minipage}{0.48\textwidth}
    \centering
    \includegraphics[width=\textwidth]{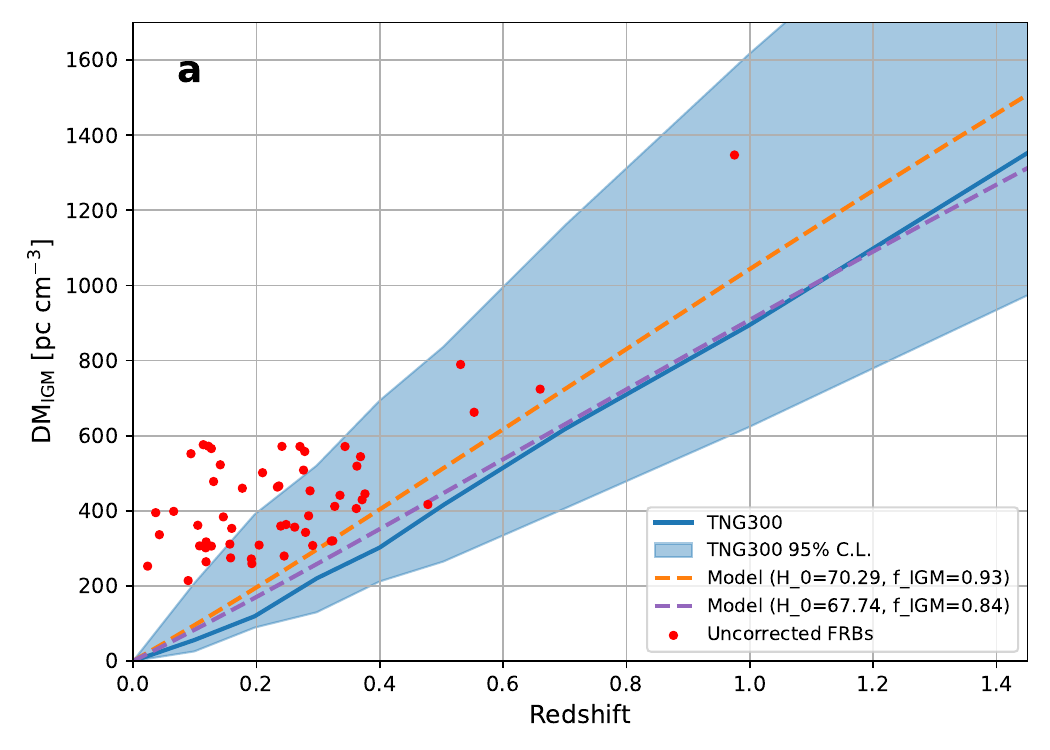}
\end{minipage}
\begin{minipage}{0.48\textwidth}
    \centering
    \includegraphics[width=\textwidth]{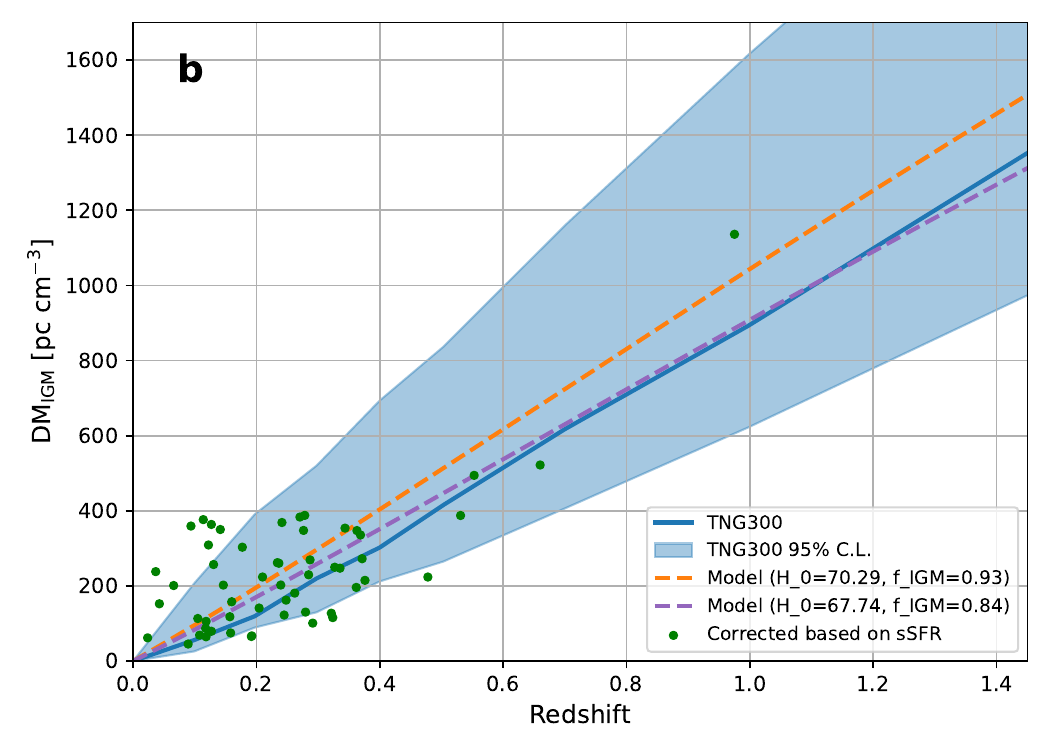}
\end{minipage}
\caption{The $\rm DM_{IGM}$–$z$ relation for 55 FRBs. The blue solid curve shows the mean $\langle\rm DM_{\rm IGM}\rangle$ derived from the IllustrisTNG-300 cosmological simulation \citep{ZhangZJ2021}. The amber dashed line corresponds to the analytic model described by Equation~(\ref{equ:5}) with $H_0 = 69.40~\rm{}km~s^{-1}~Mpc^{-1}$ and $f_{\rm IGM} = 0.93$ \citep{Gao2024,Connor2024}, while the violet dashed line shows the model from Equation~(\ref{equ:5}) with $H_0 = 68.81~\rm{}km~s^{-1}~Mpc^{-1}$ and $f_{\rm IGM} = 0.84$ \citep{Shull2012,Wu2022}.
Panel (a): Red circles represent $\rm DM_{IGM}$ values inferred  directly from the observed $\rm DM_{exc}$, without correcting for host galaxy contributions.
Panel (b): The calibrated $\rm DM_{IGM}$–$z$ relation. Green circles denote $\rm DM_{IGM}$ values obtained by subtracting the $\rm DM_{host}$ estimated via the $\rm sSFR$–$\rm DM_{host}$ correlation (Equation~(\ref{equ:9})) from the total excess dispersion measure $\rm DM_{exc}$.
}
\label{Fig5}
\end{figure}

The calibration results for 55 FRBs are shown in Figure \ref{Fig5}. To evaluate the improvement of the $\rm{DM_{IGM}}-\mathit{z}$ relation after correcting for host contribution, we compute the mean squared errors (MSEs) of the $\rm DM_{IGM}$ values with respect to three reference models: (i) the analytical prediction from Equation (\ref{equ:5}) evaluated with $H_0 = 70.41~\rm{}km~s^{-1}~Mpc^{-1}$ and $f_{\rm IGM} = 0.93$, (ii) the same analytical model evaluated with $H_0 = 67.74~\rm{}km~s^{-1}~Mpc^{-1}$ and $f_{\rm IGM} = 0.84$, and (iii) the prediction from the IllustrisTNG 300 cosmological simulation \citep{ZhangZJ2021}. After the correction, the MSEs are significantly reduced for the three models. These reductions correspond to relative improvements of $69.21\%$, $76.82\%$, and $78.88\%$, respectively. Furthermore, the fraction of FRBs falling within the $95\%$ confidence interval of the TNG300 simulation increases from $50.91\%$ to $72.73\%$ after correction. These results demonstrate that the $\rm DM_{host}$ correction based on the sSFR-DM$_{\rm IGM}$ correlation substantially improves the consistency between the observed FRB data and theoretical expectations, reinforcing the potential of FRBs as reliable cosmological probes.

\section{Discussion} \label{sec:dis}

While our correlation analysis incorporates robust linear fitting (via LtsFit), measurement uncertainties, intrinsic scatter estimation and non-parametric correlation tests, several limitations remain. First, the assumption of a linear relation in logarithmic space may oversimplify the underlying astrophysical dependencies, particularly for parameters with complex or non-monotonic behavior. Second, the small sample size (tens of FRBs) limits statistical power and may lead to unstable estimates, especially when outliers influence the correlation structure. Although LtsFit mitigates this via clipping, the results can still be sensitive to the clipping threshold. Moreover, the uncertainties of host properties (e.g., sSFR and stellar mass) are model-dependent, and potential systematics from SED fitting or inclination correction are not fully captured. 

The haloes of MW and intervening galaxies can contribute non-negligible DMs, denoted as $\rm DM_{MW,halo}$ and $\rm DM_{halo,int}$, respectively \citep{Faber2024, ZhangZ2025, Anna-Thomas2025}. Both simulations and observations suggest that the DM contribution from intervening galaxies increases statistically with redshift \citep{Connor2022, ZhangZ2025, Hussaini2025}. However, this contribution can vary significantly from one line of sight to another, due to differences in the intersected structures. Moreover, there is no reliable quantitative method to subtract the contribution from these halos currently. Incorporating this effect into our analysis would introduce additional uncertainties. The apparent excess in $\rm DM_{IGM}$ for some FRBs in Figure~\ref{Fig5}, even after subtracting the estimated host galaxy contribution, may be attributed to the influence of such intervening matter. 

Additionally, some FRBs are found to reside in dynamic magneto-ionic environments, where the circumburst medium can also contribute significantly to the observed DM, commonly referred to as $\rm DM_{source}$ \citep{Yang2017, 86, 90, Piro2018,Yang2019, 88,Zhao2021, Wang2022, Anna-Thomas2023, Zhao2023, LiR2023, LiR2025}. In this work, we collectively denote all such contributions as $\rm DM_{extra}$, as defined in Equation (\ref{equ:1}).
However, $\rm DM_{extra}$ varies significantly across different FRB sources and cannot be treated as a simple constant offset. Although it is in principle possible to estimate these components using other pulse properties—such as the scattering timescale and rotation measure (RM)-the associated uncertainties are substantial, leading to additional systematic errors in our analysis. Since these contributions are uncorrelated with redshift or sSFR, they introduce considerable scatter in the correlations shown in Figure \ref{Fig3}. 

The gas density of a specific galaxy is evolving with time, which results in relations between sSFR, stellar mass and redshift \citep{Carilli2013, Tacconi2018, Popesso2023}. Nonetheless, the sSFR - $\rm DM_{exc}$ relation shown in Figure \ref{Fig1} fundamentally captures the relation between the host DM contribution at the time the FRB is observed and the ionized gas density of the galaxy at that epoch.

\begin{figure}[htbp]
\centering
\begin{minipage}{0.49\textwidth}
    \centering
    \includegraphics[width=\textwidth]{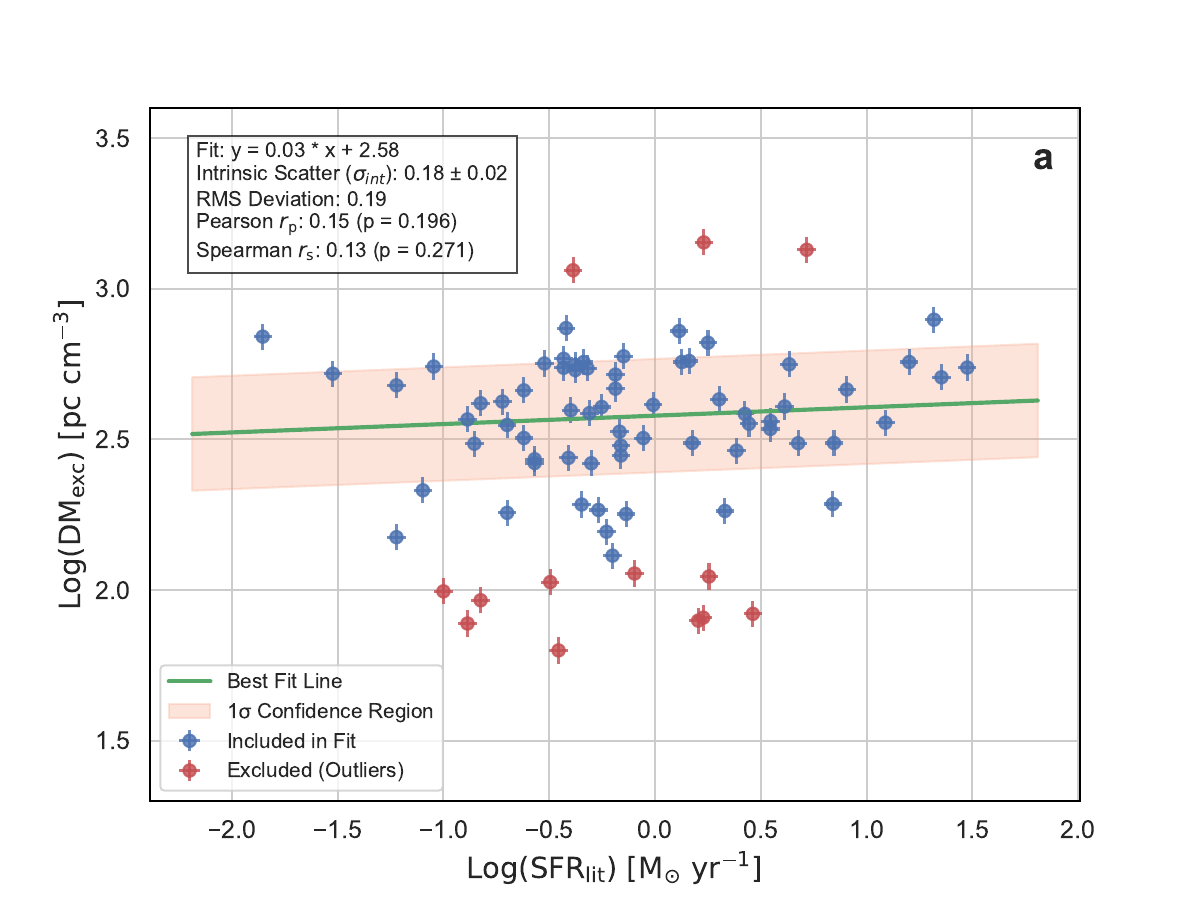}
\end{minipage}
\begin{minipage}{0.49\textwidth}
    \centering
    \includegraphics[width=\textwidth]{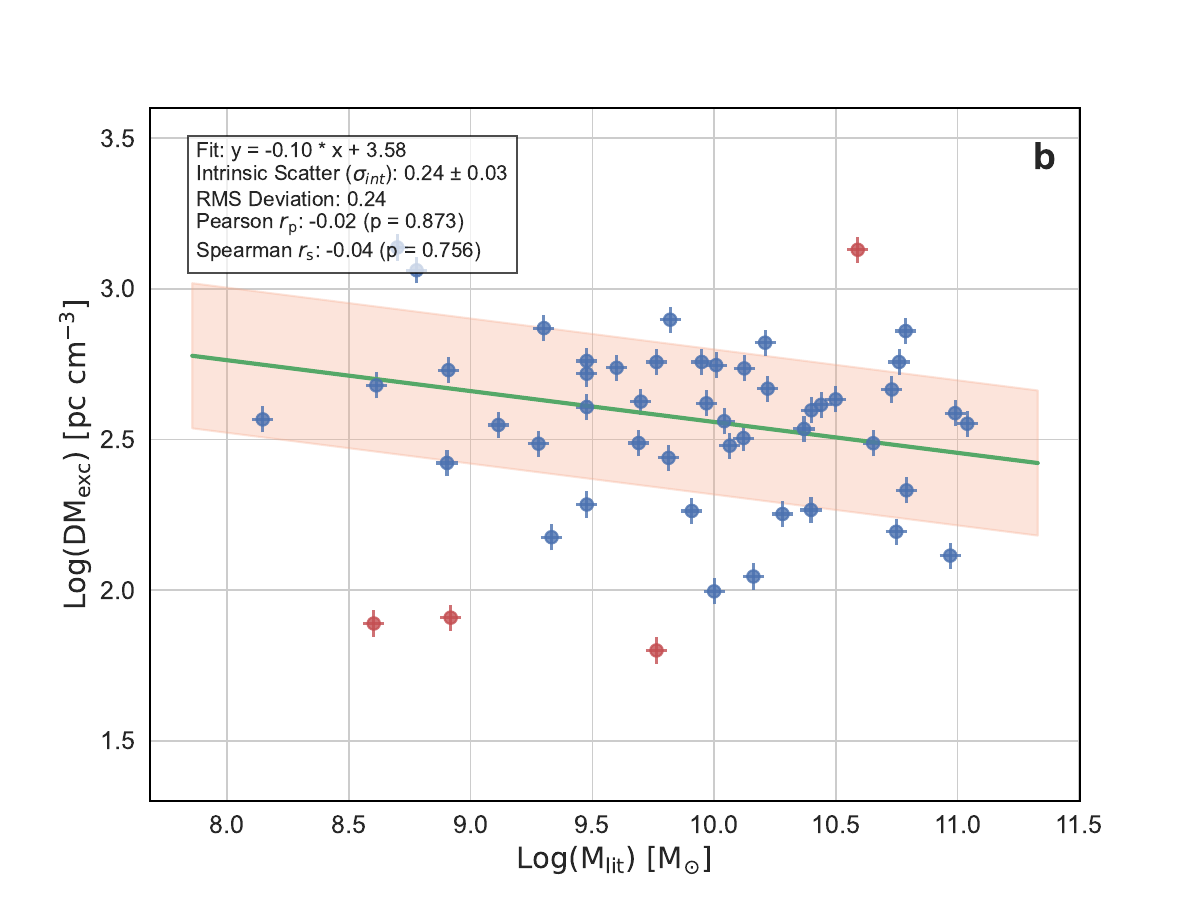}
\end{minipage}
\begin{minipage}{0.49\textwidth}
    \centering
    \includegraphics[width=\textwidth]{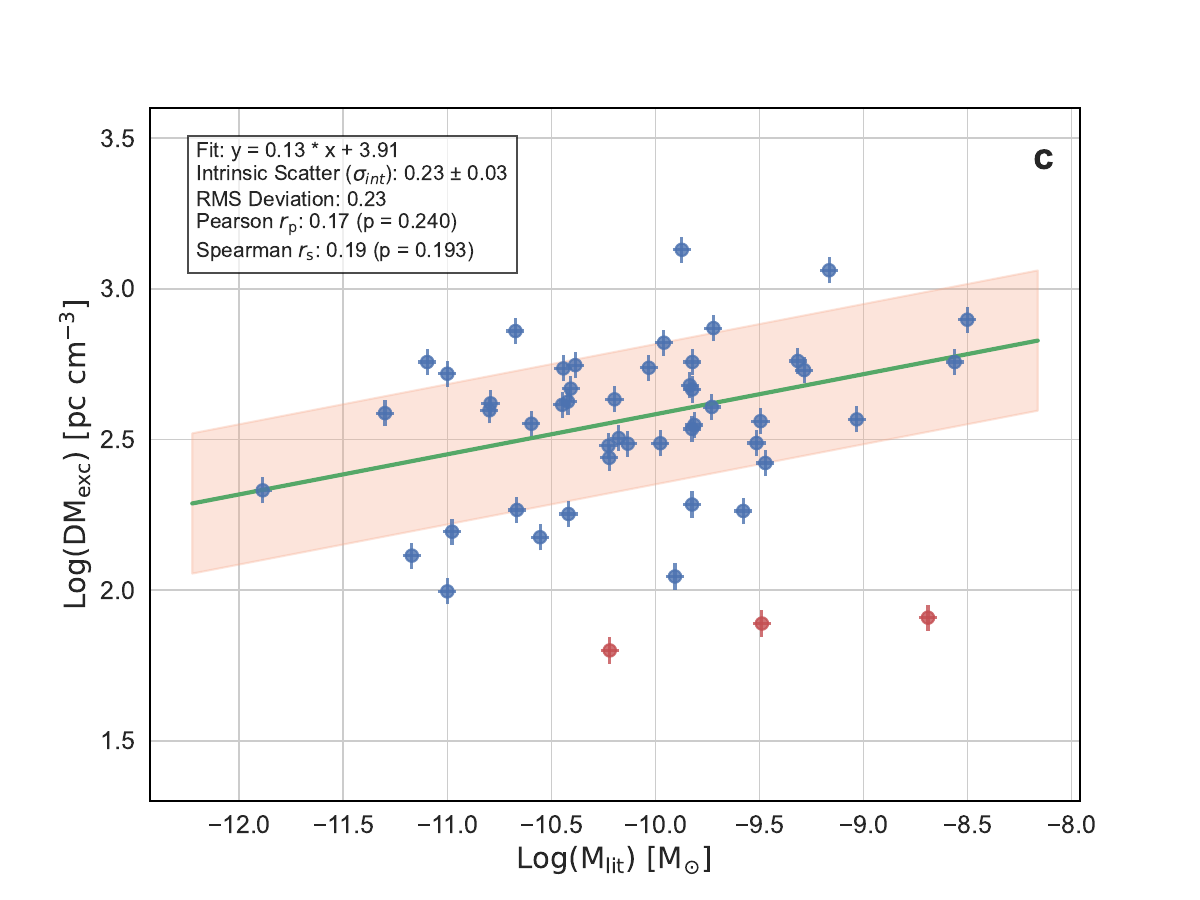}
\end{minipage}

\caption{Panel (a): Correlation fitting between $\rm \log(SFR)$ and $\rm \log(DM_{exc})$ for 78 FRBs. Blue circles with error bars represent data points included in the fit, while red circles denote outliers excluded during the fitting process. The green solid line shows the best-fit linear relation, and the shaded area represents the $1\sigma$ confidence interval based on the RMS deviation.
The fitted parameters are: $a = 0.03\pm0.03$, $b = 2.58\pm0.03$.
Panel (b): Same as the panel (a), but for $\rm \log(M)$ versus $\rm \log(DM_{exc})$ using 32 FRBs. 
The fitted parameters are: $a = -0.10\pm0.05$, $b = 3.58\pm0.51$.
Panel (c): Correlation between $\rm \log(sSFR)$ and $\rm \log(DM_{exc})$ for 50 FRBs. 
The fitted parameters are: $a = 0.13\pm0.05$, $b = 3.92\pm0.50$.
}
\label{Fig6}
\end{figure}

Although adopting a unified method to estimate the sSFR effectively mitigates systematic errors caused by methodological inconsistencies in statistical analyses, it is worth noting that some studies enhance the SFR measurements by incorporating additional follow-up observations, such as optical spectroscopy \citep{Sharma2024} or deep imaging data \citep{Bhandari2022}, in addition to publicly available photometric survey data. These efforts can improve the accuracy of individual SFR estimation. In this work, we compile SFR data for 78 FRB host galaxies and stellar mass data for 51 FRB host galaxies from the literature, as summarized in Table \ref{tab:data1}. The resulting correlation analysis is shown in Figure \ref{Fig6}. Compared to the unified-fitting sample, the correlations are significantly reduced. Part of the $\rm{}SFR_{lit}$ is calculated using $H_\alpha$ flux, which traces star formation activity over a timescale of $\sim10$ Myr. The degradation in correlation likely results from inconsistencies in the fitting parameters and models adopted by different SED-fitting software packages, introducing additional uncertainties into the analysis.

\begin{figure} 
\centering
\includegraphics[width=150 mm]{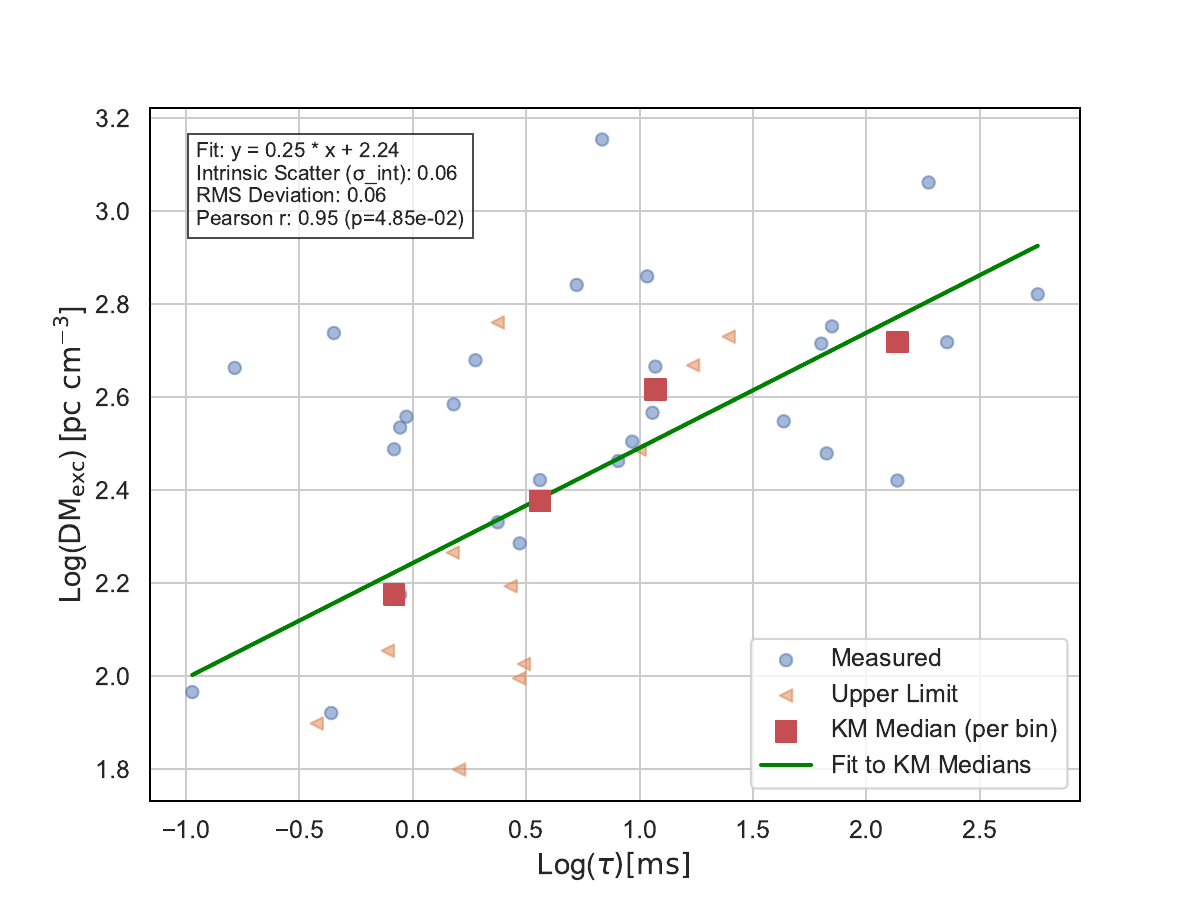}
\caption{The Kaplan–Meier estimator–based median fitting for $\tau_{\rm{}exc}$ and $\rm{}DM_{exc}$. The blue circles represent the measured values and orange inverted triangles indicate the upper limits. The data are binned along the x-axis. In each bin, the median of the y-values is estimated using the Kaplan–Meier estimator to account for upper-limit data. These binned medians are shown as red squares. The green solid line represents the linear fitting of the binned medians. The fitted parameters are $a = 0.25 \pm 0.06$ and $b = 2.24 \pm 0.07$.
}
\label{Fig7}
\end{figure}

The scattering timescale $\tau$ is thought to be primarily contributed by the host galaxy and may be proportional to the observed DM \citep{Cordes2022, Ocker2022}. However, no clear correlation between $\tau$ and $\rm DM_{obs}$ was found by \citet{Acharya2025}. In this work, we apply the Kaplan–Meier estimator–based median fitting method to robustly fit $\rm DM_{exc}$ against $\log(\tau_{\rm{}exc})$, where $\tau_{\rm{}exc}$ represents the scattering timescale with the Milky Way contribution subtracted. As shown in Figure \ref{Fig7}, a tight linear correlation between $\log(\tau_{\rm{}exc})$ and $\rm DM_{exc}$ is revealed, which is inconsistent with the conclusion of \citet{Acharya2025}. Given the limited number of scattering measurements for localized FRBs and the possibility that scattering timescales evolve over time \citep{Ocker2023}, caution should be exercised when using scattering to directly estimate $\rm DM_{host}$.

\begin{figure}[htbp]
\centering
\begin{minipage}{0.48\textwidth}
    \centering
    \includegraphics[width=\textwidth]{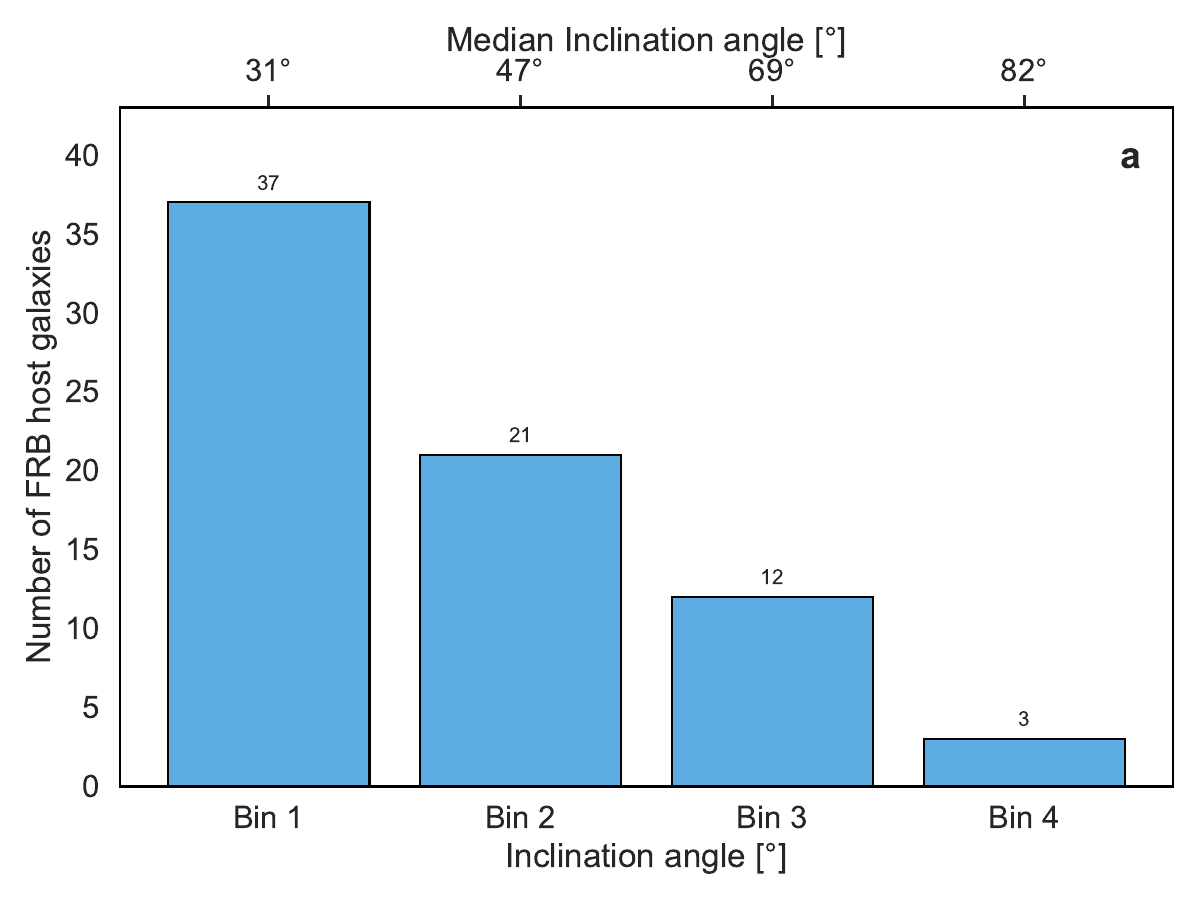}
\end{minipage}
\begin{minipage}{0.48\textwidth}
    \centering
    \includegraphics[width=\textwidth]{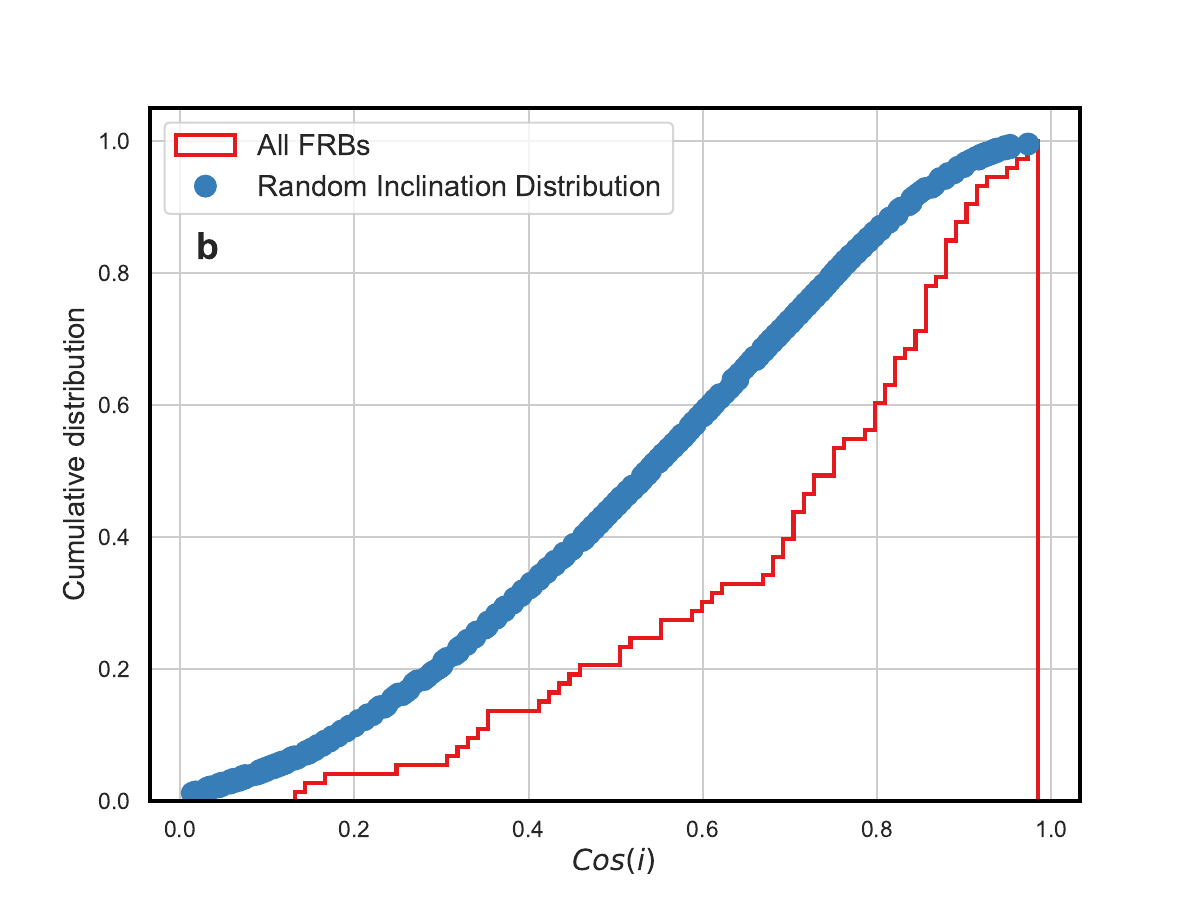}
\end{minipage}

\caption{Panel (a): Histogram of inclination angle of FRB host galaxies. We adopt the same binning strategy as used in \cite{Bhardwaj2024b}. 
Panel (b): CDF of inclination angles for FRB hosts and SDSS galaxies. The CDF 
of the cosine of the inclination angles for FRB host galaxies in our sample 
(solid red line) is compared with the CDF of randomly sampled disk-dominated 
late-type galaxies from SDSS (solid blue line) \citep{Bhardwaj2024b}.
}
\label{Fig8}
\end{figure}

Moreover, $\rm DM_{exc}$ is theoretically expected to increase with the host galaxy’s inclination angle and the projected offset of the FRB source \citep{Wada2007, Leonel2010, Ocker2020, Bhardwaj2024b}. However, as shown in panels (b) and (c) of Figure \ref{Fig2}, we find anti-correlations between them. This discrepancy may arise from selection effects. For FRBs located in highly inclined galaxies or close to the galactic center, only those with relatively low $\rm DM_{host}$ may be detectable due to observational biases. To verify whether such a selection bias exists in this sample, we present the distribution of inclination angles in Figure \ref{Fig8}, which closely matches the selection bias reported by \cite{Bhardwaj2024b}.


\section{Summary} \label{sec:con}
In this work, we investigated the correlations between various host galaxy parameters and $\rm DM_{exc}$ for a sample of localized FRBs. The SFRs, sSFRs and stellar masses for 76 FRBs are derived from unified SED fitting results, while inclination angles for 73 FRBs are obtained through Sérsic model fitting of host galaxy images. In addition, the offsets and galaxy areas are compiled from previously published data. 
As shown in Section \ref{sec:correlation analysis}, we found a tight correlation between the sSFR of host galaxies and $\rm DM_{exc}$. This result is consistent with the assumption that host galaxies with higher sSFRs are expected to contribute larger $\rm DM_{host}$ value.

The DMs from the MW halo, intervening halos and the circumburst medium are not expected to correlate with the sSFR of the host galaxy. Therefore, the derived $\rm sSFR - \rm DM_{exc}$ correlation can be used to calibrate the $\rm DM_{IGM}-z$ relation. Assuming a linear positive dependence of $\rm DM_{IGM}$ on redshift $z$, we derived Equation (\ref{equ:7}) to directly fit $\rm DM_{exc}$ as a function of both sSFR and $z$. By subtracting the DM halo and extragalactic contributions from the total DM, the value of $\rm DM_{IGM}$ can be obtained. After this correction, the MSEs are significantly reduced across all three models, corresponding to relative improvements of $69.21\%$, $76.82\%$, and $78.88\%$, respectively. Furthermore, the fraction of FRBs falling within the $95\%$ confidence interval of the $\rm DM_{IGM}$-$z$ relation from the TNG300 simulation increases from $50.91\%$ to $72.73\%$ after the correction.

As discussed in Section \ref{sec:dis}, the variation in $\rm DM_{extra}$ among different FRBs, along with several overestimated sSFR values in our sample, leads to non-negligible scatter in the $\rm DM_{exc}$–sSFR–$z$ fitting. We also found a weaker correlation with $\rm DM_{exc}$ when using published SFR and stellar mass data, as the use of different methods for estimation and fitting introduces significant uncertainties and systematic biases. The scattering timescale is considered a property of the host galaxy and is found to correlate strongly with $\rm DM_{exc}$, as expected by \citet{Cordes2022,Ocker2022}. Furthermore, a comparison between our sample and that of \citet{Bhardwaj2024b} indicates that inclination-related selection bias remains present in a larger sample of localized FRBs.

The demand for more FRB localizations continues to rise as statistical studies of FRB host galaxies and progenitor populations become more extensive. High-resolution and in-depth observations of FRB host galaxies are expected to play a key role in uncovering the physical origins of FRBs and enabling their application as cosmological probes in future.

\section*{Acknowledgments}
We would like to thank the anonymous referee for helpful
comments. This work was supported by the National Natural Science Foundation of China (grant Nos. 12494575 and 12273009), the National
SKA Program of China (grant No. 2022SKA0130100), and Postgraduate Research \& Practice Innovation Program of Jiangsu Province (KYCX24\_0184).

\bibliographystyle{aasjournal}
\bibliography{ref1}

\begin{thebibliography}{}
\expandafter\ifx\csname natexlab\endcsname\relax\def\natexlab#1{#1}\fi
\providecommand{\url}[1]{\href{#1}{#1}}
\providecommand{\dodoi}[1]{doi:~\href{http://doi.org/#1}{\nolinkurl{#1}}}
\providecommand{\doeprint}[1]{\href{http://ascl.net/#1}{\nolinkurl{http://ascl.net/#1}}}
\providecommand{\doarXiv}[1]{\href{https://arxiv.org/abs/#1}{\nolinkurl{https://arxiv.org/abs/#1}}}

\bibitem[{{Acharya} \& {Beniamini}(2025)}]{Acharya2025}
{Acharya}, S.~K., \& {Beniamini}, P. 2025, arXiv e-prints, arXiv:2503.08441,
  \dodoi{10.48550/arXiv.2503.08441}

\bibitem[{{Alam} {et~al.}(2015){Alam}, {Albareti}, {Allende Prieto}, {Anders},
  {Anderson}, {Anderton}, {Andrews}, {Armengaud}, {Aubourg}, {Bailey}, {Basu},
  {Bautista}, {Beaton}, {Beers}, {Bender}, {Berlind}, {Beutler}, {Bhardwaj},
  {Bird}, {Bizyaev}, {Blake}, {Blanton}, {Blomqvist}, {Bochanski}, {Bolton},
  {Bovy}, {Shelden Bradley}, {Brandt}, {Brauer}, {Brinkmann}, {Brown},
  {Brownstein}, {Burden}, {Burtin}, {Busca}, {Cai}, {Capozzi}, {Carnero
  Rosell}, {Carr}, {Carrera}, {Chambers}, {Chaplin}, {Chen}, {Chiappini},
  {Chojnowski}, {Chuang}, {Clerc}, {Comparat}, {Covey}, {Croft}, {Cuesta},
  {Cunha}, {da Costa}, {Da Rio}, {Davenport}, {Dawson}, {De Lee}, {Delubac},
  {Deshpande}, {Dhital}, {Dutra-Ferreira}, {Dwelly}, {Ealet}, {Ebelke},
  {Edmondson}, {Eisenstein}, {Ellsworth}, {Elsworth}, {Epstein}, {Eracleous},
  {Escoffier}, {Esposito}, {Evans}, {Fan}, {Fern{\'a}ndez-Alvar}, {Feuillet},
  {Filiz Ak}, {Finley}, {Finoguenov}, {Flaherty}, {Fleming}, {Font-Ribera},
  {Foster}, {Frinchaboy}, {Galbraith-Frew}, {Garc{\'\i}a},
  {Garc{\'\i}a-Hern{\'a}ndez}, {Garc{\'\i}a P{\'e}rez}, {Gaulme}, {Ge},
  {G{\'e}nova-Santos}, {Georgakakis}, {Ghezzi}, {Gillespie}, {Girardi},
  {Goddard}, {Gontcho}, {Gonz{\'a}lez Hern{\'a}ndez}, {Grebel}, {Green},
  {Grieb}, {Grieves}, {Gunn}, {Guo}, {Harding}, {Hasselquist}, {Hawley},
  {Hayden}, {Hearty}, {Hekker}, {Ho}, {Hogg}, {Holley-Bockelmann}, {Holtzman},
  {Honscheid}, {Huber}, {Huehnerhoff}, {Ivans}, {Jiang}, {Johnson},
  {Kinemuchi}, {Kirkby}, {Kitaura}, {Klaene}, {Knapp}, {Kneib}, {Koenig},
  {Lam}, {Lan}, {Lang}, {Laurent}, {Le Goff}, {Leauthaud}, {Lee}, {Lee},
  {Licquia}, {Liu}, {Long}, {L{\'o}pez-Corredoira}, {Lorenzo-Oliveira},
  {Lucatello}, {Lundgren}, {Lupton}, {Mack}, {Mahadevan}, {Maia}, {Majewski},
  {Malanushenko}, {Malanushenko}, {Manchado}, {Manera}, {Mao}, {Maraston},
  {Marchwinski}, {Margala}, {Martell}, {Martig}, {Masters}, {Mathur},
  {McBride}, {McGehee}, {McGreer}, {McMahon}, {M{\'e}nard}, {Menzel},
  {Merloni}, {M{\'e}sz{\'a}ros}, {Miller}, {Miralda-Escud{\'e}}, {Miyatake},
  {Montero-Dorta}, {More}, {Morganson}, {Morice-Atkinson}, {Morrison},
  {Mosser}, {Muna}, {Myers}, {Nandra}, {Newman}, {Neyrinck}, {Nguyen},
  {Nichol}, {Nidever}, {Noterdaeme}, {Nuza}, {O'Connell}, {O'Connell},
  {O'Connell}, {Ogando}, {Olmstead}, {Oravetz}, {Oravetz}, {Osumi}, {Owen},
  {Padgett}, {Padmanabhan}, {Paegert}, {Palanque-Delabrouille}, \&
  {Pan}}]{SDSS}
{Alam}, S., {Albareti}, F.~D., {Allende Prieto}, C., {et~al.} 2015, \apjs, 219,
  12, \dodoi{10.1088/0067-0049/219/1/12}

\bibitem[{{Amiri} {et~al.}(2025){Amiri}, {Amouyal}, {Andersen}, {Andrew},
  {Bandura}, {Bhardwaj}, {Boyle}, {Brar}, {Cassity}, {Chatterjee}, {Curtin},
  {Dobbs}, {Dong}, {Dong}, {Eadie}, {Eftekhari}, {Fong}, {Fonseca}, {Gaensler},
  {Halpern}, {Hessels}, {Hopkins}, {Ibik}, {Joseph}, {Kaczmarek}, {Kahinga},
  {Kaspi}, {Khairy}, {Kilpatrick}, {Lanman}, {Lazda}, {Leung}, {Main},
  {Mas-Ribas}, {Masui}, {Mckinven}, {Mena-Parra}, {Meyers}, {Michilli},
  {Milutinovic}, {Nimmo}, {Noble}, {Pandhi}, {Shivraj Patil}, {Pearlman},
  {Petroff}, {Pleunis}, {Prochaska}, {Rafiei-Ravandi}, {Rahman}, {Renard},
  {Sammons}, {Sand}, {Scholz}, {Shah}, {Shin}, {Siegel}, {Simha}, {Smith},
  {Stairs}, {Vanderlinde}, {Wang}, {Wulf}, \& {Zegmott}}]{Amiri2025}
{Amiri}, M., {Amouyal}, D., {Andersen}, B.~C., {et~al.} 2025, arXiv e-prints,
  arXiv:2502.11217, \dodoi{10.48550/arXiv.2502.11217}

\bibitem[{{Anna-Thomas} {et~al.}(2023){Anna-Thomas}, {Connor}, {Dai}, {Feng},
  {Burke-Spolaor}, {Beniamini}, {Yang}, {Zhang}, {Aggarwal}, {Law}, {Li},
  {Niu}, {Chatterjee}, {Cruces}, {Duan}, {Filipovic}, {Hobbs}, {Lynch}, {Miao},
  {Niu}, {Ocker}, {Tsai}, {Wang}, {Xue}, {Yao}, {Yu}, {Zhang}, {Zhang}, {Zhu},
  \& {Zhu}}]{Anna-Thomas2023}
{Anna-Thomas}, R., {Connor}, L., {Dai}, S., {et~al.} 2023, Science, 380, 599,
  \dodoi{10.1126/science.abo6526}

\bibitem[{{Anna-Thomas} {et~al.}(2025){Anna-Thomas}, {Law}, {Koch}, {Gordon},
  {Sharma}, {Williams}, {Pingel}, {Burke-Spolaor}, {Chen}, {Stanley}, {Dear},
  {Verdi}, {Prochaska}, {Bower}, {Chomiuk}, {Connor}, {Demorest}, {Nugent}, \&
  {Walter}}]{Anna-Thomas2025}
{Anna-Thomas}, R., {Law}, C.~J., {Koch}, E.~W., {et~al.} 2025, arXiv e-prints,
  arXiv:2503.02947, \dodoi{10.48550/arXiv.2503.02947}

\bibitem[{{Bannister} {et~al.}(2019){Bannister}, {Deller}, {Phillips},
  {Macquart}, {Prochaska}, {Tejos}, {Ryder}, {Sadler}, {Shannon}, {Simha},
  {Day}, {McQuinn}, {North-Hickey}, {Bhandari}, {Arcus}, {Bennert}, {Burchett},
  {Bouwhuis}, {Dodson}, {Ekers}, {Farah}, {Flynn}, {James}, {Kerr}, {Lenc},
  {Mahony}, {O'Meara}, {Os{\l}owski}, {Qiu}, {Treu}, {U}, {Bateman}, {Bock},
  {Bolton}, {Brown}, {Bunton}, {Chippendale}, {Cooray}, {Cornwell}, {Gupta},
  {Hayman}, {Kesteven}, {Koribalski}, {MacLeod}, {McClure-Griffiths},
  {Neuhold}, {Norris}, {Pilawa}, {Qiao}, {Reynolds}, {Roxby}, {Shimwell},
  {Voronkov}, \& {Wilson}}]{Bannister2019}
{Bannister}, K.~W., {Deller}, A.~T., {Phillips}, C., {et~al.} 2019, Science,
  365, 565, \dodoi{10.1126/science.aaw5903}

\bibitem[{{Batten} {et~al.}(2021){Batten}, {Duffy}, {Wijers}, {Gupta}, {Flynn},
  {Schaye}, \& {Ryan-Weber}}]{Batten2021}
{Batten}, A.~J., {Duffy}, A.~R., {Wijers}, N.~A., {et~al.} 2021, \mnras, 505,
  5356, \dodoi{10.1093/mnras/stab1528}

\bibitem[{{Beniamini} {et~al.}(2021){Beniamini}, {Kumar}, {Ma}, \&
  {Quataert}}]{Beniamini2021}
{Beniamini}, P., {Kumar}, P., {Ma}, X., \& {Quataert}, E. 2021, \mnras, 502,
  5134, \dodoi{10.1093/mnras/stab309}

\bibitem[{{Bhandari} {et~al.}(2020){Bhandari}, {Sadler}, {Prochaska}, {Simha},
  {Ryder}, {Marnoch}, {Bannister}, {Macquart}, {Flynn}, {Shannon}, {Tejos},
  {Corro-Guerra}, {Day}, {Deller}, {Ekers}, {Lopez}, {Mahony}, {Nu{\~n}ez}, \&
  {Phillips}}]{Bhandari2020}
{Bhandari}, S., {Sadler}, E.~M., {Prochaska}, J.~X., {et~al.} 2020, \apjl, 895,
  L37, \dodoi{10.3847/2041-8213/ab672e}

\bibitem[{{Bhandari} {et~al.}(2022){Bhandari}, {Heintz}, {Aggarwal}, {Marnoch},
  {Day}, {Sydnor}, {Burke-Spolaor}, {Law}, {Xavier Prochaska}, {Tejos},
  {Bannister}, {Butler}, {Deller}, {Ekers}, {Flynn}, {Fong}, {James}, {Lazio},
  {Luo}, {Mahony}, {Ryder}, {Sadler}, {Shannon}, {Han}, {Lee}, \&
  {Zhang}}]{Bhandari2022}
{Bhandari}, S., {Heintz}, K.~E., {Aggarwal}, K., {et~al.} 2022, \aj, 163, 69,
  \dodoi{10.3847/1538-3881/ac3aec}

\bibitem[{{Bhardwaj} {et~al.}(2024{\natexlab{a}}){Bhardwaj}, {Lee}, \&
  {Ji}}]{Bhardwaj2024b}
{Bhardwaj}, M., {Lee}, J., \& {Ji}, K. 2024{\natexlab{a}}, \nat, 634, 1065,
  \dodoi{10.1038/s41586-024-08065-w}

\bibitem[{{Bhardwaj} {et~al.}(2021{\natexlab{a}}){Bhardwaj}, {Kirichenko},
  {Michilli}, {Mayya}, {Kaspi}, {Gaensler}, {Rahman}, {Tendulkar}, {Fonseca},
  {Josephy}, {Leung}, {Merryfield}, {Petroff}, {Pleunis}, {Sanghavi}, {Scholz},
  {Shin}, {Smith}, \& {Stairs}}]{Bhardwaj2021}
{Bhardwaj}, M., {Kirichenko}, A.~Y., {Michilli}, D., {et~al.}
  2021{\natexlab{a}}, \apjl, 919, L24, \dodoi{10.3847/2041-8213/ac223b}

\bibitem[{{Bhardwaj} {et~al.}(2021{\natexlab{b}}){Bhardwaj}, {Gaensler},
  {Kaspi}, {Landecker}, {Mckinven}, {Michilli}, {Pleunis}, {Tendulkar},
  {Andersen}, {Boyle}, {Cassanelli}, {Chawla}, {Cook}, {Dobbs}, {Fonseca},
  {Kaczmarek}, {Leung}, {Masui}, {Mnchmeyer}, {Ng}, {Rafiei-Ravandi}, {Scholz},
  {Shin}, {Smith}, {Stairs}, \& {Zwaniga}}]{Bhardwaj2021b}
{Bhardwaj}, M., {Gaensler}, B.~M., {Kaspi}, V.~M., {et~al.} 2021{\natexlab{b}},
  \apjl, 910, L18, \dodoi{10.3847/2041-8213/abeaa6}

\bibitem[{{Bhardwaj} {et~al.}(2024{\natexlab{b}}){Bhardwaj}, {Michilli},
  {Kirichenko}, {Modilim}, {Shin}, {Kaspi}, {Andersen}, {Cassanelli}, {Brar},
  {Chatterjee}, {Cook}, {Dong}, {Fonseca}, {Gaensler}, {Ibik}, {Kaczmarek},
  {Lanman}, {Leung}, {Masui}, {Pandhi}, {Pearlman}, {Petroff}, {Pleunis},
  {Prochaska}, {Rafiei-Ravandi}, {Sand}, {Scholz}, \& {Smith}}]{Bhardwaj2024}
{Bhardwaj}, M., {Michilli}, D., {Kirichenko}, A.~Y., {et~al.}
  2024{\natexlab{b}}, \apjl, 971, L51, \dodoi{10.3847/2041-8213/ad64d1}

\bibitem[{{Bruzual} \& {Charlot}(2003)}]{Bruzual2003}
{Bruzual}, G., \& {Charlot}, S. 2003, \mnras, 344, 1000,
  \dodoi{10.1046/j.1365-8711.2003.06897.x}

\bibitem[{{Byler} {et~al.}(2017){Byler}, {Dalcanton}, {Conroy}, \&
  {Johnson}}]{Byler2017}
{Byler}, N., {Dalcanton}, J.~J., {Conroy}, C., \& {Johnson}, B.~D. 2017, \apj,
  840, 44, \dodoi{10.3847/1538-4357/aa6c66}

\bibitem[{{Caleb} {et~al.}(2023){Caleb}, {Driessen}, {Gordon}, {Tejos},
  {Bernales}, {Qiu}, {Chibueze}, {Stappers}, {Rajwade}, {Cavallaro}, {Wang},
  {Kumar}, {Majid}, {Wharton}, {Naudet}, {Bezuidenhout}, {Jankowski},
  {Malenta}, {Morello}, {Sanidas}, {Surnis}, {Barr}, {Chen}, {Kramer}, {Fong},
  {Kilpatrick}, {Prochaska}, {Simha}, {Venter}, {Heywood}, {Kundu}, \&
  {Schussler}}]{Caleb2023}
{Caleb}, M., {Driessen}, L.~N., {Gordon}, A.~C., {et~al.} 2023, \mnras, 524,
  2064, \dodoi{10.1093/mnras/stad1839}

\bibitem[{{Calzetti} {et~al.}(2000){Calzetti}, {Armus}, {Bohlin}, {Kinney},
  {Koornneef}, \& {Storchi-Bergmann}}]{Calzetti2020}
{Calzetti}, D., {Armus}, L., {Bohlin}, R.~C., {et~al.} 2000, \apj, 533, 682,
  \dodoi{10.1086/308692}

\bibitem[{{Cappellari}(2014)}]{Cappellari2014}
{Cappellari}, M. 2014, {LTS\_LINEFIT \& LTS\_PLANEFIT: LTS fit of lines or
  planes}, Astrophysics Source Code Library, record ascl:1404.001

\bibitem[{{Cappellari} {et~al.}(2013){Cappellari}, {Scott}, {Alatalo}, {Blitz},
  {Bois}, {Bournaud}, {Bureau}, {Crocker}, {Davies}, {Davis}, {de Zeeuw},
  {Duc}, {Emsellem}, {Khochfar}, {Krajnovi{\'c}}, {Kuntschner}, {McDermid},
  {Morganti}, {Naab}, {Oosterloo}, {Sarzi}, {Serra}, {Weijmans}, \&
  {Young}}]{Cappellari2013}
{Cappellari}, M., {Scott}, N., {Alatalo}, K., {et~al.} 2013, \mnras, 432, 1709,
  \dodoi{10.1093/mnras/stt562}

\bibitem[{{Cardelli} {et~al.}(1989){Cardelli}, {Clayton}, \&
  {Mathis}}]{Cardelli1989}
{Cardelli}, J.~A., {Clayton}, G.~C., \& {Mathis}, J.~S. 1989, \apj, 345, 245,
  \dodoi{10.1086/167900}

\bibitem[{{Carilli} \& {Walter}(2013)}]{Carilli2013}
{Carilli}, C.~L., \& {Walter}, F. 2013, \araa, 51, 105,
  \dodoi{10.1146/annurev-astro-082812-140953}

\bibitem[{{Carnall} {et~al.}(2018){Carnall}, {McLure}, {Dunlop}, \&
  {Dav{\'e}}}]{Carnall2018}
{Carnall}, A.~C., {McLure}, R.~J., {Dunlop}, J.~S., \& {Dav{\'e}}, R. 2018,
  \mnras, 480, 4379, \dodoi{10.1093/mnras/sty2169}

\bibitem[{{Cassanelli} {et~al.}(2024){Cassanelli}, {Leung}, {Sanghavi},
  {Mena-Parra}, {Cary}, {Mckinven}, {Bhardwaj}, {Masui}, {Michilli}, {Bandura},
  {Chatterjee}, {Peterson}, {Kaczmarek}, {Rahman}, {Shin}, {Vanderlinde},
  {Berger}, {Brar}, {Boyle}, {Breitman}, {Chawla}, {Curtin}, {Dobbs}, {Dong},
  {Fonseca}, {Gaensler}, {Ibik}, {Kaspi}, {Khairy}, {Lanman}, {Lazda}, {Lin},
  {Luo}, {Meyers}, {Milutinovic}, {Ng}, {Noble}, {Pearlman}, {Pen}, {Petroff},
  {Pleunis}, {Quine}, {Rafiei-Ravandi}, {Renard}, {Sand}, {Schoen}, {Scholz},
  {Smith}, {Stairs}, \& {Tendulkar}}]{Cassanelli2024}
{Cassanelli}, T., {Leung}, C., {Sanghavi}, P., {et~al.} 2024, Nature Astronomy,
  8, 1429, \dodoi{10.1038/s41550-024-02357-x}

\bibitem[{{Chambers} {et~al.}(2016){Chambers}, {Magnier}, {Metcalfe},
  {Flewelling}, {Huber}, {Waters}, {Denneau}, {Draper}, {Farrow}, {Finkbeiner},
  {Holmberg}, {Koppenhoefer}, {Price}, {Rest}, {Saglia}, {Schlafly}, {Smartt},
  {Sweeney}, {Wainscoat}, {Burgett}, {Chastel}, {Grav}, {Heasley}, {Hodapp},
  {Jedicke}, {Kaiser}, {Kudritzki}, {Luppino}, {Lupton}, {Monet}, {Morgan},
  {Onaka}, {Shiao}, {Stubbs}, {Tonry}, {White}, {Ba{\~n}ados}, {Bell},
  {Bender}, {Bernard}, {Boegner}, {Boffi}, {Botticella}, {Calamida},
  {Casertano}, {Chen}, {Chen}, {Cole}, {Deacon}, {Frenk}, {Fitzsimmons},
  {Gezari}, {Gibbs}, {Goessl}, {Goggia}, {Gourgue}, {Goldman}, {Grant},
  {Grebel}, {Hambly}, {Hasinger}, {Heavens}, {Heckman}, {Henderson}, {Henning},
  {Holman}, {Hopp}, {Ip}, {Isani}, {Jackson}, {Keyes}, {Koekemoer}, {Kotak},
  {Le}, {Liska}, {Long}, {Lucey}, {Liu}, {Martin}, {Masci}, {McLean}, {Mindel},
  {Misra}, {Morganson}, {Murphy}, {Obaika}, {Narayan}, {Nieto-Santisteban},
  {Norberg}, {Peacock}, {Pier}, {Postman}, {Primak}, {Rae}, {Rai}, {Riess},
  {Riffeser}, {Rix}, {R{\"o}ser}, {Russel}, {Rutz}, {Schilbach}, {Schultz},
  {Scolnic}, {Strolger}, {Szalay}, {Seitz}, {Small}, {Smith}, {Soderblom},
  {Taylor}, {Thomson}, {Taylor}, {Thakar}, {Thiel}, {Thilker}, {Unger},
  {Urata}, {Valenti}, {Wagner}, {Walder}, {Walter}, {Watters}, {Werner},
  {Wood-Vasey}, \& {Wyse}}]{Panstarr}
{Chambers}, K.~C., {Magnier}, E.~A., {Metcalfe}, N., {et~al.} 2016, arXiv
  e-prints, arXiv:1612.05560, \dodoi{10.48550/arXiv.1612.05560}

\bibitem[{{Champati} \& {Petrosian}(2025)}]{Champati2025}
{Champati}, S., \& {Petrosian}, V. 2025, arXiv e-prints, arXiv:2504.13343,
  \dodoi{10.48550/arXiv.2504.13343}

\bibitem[{{Chatterjee} {et~al.}(2017){Chatterjee}, {Law}, {Wharton},
  {Burke-Spolaor}, {Hessels}, {Bower}, {Cordes}, {Tendulkar}, {Bassa},
  {Demorest}, {Butler}, {Seymour}, {Scholz}, {Abruzzo}, {Bogdanov}, {Kaspi},
  {Keimpema}, {Lazio}, {Marcote}, {McLaughlin}, {Paragi}, {Ransom}, {Rupen},
  {Spitler}, \& {van Langevelde}}]{Chatterjee2017}
{Chatterjee}, S., {Law}, C.~J., {Wharton}, R.~S., {et~al.} 2017, \nat, 541, 58,
  \dodoi{10.1038/nature20797}

\bibitem[{{Chawla} {et~al.}(2022){Chawla}, {Kaspi}, {Ransom}, {Bhardwaj},
  {Boyle}, {Breitman}, {Cassanelli}, {Cubranic}, {Dong}, {Fonseca}, {Gaensler},
  {Giri}, {Josephy}, {Kaczmarek}, {Leung}, {Masui}, {Mena-Parra}, {Merryfield},
  {Michilli}, {M{\"u}nchmeyer}, {Ng}, {Patel}, {Pearlman}, {Petroff},
  {Pleunis}, {Rahman}, {Sanghavi}, {Shin}, {Smith}, {Stairs}, \&
  {Tendulkar}}]{Chawla2022}
{Chawla}, P., {Kaspi}, V.~M., {Ransom}, S.~M., {et~al.} 2022, \apj, 927, 35,
  \dodoi{10.3847/1538-4357/ac49e1}

\bibitem[{{Chen} {et~al.}(2024){Chen}, {Jia}, {Dong}, \& {Wang}}]{Chen2024}
{Chen}, J.~H., {Jia}, X.~D., {Dong}, X.~F., \& {Wang}, F.~Y. 2024, \apjl, 973,
  L54, \dodoi{10.3847/2041-8213/ad7b39}

\bibitem[{{Chen} {et~al.}(2025){Chen}, {Tsai}, {Li}, {Wang}, {Feng}, {Zhang},
  {Li}, {Zhang}, {Bao}, {Liao}, {Zhang}, {Zuo}, {Bao}, {Niu}, {Luo}, {Zhu},
  {Zou}, {Xue}, \& {Zhang}}]{Chen2025}
{Chen}, X.-L., {Tsai}, C.-W., {Li}, D., {et~al.} 2025, \apjl, 980, L24,
  \dodoi{10.3847/2041-8213/adadfd}

\bibitem[{{Chittidi} {et~al.}(2021){Chittidi}, {Simha}, {Mannings},
  {Prochaska}, {Ryder}, {Rafelski}, {Neeleman}, {Macquart}, {Tejos},
  {Jorgenson}, {Day}, {Marnoch}, {Bhandari}, {Deller}, {Qiu}, {Bannister},
  {Shannon}, \& {Heintz}}]{Chittidi2021}
{Chittidi}, J.~S., {Simha}, S., {Mannings}, A., {et~al.} 2021, \apj, 922, 173,
  \dodoi{10.3847/1538-4357/ac2818}

\bibitem[{{Connor} \& {Ravi}(2022)}]{Connor2022}
{Connor}, L., \& {Ravi}, V. 2022, Nature Astronomy, 6, 1035,
  \dodoi{10.1038/s41550-022-01719-7}

\bibitem[{{Connor} {et~al.}(2024){Connor}, {Ravi}, {Sharma}, {Ocker}, {Faber},
  {Hallinan}, {Harnach}, {Hellbourg}, {Hobbs}, {Hodge}, {Hodges}, {Kosogorov},
  {Lamb}, {Law}, {Rasmussen}, {Sherman}, {Somalwar}, {Weinreb}, \&
  {Woody}}]{Connor2024}
{Connor}, L., {Ravi}, V., {Sharma}, K., {et~al.} 2024, arXiv e-prints,
  arXiv:2409.16952, \dodoi{10.48550/arXiv.2409.16952}

\bibitem[{{Cooke} {et~al.}(2018){Cooke}, {Pettini}, \& {Steidel}}]{Cooke2018}
{Cooke}, R.~J., {Pettini}, M., \& {Steidel}, C.~C. 2018, \apj, 855, 102,
  \dodoi{10.3847/1538-4357/aaab53}

\bibitem[{{Cordes} \& {Lazio}(2002)}]{ne2001}
{Cordes}, J.~M., \& {Lazio}, T.~J.~W. 2002, arXiv e-prints, astro,
  \dodoi{10.48550/arXiv.astro-ph/0207156}

\bibitem[{{Cordes} {et~al.}(2022){Cordes}, {Ocker}, \&
  {Chatterjee}}]{Cordes2022}
{Cordes}, J.~M., {Ocker}, S.~K., \& {Chatterjee}, S. 2022, \apj, 931, 88,
  \dodoi{10.3847/1538-4357/ac6873}

\bibitem[{{Cordes} {et~al.}(2016){Cordes}, {Wharton}, {Spitler}, {Chatterjee},
  \& {Wasserman}}]{Cordes2016}
{Cordes}, J.~M., {Wharton}, R.~S., {Spitler}, L.~G., {Chatterjee}, S., \&
  {Wasserman}, I. 2016, arXiv e-prints, arXiv:1605.05890,
  \dodoi{10.48550/arXiv.1605.05890}

\bibitem[{{Cutri} {et~al.}(2021){Cutri}, {Wright}, {Conrow}, {Fowler},
  {Eisenhardt}, {Grillmair}, {Kirkpatrick}, {Masci}, {McCallon}, {Wheelock},
  {Fajardo-Acosta}, {Yan}, {Benford}, {Harbut}, {Jarrett}, {Lake}, {Leisawitz},
  {Ressler}, {Stanford}, {Tsai}, {Liu}, {Helou}, {Mainzer}, {Gettngs},
  {Gonzalez}, {Hoffman}, {Marsh}, {Padgett}, {Skrutskie}, {Beck}, {Papin}, \&
  {Wittman}}]{WISE}
{Cutri}, R.~M., {Wright}, E.~L., {Conrow}, T., {et~al.} 2021, {VizieR Online
  Data Catalog: AllWISE Data Release (Cutri+ 2013)}, VizieR On-line Data
  Catalog: II/328. Originally published in: IPAC/Caltech (2013)

\bibitem[{{Deng} \& {Zhang}(2014)}]{Deng2014}
{Deng}, W., \& {Zhang}, B. 2014, \apjl, 783, L35,
  \dodoi{10.1088/2041-8205/783/2/L35}

\bibitem[{{DESI Collaboration} {et~al.}(2016){DESI Collaboration}, {Aghamousa},
  {Aguilar}, {Ahlen}, {Alam}, {Allen}, {Allende Prieto}, {Annis}, {Bailey},
  {Balland}, {Ballester}, {Baltay}, {Beaufore}, {Bebek}, {Beers}, {Bell},
  {Bernal}, {Besuner}, {Beutler}, {Blake}, {Bleuler}, {Blomqvist}, {Blum},
  {Bolton}, {Briceno}, {Brooks}, {Brownstein}, {Buckley-Geer}, {Burden},
  {Burtin}, {Busca}, {Cahn}, {Cai}, {Cardiel-Sas}, {Carlberg}, {Carton},
  {Casas}, {Castander}, {Cervantes-Cota}, {Claybaugh}, {Close}, {Coker},
  {Cole}, {Comparat}, {Cooper}, {Cousinou}, {Crocce}, {Cuby}, {Cunningham},
  {Davis}, {Dawson}, {de la Macorra}, {De Vicente}, {Delubac}, {Derwent},
  {Dey}, {Dhungana}, {Ding}, {Doel}, {Duan}, {Ealet}, {Edelstein},
  {Eftekharzadeh}, {Eisenstein}, {Elliott}, {Escoffier}, {Evatt}, {Fagrelius},
  {Fan}, {Fanning}, {Farahi}, {Farihi}, {Favole}, {Feng}, {Fernandez},
  {Findlay}, {Finkbeiner}, {Fitzpatrick}, {Flaugher}, {Flender}, {Font-Ribera},
  {Forero-Romero}, {Fosalba}, {Frenk}, {Fumagalli}, {Gaensicke}, {Gallo},
  {Garcia-Bellido}, {Gaztanaga}, {Pietro Gentile Fusillo}, {Gerard},
  {Gershkovich}, {Giannantonio}, {Gillet}, {Gonzalez-de-Rivera},
  {Gonzalez-Perez}, {Gott}, {Graur}, {Gutierrez}, {Guy}, {Habib}, {Heetderks},
  {Heetderks}, {Heitmann}, {Hellwing}, {Herrera}, {Ho}, {Holland}, {Honscheid},
  {Huff}, {Hutchinson}, {Huterer}, {Hwang}, {Illa Laguna}, {Ishikawa},
  {Jacobs}, {Jeffrey}, {Jelinsky}, {Jennings}, {Jiang}, {Jimenez}, {Johnson},
  {Joyce}, {Jullo}, {Juneau}, {Kama}, {Karcher}, {Karkar}, {Kehoe}, {Kennamer},
  {Kent}, {Kilbinger}, {Kim}, {Kirkby}, {Kisner}, {Kitanidis}, {Kneib},
  {Koposov}, {Kovacs}, {Koyama}, {Kremin}, {Kron}, {Kronig}, {Kueter-Young},
  {Lacey}, {Lafever}, {Lahav}, {Lambert}, {Lampton}, {Landriau}, {Lang},
  {Lauer}, {Le Goff}, {Le Guillou}, {Le Van Suu}, {Lee}, {Lee}, {Leitner},
  {Lesser}, {Levi}, {L'Huillier}, {Li}, {Liang}, {Lin}, {Linder}, {Loebman},
  {Luki{\'c}}, {Ma}, {MacCrann}, {Magneville}, {Makarem}, {Manera}, {Manser},
  {Marshall}, {Martini}, {Massey}, {Matheson}, {McCauley}, {McDonald},
  {McGreer}, {Meisner}, {Metcalfe}, {Miller}, {Miquel}, {Moustakas}, {Myers},
  {Naik}, {Newman}, {Nichol}, {Nicola}, {Nicolati da Costa}, {Nie}, {Niz},
  {Norberg}, {Nord}, {Norman}, {Nugent}, {O'Brien}, {Oh}, \& {Olsen}}]{DESI}
{DESI Collaboration}, {Aghamousa}, A., {Aguilar}, J., {et~al.} 2016, arXiv
  e-prints, arXiv:1611.00036, \dodoi{10.48550/arXiv.1611.00036}

\bibitem[{{Dey} {et~al.}(2019){Dey}, {Schlegel}, {Lang}, {Blum}, {Burleigh},
  {Fan}, {Findlay}, {Finkbeiner}, {Herrera}, {Juneau}, {Landriau}, {Levi},
  {McGreer}, {Meisner}, {Myers}, {Moustakas}, {Nugent}, {Patej}, {Schlafly},
  {Walker}, {Valdes}, {Weaver}, {Y{\`e}che}, {Zou}, {Zhou}, {Abareshi},
  {Abbott}, {Abolfathi}, {Aguilera}, {Alam}, {Allen}, {Alvarez}, {Annis},
  {Ansarinejad}, {Aubert}, {Beechert}, {Bell}, {BenZvi}, {Beutler}, {Bielby},
  {Bolton}, {Brice{\~n}o}, {Buckley-Geer}, {Butler}, {Calamida}, {Carlberg},
  {Carter}, {Casas}, {Castander}, {Choi}, {Comparat}, {Cukanovaite}, {Delubac},
  {DeVries}, {Dey}, {Dhungana}, {Dickinson}, {Ding}, {Donaldson}, {Duan},
  {Duckworth}, {Eftekharzadeh}, {Eisenstein}, {Etourneau}, {Fagrelius},
  {Farihi}, {Fitzpatrick}, {Font-Ribera}, {Fulmer}, {G{\"a}nsicke},
  {Gaztanaga}, {George}, {Gerdes}, {Gontcho}, {Gorgoni}, {Green}, {Guy},
  {Harmer}, {Hernandez}, {Honscheid}, {Huang}, {James}, {Jannuzi}, {Jiang},
  {Joyce}, {Karcher}, {Karkar}, {Kehoe}, {Kneib}, {Kueter-Young}, {Lan},
  {Lauer}, {Le Guillou}, {Le Van Suu}, {Lee}, {Lesser}, {Perreault Levasseur},
  {Li}, {Mann}, {Marshall}, {Mart{\'\i}nez-V{\'a}zquez}, {Martini}, {du Mas des
  Bourboux}, {McManus}, {Meier}, {M{\'e}nard}, {Metcalfe},
  {Mu{\~n}oz-Guti{\'e}rrez}, {Najita}, {Napier}, {Narayan}, {Newman}, {Nie},
  {Nord}, {Norman}, {Olsen}, {Paat}, {Palanque-Delabrouille}, {Peng},
  {Poppett}, {Poremba}, {Prakash}, {Rabinowitz}, {Raichoor}, {Rezaie},
  {Robertson}, {Roe}, {Ross}, {Ross}, {Rudnick}, {Safonova}, {Saha},
  {S{\'a}nchez}, {Savary}, {Schweiker}, {Scott}, {Seo}, {Shan}, {Silva},
  {Slepian}, {Soto}, {Sprayberry}, {Staten}, {Stillman}, {Stupak}, {Summers},
  {Sien Tie}, {Tirado}, {Vargas-Maga{\~n}a}, {Vivas}, {Wechsler}, {Williams},
  {Yang}, {Yang}, {Yapici}, {Zaritsky}, {Zenteno}, {Zhang}, {Zhang}, {Zhou}, \&
  {Zhou}}]{Dey2019}
{Dey}, A., {Schlegel}, D.~J., {Lang}, D., {et~al.} 2019, \aj, 157, 168,
  \dodoi{10.3847/1538-3881/ab089d}

\bibitem[{{Dial} {et~al.}(2025){Dial}, {Deller}, {Uttarkar}, {Lower},
  {Shannon}, {Gourdji}, {Marnoch}, {Bera}, {Ryder}, {Glowacki}, \&
  {Prochaska}}]{Dial2025}
{Dial}, T., {Deller}, A.~T., {Uttarkar}, P.~A., {et~al.} 2025, \mnras, 536,
  3220, \dodoi{10.1093/mnras/stae2756}

\bibitem[{{Driessen} {et~al.}(2024){Driessen}, {Barr}, {Buckley}, {Caleb},
  {Chen}, {Chen}, {Gromadzki}, {Jankowski}, {Kraan-Korteweg}, {Palmerio},
  {Rajwade}, {Tremou}, {Kramer}, {Stappers}, {Vergani}, {Woudt},
  {Bezuidenhout}, {Malenta}, {Morello}, {Sanidas}, {Surnis}, \&
  {Fender}}]{Driessen2024}
{Driessen}, L.~N., {Barr}, E.~D., {Buckley}, D.~A.~H., {et~al.} 2024, \mnras,
  527, 3659, \dodoi{10.1093/mnras/stad3329}

\bibitem[{{Eftekhari} {et~al.}(2025){Eftekhari}, {Dong}, {Fong}, {Shah},
  {Simha}, {Andersen}, {Andrew}, {Bhardwaj}, {Cassanelli}, {Chatterjee},
  {Coulter}, {Fonseca}, {Gaensler}, {Gordon}, {Hessels}, {Ibik}, {Joseph},
  {Kahinga}, {Kaspi}, {Kharel}, {Kilpatrick}, {Lanman}, {Lazda}, {Leung},
  {Liu}, {Mas-Ribas}, {Masui}, {Mckinven}, {Mena-Parra}, {Miller}, {Nimmo},
  {Pandhi}, {Patil}, {Pearlman}, {Pleunis}, {Prochaska}, {Rafiei-Ravandi},
  {Sammons}, {Scholz}, {Shin}, {Smith}, \& {Stairs}}]{Eftekhari2025}
{Eftekhari}, T., {Dong}, Y., {Fong}, W., {et~al.} 2025, \apjl, 979, L22,
  \dodoi{10.3847/2041-8213/ad9de2}

\bibitem[{{Faber} {et~al.}(2024){Faber}, {Ravi}, {Ocker}, {Sherman}, {Sharma},
  {Connor}, {Law}, {Kosogorov}, {Hallinan}, {Harnach}, {Hellbourg}, {Hobbs},
  {Hodge}, {Hodges}, {Lamb}, {Rasmussen}, {Somalwar}, {Weinreb}, \&
  {Woody}}]{Faber2024}
{Faber}, J.~T., {Ravi}, V., {Ocker}, S.~K., {et~al.} 2024, arXiv e-prints,
  arXiv:2405.14182, \dodoi{10.48550/arXiv.2405.14182}

\bibitem[{{Ferland} {et~al.}(2017){Ferland}, {Chatzikos}, {Guzm{\'a}n},
  {Lykins}, {van Hoof}, {Williams}, {Abel}, {Badnell}, {Keenan}, {Porter}, \&
  {Stancil}}]{Ferland2017}
{Ferland}, G.~J., {Chatzikos}, M., {Guzm{\'a}n}, F., {et~al.} 2017, \rmxaa, 53,
  385, \dodoi{10.48550/arXiv.1705.10877}

\bibitem[{{Gao} {et~al.}(2024){Gao}, {Wu}, {Hu}, {Yi}, {Zhou}, \&
  {Wang}}]{Gao2024}
{Gao}, D.~H., {Wu}, Q., {Hu}, J.~P., {et~al.} 2024, arXiv e-prints,
  arXiv:2410.03994, \dodoi{10.48550/arXiv.2410.03994}

\bibitem[{{Gao} {et~al.}(2025){Gao}, {Gao}, {Li}, \& {Yang}}]{Gao2025}
{Gao}, R., {Gao}, H., {Li}, Z., \& {Yang}, Y.-P. 2025, arXiv e-prints,
  arXiv:2504.15119, \dodoi{10.48550/arXiv.2504.15119}

\bibitem[{{Geda} {et~al.}(2022){Geda}, {Crawford}, {Hunt}, {Bershady},
  {Tollerud}, \& {Randriamampandry}}]{Geda2022}
{Geda}, R., {Crawford}, S.~M., {Hunt}, L., {et~al.} 2022, \aj, 163, 202,
  \dodoi{10.3847/1538-3881/ac5908}

\bibitem[{{Glowacki} {et~al.}(2024){Glowacki}, {Bera}, {Lee-Waddell}, {Deller},
  {Dial}, {Gourdji}, {Simha}, {Caleb}, {Marnoch}, {Prochaska}, {Ryder},
  {Shannon}, \& {Tejos}}]{Glowacki2024}
{Glowacki}, M., {Bera}, A., {Lee-Waddell}, K., {et~al.} 2024, \apjl, 962, L13,
  \dodoi{10.3847/2041-8213/ad1f62}

\bibitem[{{Gordon} {et~al.}(2023){Gordon}, {Fong}, {Kilpatrick}, {Eftekhari},
  {Leja}, {Prochaska}, {Nugent}, {Bhandari}, {Blanchard}, {Caleb}, {Day},
  {Deller}, {Dong}, {Glowacki}, {Gourdji}, {Mannings}, {Mahoney}, {Marnoch},
  {Miller}, {Paterson}, {Rastinejad}, {Ryder}, {Sadler}, {Scott}, {Sears},
  {Shannon}, {Simha}, {Stappers}, \& {Tejos}}]{Gordon2023}
{Gordon}, A.~C., {Fong}, W.-f., {Kilpatrick}, C.~D., {et~al.} 2023, \apj, 954,
  80, \dodoi{10.3847/1538-4357/ace5aa}

\bibitem[{{Gordon} {et~al.}(2025){Gordon}, {Fong}, {Deller}, {Marnoch}, {Lim},
  {Peng}, {Bannister}, {Bera}, {Bhat}, {Dial}, {Dong}, {Eftekhari}, {Glowacki},
  {Gourdji}, {Gupta}, {Jahns-Schindler}, {Jaini}, {Kilpatrick}, {Liu},
  {Prochaska}, {Ryder}, {Shannon}, {Simha}, {Tejos}, {Wang}, \&
  {Wang}}]{Gordon2025}
{Gordon}, A.~C., {Fong}, W.-f., {Deller}, A.~T., {et~al.} 2025, arXiv e-prints,
  arXiv:2506.06453, \dodoi{10.48550/arXiv.2506.06453}

\bibitem[{{Guti{\'e}rrez} \& {Beckman}(2010)}]{Leonel2010}
{Guti{\'e}rrez}, L., \& {Beckman}, J.~E. 2010, \apjl, 710, L44,
  \dodoi{10.1088/2041-8205/710/1/L44}

\bibitem[{{Hagstotz} {et~al.}(2022){Hagstotz}, {Reischke}, \&
  {Lilow}}]{Hagstotz2022}
{Hagstotz}, S., {Reischke}, R., \& {Lilow}, R. 2022, \mnras, 511, 662,
  \dodoi{10.1093/mnras/stac077}

\bibitem[{{Heintz} {et~al.}(2020){Heintz}, {Prochaska}, {Simha}, {Platts},
  {Fong}, {Tejos}, {Ryder}, {Aggerwal}, {Bhandari}, {Day}, {Deller},
  {Kilpatrick}, {Law}, {Macquart}, {Mannings}, {Marnoch}, {Sadler}, \&
  {Shannon}}]{Heintz2020}
{Heintz}, K.~E., {Prochaska}, J.~X., {Simha}, S., {et~al.} 2020, \apj, 903,
  152, \dodoi{10.3847/1538-4357/abb6fb}

\bibitem[{{Herrera-Camus} {et~al.}(2016){Herrera-Camus}, {Bolatto}, {Smith},
  {Draine}, {Pellegrini}, {Wolfire}, {Croxall}, {de Looze}, {Calzetti},
  {Kennicutt}, {Crocker}, {Armus}, {van der Werf}, {Sandstrom}, {Galametz},
  {Brandl}, {Groves}, {Rigopoulou}, {Walter}, {Leroy}, {Boquien}, {Tabatabaei},
  \& {Beirao}}]{Herrera2016}
{Herrera-Camus}, R., {Bolatto}, A., {Smith}, J.~D., {et~al.} 2016, \apj, 826,
  175, \dodoi{10.3847/0004-637X/826/2/175}

\bibitem[{{Hewitt} {et~al.}(2024){Hewitt}, {Bhardwaj}, {Gordon}, {Kirichenko},
  {Nimmo}, {Bhandari}, {Cognard}, {Fong}, {Gil de Paz}, {Gopinath}, {Hessels},
  {Kirsten}, {Marcote}, {Bezrukovs}, {Blaauw}, {Bray}, {Buttaccio},
  {Cassanelli}, {Chawla}, {Corongiu}, {Deng}, {Didehbani}, {Dong},
  {Gawro{\'n}ski}, {Giroletti}, {Guillemot}, {Huang}, {Ivanov}, {Joseph},
  {Kaspi}, {Kharinov}, {Lazda}, {Lindqvist}, {Maccaferri}, {Mas-Ribas},
  {Masui}, {Mckinven}, {Melnikov}, {Michilli}, {Mikhailov}, {Nugent},
  {Ould-Boukattine}, {Paragi}, {Pearlman}, {Pen}, {Pleunis}, {Sand}, {Shah},
  {Shin}, {Snelders}, {Venturi}, {Wang}, {Williams-Baldwin}, {Yang}, \&
  {Yuan}}]{Hewitt2024}
{Hewitt}, D.~M., {Bhardwaj}, M., {Gordon}, A.~C., {et~al.} 2024, \apjl, 977,
  L4, \dodoi{10.3847/2041-8213/ad8ce1}

\bibitem[{{Hubble}(1926)}]{Hubble1926}
{Hubble}, E.~P. 1926, \apj, 64, 321, \dodoi{10.1086/143018}

\bibitem[{{Hussaini} {et~al.}(2025){Hussaini}, {Connor}, {Konietzka}, {Ravi},
  {Faber}, {Sharma}, \& {Sherman}}]{Hussaini2025}
{Hussaini}, M., {Connor}, L., {Konietzka}, R.~M., {et~al.} 2025, arXiv
  e-prints, arXiv:2506.04186, \dodoi{10.48550/arXiv.2506.04186}

\bibitem[{{Ibik} {et~al.}(2024){Ibik}, {Drout}, {Gaensler}, {Scholz},
  {Michilli}, {Bhardwaj}, {Kaspi}, {Pleunis}, {Cassanelli}, {Cook}, {Dong},
  {Kaczmarek}, {Leung}, {Lu}, {Masui}, {Pearlman}, {Rafiei-Ravandi}, {Sand},
  {Shin}, {Smith}, \& {Stairs}}]{Ibik2024}
{Ibik}, A.~L., {Drout}, M.~R., {Gaensler}, B.~M., {et~al.} 2024, \apj, 961, 99,
  \dodoi{10.3847/1538-4357/ad0893}

\bibitem[{{Isobe} {et~al.}(2023){Isobe}, {Ouchi}, {Nakajima}, {Harikane},
  {Ono}, {Xu}, {Zhang}, \& {Umeda}}]{Isobe2023}
{Isobe}, Y., {Ouchi}, M., {Nakajima}, K., {et~al.} 2023, \apj, 956, 139,
  \dodoi{10.3847/1538-4357/acf376}

\bibitem[{{James} {et~al.}(2022{\natexlab{a}}){James}, {Prochaska}, {Macquart},
  {North-Hickey}, {Bannister}, \& {Dunning}}]{James2022b}
{James}, C.~W., {Prochaska}, J.~X., {Macquart}, J.~P., {et~al.}
  2022{\natexlab{a}}, \mnras, 510, L18, \dodoi{10.1093/mnrasl/slab117}

\bibitem[{{James} {et~al.}(2022{\natexlab{b}}){James}, {Ghosh}, {Prochaska},
  {Bannister}, {Bhandari}, {Day}, {Deller}, {Glowacki}, {Gordon}, {Heintz},
  {Marnoch}, {Ryder}, {Scott}, {Shannon}, \& {Tejos}}]{James2022}
{James}, C.~W., {Ghosh}, E.~M., {Prochaska}, J.~X., {et~al.}
  2022{\natexlab{b}}, \mnras, 516, 4862, \dodoi{10.1093/mnras/stac2524}

\bibitem[{{Kaasinen} {et~al.}(2017){Kaasinen}, {Bian}, {Groves}, {Kewley}, \&
  {Gupta}}]{Kaasinen2017}
{Kaasinen}, M., {Bian}, F., {Groves}, B., {Kewley}, L.~J., \& {Gupta}, A. 2017,
  \mnras, 465, 3220, \dodoi{10.1093/mnras/stw2827}

\bibitem[{{Kalita} {et~al.}(2024){Kalita}, {Bhatporia}, \&
  {Weltman}}]{Kalita2024}
{Kalita}, S., {Bhatporia}, S., \& {Weltman}, A. 2024, arXiv e-prints,
  arXiv:2410.01974, \dodoi{10.48550/arXiv.2410.01974}

\bibitem[{Katz(2020)}]{88}
Katz, J.~I. 2020, Monthly Notices of the Royal Astronomical Society: Letters,
  501, L76, \dodoi{10.1093/mnrasl/slaa202}

\bibitem[{{Kennicutt}(1998)}]{Kennicutt1998}
{Kennicutt}, Jr., R.~C. 1998, \araa, 36, 189,
  \dodoi{10.1146/annurev.astro.36.1.189}

\bibitem[{{Kennicutt} {et~al.}(1994){Kennicutt}, {Tamblyn}, \&
  {Congdon}}]{Kennicutt1994}
{Kennicutt}, Jr., R.~C., {Tamblyn}, P., \& {Congdon}, C.~E. 1994, \apj, 435,
  22, \dodoi{10.1086/174790}

\bibitem[{{Kumar} {et~al.}(2023){Kumar}, {Luo}, {Price}, {Shannon}, {Deller},
  {Bhandari}, {Feng}, {Flynn}, {Jiang}, {Uttarkar}, {Wang}, \&
  {Zhang}}]{Kumar2023}
{Kumar}, P., {Luo}, R., {Price}, D.~C., {et~al.} 2023, \mnras, 526, 3652,
  \dodoi{10.1093/mnras/stad2969}

\bibitem[{{Law} {et~al.}(2020){Law}, {Butler}, {Prochaska}, {Zackay},
  {Burke-Spolaor}, {Mannings}, {Tejos}, {Josephy}, {Andersen}, {Chawla},
  {Heintz}, {Aggarwal}, {Bower}, {Demorest}, {Kilpatrick}, {Lazio}, {Linford},
  {Mckinven}, {Tendulkar}, \& {Simha}}]{Law2020}
{Law}, C.~J., {Butler}, B.~J., {Prochaska}, J.~X., {et~al.} 2020, ApJ, 899,
  161, \dodoi{10.3847/1538-4357/aba4ac}

\bibitem[{{Law} {et~al.}(2024){Law}, {Sharma}, {Ravi}, {Chen}, {Catha},
  {Connor}, {Faber}, {Hallinan}, {Harnach}, {Hellbourg}, {Hobbs}, {Hodge},
  {Hodges}, {Lamb}, {Rasmussen}, {Sherman}, {Shi}, {Simard}, {Squillace},
  {Weinreb}, {Woody}, \& {Yurk}}]{Law2024}
{Law}, C.~J., {Sharma}, K., {Ravi}, V., {et~al.} 2024, \apj, 967, 29,
  \dodoi{10.3847/1538-4357/ad3736}

\bibitem[{{Lee} {et~al.}(2022){Lee}, {Ata}, {Khrykin}, {Huang}, {Prochaska},
  {Cooke}, {Zhang}, \& {Batten}}]{Lee2022}
{Lee}, K.-G., {Ata}, M., {Khrykin}, I.~S., {et~al.} 2022, \apj, 928, 9,
  \dodoi{10.3847/1538-4357/ac4f62}

\bibitem[{{Li} {et~al.}(2023){Li}, {Zhao}, {Gao}, \& {Wang}}]{LiR2023}
{Li}, R.-N., {Zhao}, Z.-Y., {Gao}, Z., \& {Wang}, F.-Y. 2023, \apjl, 956, L2,
  \dodoi{10.3847/2041-8213/acfa9e}

\bibitem[{{Li} {et~al.}(2025{\natexlab{a}}){Li}, {Zhao}, {Wu}, {Yi}, \&
  {Wang}}]{LiR2025}
{Li}, R.-N., {Zhao}, Z.-Y., {Wu}, Q., {Yi}, S.-X., \& {Wang}, F.-Y.
  2025{\natexlab{a}}, \apjl, 979, L41, \dodoi{10.3847/2041-8213/adabc2}

\bibitem[{{Li} {et~al.}(2025{\natexlab{b}}){Li}, {Wang}, {Chen}, {Jones},
  {Treu}, {Glazebrook}, {He}, {Henry}, {Meng}, {Morishita}, {Roberts-Borsani},
  {Yang}, {Yu}, {Calabr{\`o}}, {Castellano}, {Leethochawalit}, {Metha},
  {Nanayakkara}, {Roy}, \& {Vulcani}}]{Li2025SFR}
{Li}, S., {Wang}, X., {Chen}, Y., {et~al.} 2025{\natexlab{b}}, \apjl, 979, L13,
  \dodoi{10.3847/2041-8213/ad9eac}

\bibitem[{{Li} {et~al.}(2025{\natexlab{c}}){Li}, {Zhang}, {Yang}, {Tsai},
  {Yang}, {Law}, {Anna-Thomas}, {Chen}, {Lee}, {Tang}, {Xiao}, {Xu}, {Yang},
  {Chen}, {Feng}, {Li}, {Mckinven}, {Niu}, {Shin}, {Wang}, {Zhang}, {Zhang},
  {Zhou}, {Zhu}, {Dai}, {Chang}, {Geng}, {Han}, {Hu}, {Li}, {Luo}, {Niu},
  {Shi}, {Sun}, {Wu}, {Zhu}, {Jiang}, \& {Zhang}}]{LiYe2025}
{Li}, Y., {Zhang}, S.~B., {Yang}, Y.~P., {et~al.} 2025{\natexlab{c}}, arXiv
  e-prints, arXiv:2503.04727, \dodoi{10.48550/arXiv.2503.04727}

\bibitem[{{Li} {et~al.}(2025{\natexlab{d}}){Li}, {Zhang}, {Yang}, {Tsai},
  {Yang}, {Law}, {Anna-Thomas}, {Chen}, {Lee}, {Tang}, {Xiao}, {Xu}, {Yang},
  {Chen}, {Feng}, {Li}, {Mckinven}, {Niu}, {Shin}, {Wang}, {Zhang}, {Zhang},
  {Zhou}, {Zhu}, {Dai}, {Chang}, {Geng}, {Han}, {Hu}, {Li}, {Luo}, {Niu},
  {Shi}, {Sun}, {Wu}, {Zhu}, {Jiang}, \& {Zhang}}]{LiY2025}
---. 2025{\natexlab{d}}, arXiv e-prints, arXiv:2503.04727,
  \dodoi{10.48550/arXiv.2503.04727}

\bibitem[{{Li} {et~al.}(2019){Li}, {Gao}, {Wei}, {Yang}, {Zhang}, \&
  {Zhu}}]{LiZX2019}
{Li}, Z., {Gao}, H., {Wei}, J.-J., {et~al.} 2019, \apj, 876, 146,
  \dodoi{10.3847/1538-4357/ab18fe}

\bibitem[{{Li} {et~al.}(2020){Li}, {Gao}, {Wei}, {Yang}, {Zhang}, \&
  {Zhu}}]{li2020b}
{Li}, Z., {Gao}, H., {Wei}, J.~J., {et~al.} 2020, \mnras, 496, L28,
  \dodoi{10.1093/mnrasl/slaa070}

\bibitem[{{Lin} \& {Zou}(2023)}]{Lin2023}
{Lin}, H.-N., \& {Zou}, R. 2023, \mnras, 520, 6237,
  \dodoi{10.1093/mnras/stad509}

\bibitem[{{Lorimer} {et~al.}(2007){Lorimer}, {Bailes}, {McLaughlin},
  {Narkevic}, \& {Crawford}}]{Lorimer2007}
{Lorimer}, D.~R., {Bailes}, M., {McLaughlin}, M.~A., {Narkevic}, D.~J., \&
  {Crawford}, F. 2007, Science, 318, 777, \dodoi{10.1126/science.1147532}

\bibitem[{{Luo} {et~al.}(2018){Luo}, {Lee}, {Lorimer}, \& {Zhang}}]{Luo2018}
{Luo}, R., {Lee}, K., {Lorimer}, D.~R., \& {Zhang}, B. 2018, \mnras, 481, 2320,
  \dodoi{10.1093/mnras/sty2364}

\bibitem[{{Macquart} {et~al.}(2020){Macquart}, {Prochaska}, {McQuinn},
  {Bannister}, {Bhandari}, {Day}, {Deller}, {Ekers}, {James}, {Marnoch},
  {Os{\l}owski}, {Phillips}, {Ryder}, {Scott}, {Shannon}, \&
  {Tejos}}]{Macquart2020}
{Macquart}, J.~P., {Prochaska}, J.~X., {McQuinn}, M., {et~al.} 2020, \nat, 581,
  391, \dodoi{10.1038/s41586-020-2300-2}

\bibitem[{{Mahony} {et~al.}(2018){Mahony}, {Ekers}, {Macquart}, {Sadler},
  {Bannister}, {Bhandari}, {Flynn}, {Koribalski}, {Prochaska}, {Ryder},
  {Shannon}, {Tejos}, {Whiting}, \& {Wong}}]{Mahony2018}
{Mahony}, E.~K., {Ekers}, R.~D., {Macquart}, J.-P., {et~al.} 2018, \apjl, 867,
  L10, \dodoi{10.3847/2041-8213/aae7cb}

\bibitem[{{Marcote} {et~al.}(2020){Marcote}, {Nimmo}, {Hessels}, {Tendulkar},
  {Bassa}, {Paragi}, {Keimpema}, {Bhardwaj}, {Karuppusamy}, {Kaspi}, {Law},
  {Michilli}, {Aggarwal}, {Andersen}, {Archibald}, {Bandura}, {Bower}, {Boyle},
  {Brar}, {Burke-Spolaor}, {Butler}, {Cassanelli}, {Chawla}, {Demorest},
  {Dobbs}, {Fonseca}, {Giri}, {Good}, {Gourdji}, {Josephy}, {Kirichenko},
  {Kirsten}, {Landecker}, {Lang}, {Lazio}, {Li}, {Lin}, {Linford}, {Masui},
  {Mena-Parra}, {Naidu}, {Ng}, {Patel}, {Pen}, {Pleunis}, {Rafiei-Ravandi},
  {Rahman}, {Renard}, {Scholz}, {Siegel}, {Smith}, {Stairs}, {Vanderlinde}, \&
  {Zwaniga}}]{Marcote2020}
{Marcote}, B., {Nimmo}, K., {Hessels}, J.~W.~T., {et~al.} 2020, \nat, 577, 190,
  \dodoi{10.1038/s41586-019-1866-z}

\bibitem[{{Margalit} \& {Metzger}(2018)}]{86}
{Margalit}, B., \& {Metzger}, B.~D. 2018, \apjl, 868, L4,
  \dodoi{10.3847/2041-8213/aaedad}

\bibitem[{{Martin} {et~al.}(2005){Martin}, {Fanson}, {Schiminovich},
  {Morrissey}, {Friedman}, {Barlow}, {Conrow}, {Grange}, {Jelinsky},
  {Milliard}, {Siegmund}, {Bianchi}, {Byun}, {Donas}, {Forster}, {Heckman},
  {Lee}, {Madore}, {Malina}, {Neff}, {Rich}, {Small}, {Surber}, {Szalay},
  {Welsh}, \& {Wyder}}]{GALEX}
{Martin}, D.~C., {Fanson}, J., {Schiminovich}, D., {et~al.} 2005, \apjl, 619,
  L1, \dodoi{10.1086/426387}

\bibitem[{{McQuinn}(2014)}]{McQuinn2014}
{McQuinn}, M. 2014, \apjl, 780, L33, \dodoi{10.1088/2041-8205/780/2/L33}

\bibitem[{{Michilli} {et~al.}(2023){Michilli}, {Bhardwaj}, {Brar}, {Gaensler},
  {Kaspi}, {Kirichenko}, {Masui}, {Mckinven}, {Ng}, {Patel}, {Sand}, {Scholz},
  {Shin}, {Siegel}, {Stairs}, {Cassanelli}, {Cook}, {Dobbs}, {Dong}, {Fonseca},
  {Ibik}, {Kaczmarek}, {Leung}, {Pearlman}, {Petroff}, {Pleunis},
  {Rafiei-Ravandi}, {Sanghavi}, {Shaw}, \& {Tendulkar}}]{Michilli2023}
{Michilli}, D., {Bhardwaj}, M., {Brar}, C., {et~al.} 2023, \apj, 950, 134,
  \dodoi{10.3847/1538-4357/accf89}

\bibitem[{{Mo} {et~al.}(2025){Mo}, {Zhu}, \& {Feng}}]{Mo2025}
{Mo}, J.-f., {Zhu}, W., \& {Feng}, L.-L. 2025, \apjs, 277, 43,
  \dodoi{10.3847/1538-4365/adb616}

\bibitem[{{Niu} {et~al.}(2022){Niu}, {Aggarwal}, {Li}, {Zhang}, {Chatterjee},
  {Tsai}, {Yu}, {Law}, {Burke-Spolaor}, {Cordes}, {Zhang}, {Ocker}, {Yao},
  {Wang}, {Feng}, {Niino}, {Bochenek}, {Cruces}, {Connor}, {Jiang}, {Dai},
  {Luo}, {Li}, {Miao}, {Niu}, {Anna-Thomas}, {Sydnor}, {Stern}, {Wang}, {Yuan},
  {Yue}, {Zhou}, {Yan}, {Zhu}, \& {Zhang}}]{Niu2022}
{Niu}, C.~H., {Aggarwal}, K., {Li}, D., {et~al.} 2022, \nat, 606, 873,
  \dodoi{10.1038/s41586-022-04755-5}

\bibitem[{{Ocker} {et~al.}(2020){Ocker}, {Cordes}, \& {Chatterjee}}]{Ocker2020}
{Ocker}, S.~K., {Cordes}, J.~M., \& {Chatterjee}, S. 2020, \apj, 897, 124,
  \dodoi{10.3847/1538-4357/ab98f9}

\bibitem[{{Ocker} {et~al.}(2023){Ocker}, {Cordes}, {Chatterjee}, {Li}, {Niu},
  {McKee}, {Law}, \& {Anna-Thomas}}]{Ocker2023}
{Ocker}, S.~K., {Cordes}, J.~M., {Chatterjee}, S., {et~al.} 2023, \mnras, 519,
  821, \dodoi{10.1093/mnras/stac3547}

\bibitem[{{Ocker} {et~al.}(2022){Ocker}, {Cordes}, {Chatterjee}, {Niu}, {Li},
  {McKee}, {Law}, {Tsai}, {Anna-Thomas}, {Yao}, \& {Cruces}}]{Ocker2022}
---. 2022, \apj, 931, 87, \dodoi{10.3847/1538-4357/ac6504}

\bibitem[{{Panda} {et~al.}(2024){Panda}, {Roy}, {Bhattacharyya}, {Dudeja}, \&
  {Kudale}}]{Panda2024}
{Panda}, U., {Roy}, J., {Bhattacharyya}, S., {Dudeja}, C., \& {Kudale}, S.
  2024, arXiv e-prints, arXiv:2405.09749, \dodoi{10.48550/arXiv.2405.09749}

\bibitem[{{Piro} \& {Gaensler}(2018)}]{Piro2018}
{Piro}, A.~L., \& {Gaensler}, B.~M. 2018, \apj, 861, 150,
  \dodoi{10.3847/1538-4357/aac9bc}

\bibitem[{{Planck Collaboration} {et~al.}(2020){Planck Collaboration},
  {Aghanim}, {Akrami}, {Ashdown}, {Aumont}, {Baccigalupi}, {Ballardini},
  {Banday}, {Barreiro}, {Bartolo}, {Basak}, {Battye}, {Benabed}, {Bernard},
  {Bersanelli}, {Bielewicz}, {Bock}, {Bond}, {Borrill}, {Bouchet}, {Boulanger},
  {Bucher}, {Burigana}, {Butler}, {Calabrese}, {Cardoso}, {Carron},
  {Challinor}, {Chiang}, {Chluba}, {Colombo}, {Combet}, {Contreras}, {Crill},
  {Cuttaia}, {de Bernardis}, {de Zotti}, {Delabrouille}, {Delouis}, {Di
  Valentino}, {Diego}, {Dor{\'e}}, {Douspis}, {Ducout}, {Dupac}, {Dusini},
  {Efstathiou}, {Elsner}, {En{\ss}lin}, {Eriksen}, {Fantaye}, {Farhang},
  {Fergusson}, {Fernandez-Cobos}, {Finelli}, {Forastieri}, {Frailis},
  {Fraisse}, {Franceschi}, {Frolov}, {Galeotta}, {Galli}, {Ganga},
  {G{\'e}nova-Santos}, {Gerbino}, {Ghosh}, {Gonz{\'a}lez-Nuevo}, {G{\'o}rski},
  {Gratton}, {Gruppuso}, {Gudmundsson}, {Hamann}, {Handley}, {Hansen},
  {Herranz}, {Hildebrandt}, {Hivon}, {Huang}, {Jaffe}, {Jones}, {Karakci},
  {Keih{\"a}nen}, {Keskitalo}, {Kiiveri}, {Kim}, {Kisner}, {Knox},
  {Krachmalnicoff}, {Kunz}, {Kurki-Suonio}, {Lagache}, {Lamarre}, {Lasenby},
  {Lattanzi}, {Lawrence}, {Le Jeune}, {Lemos}, {Lesgourgues}, {Levrier},
  {Lewis}, {Liguori}, {Lilje}, {Lilley}, {Lindholm}, {L{\'o}pez-Caniego},
  {Lubin}, {Ma}, {Mac{\'\i}as-P{\'e}rez}, {Maggio}, {Maino}, {Mandolesi},
  {Mangilli}, {Marcos-Caballero}, {Maris}, {Martin}, {Martinelli},
  {Mart{\'\i}nez-Gonz{\'a}lez}, {Matarrese}, {Mauri}, {McEwen}, {Meinhold},
  {Melchiorri}, {Mennella}, {Migliaccio}, {Millea}, {Mitra},
  {Miville-Desch{\^e}nes}, {Molinari}, {Montier}, {Morgante}, {Moss}, {Natoli},
  {N{\o}rgaard-Nielsen}, {Pagano}, {Paoletti}, {Partridge}, {Patanchon},
  {Peiris}, {Perrotta}, {Pettorino}, {Piacentini}, {Polastri}, {Polenta},
  {Puget}, {Rachen}, {Reinecke}, {Remazeilles}, {Renzi}, {Rocha}, {Rosset},
  {Roudier}, {Rubi{\~n}o-Mart{\'\i}n}, {Ruiz-Granados}, {Salvati}, {Sandri},
  {Savelainen}, {Scott}, {Shellard}, {Sirignano}, {Sirri}, {Spencer},
  {Sunyaev}, {Suur-Uski}, {Tauber}, {Tavagnacco}, {Tenti}, {Toffolatti},
  {Tomasi}, {Trombetti}, {Valenziano}, {Valiviita}, {Van Tent}, {Vibert},
  {Vielva}, {Villa}, {Vittorio}, {Wandelt}, {Wehus}, {White}, {White},
  {Zacchei}, \& {Zonca}}]{Planck2020}
{Planck Collaboration}, {Aghanim}, N., {Akrami}, Y., {et~al.} 2020, AAP, 641,
  A6, \dodoi{10.1051/0004-6361/201833910}

\bibitem[{{Popesso} {et~al.}(2023){Popesso}, {Concas}, {Cresci}, {Belli},
  {Rodighiero}, {Inami}, {Dickinson}, {Ilbert}, {Pannella}, \&
  {Elbaz}}]{Popesso2023}
{Popesso}, P., {Concas}, A., {Cresci}, G., {et~al.} 2023, \mnras, 519, 1526,
  \dodoi{10.1093/mnras/stac3214}

\bibitem[{{Prochaska} {et~al.}(2019){Prochaska}, {Macquart}, {McQuinn},
  {Simha}, {Shannon}, {Day}, {Marnoch}, {Ryder}, {Deller}, {Bannister},
  {Bhandari}, {Bordoloi}, {Bunton}, {Cho}, {Flynn}, {Mahony}, {Phillips},
  {Qiu}, \& {Tejos}}]{Prochaska2019}
{Prochaska}, J.~X., {Macquart}, J.-P., {McQuinn}, M., {et~al.} 2019, Science,
  366, 231, \dodoi{10.1126/science.aay0073}

\bibitem[{{Qiang} {et~al.}(2022){Qiang}, {Li}, \& {Wei}}]{Qiang2022}
{Qiang}, D.-C., {Li}, S.-L., \& {Wei}, H. 2022, \jcap, 2022, 040,
  \dodoi{10.1088/1475-7516/2022/01/040}

\bibitem[{{Rajwade} {et~al.}(2022){Rajwade}, {Bezuidenhout}, {Caleb},
  {Driessen}, {Jankowski}, {Malenta}, {Morello}, {Sanidas}, {Stappers},
  {Surnis}, {Barr}, {Chen}, {Kramer}, {Wu}, {Buchner}, {Serylak}, {Combes},
  {Fong}, {Gupta}, {Jagannathan}, {Kilpatrick}, {Krogager}, {Noterdaeme},
  {N{\'u}nẽz}, {Prochaska}, {Srianand}, \& {Tejos}}]{Rajwade2022}
{Rajwade}, K.~M., {Bezuidenhout}, M.~C., {Caleb}, M., {et~al.} 2022, \mnras,
  514, 1961, \dodoi{10.1093/mnras/stac1450}

\bibitem[{{Rajwade} {et~al.}(2024){Rajwade}, {Driessen}, {Barr},
  {Pastor-Marazuela}, {Berezina}, {Jankowski}, {Muller}, {Kahinga}, {Stappers},
  {Bezuidenhout}, {Caleb}, {Deller}, {Fong}, {Gordon}, {Kramer}, {Malenta},
  {Morello}, {Prochaska}, {Sanidas}, {Surnis}, {Tejos}, \&
  {Wagner}}]{Rajwade2024}
{Rajwade}, K.~M., {Driessen}, L.~N., {Barr}, E.~D., {et~al.} 2024, \mnras, 532,
  3881, \dodoi{10.1093/mnras/stae1652}

\bibitem[{{Ravi} {et~al.}(2019){Ravi}, {Catha}, {D'Addario}, {Djorgovski},
  {Hallinan}, {Hobbs}, {Kocz}, {Kulkarni}, {Shi}, {Vedantham}, {Weinreb}, \&
  {Woody}}]{Ravi2019}
{Ravi}, V., {Catha}, M., {D'Addario}, L., {et~al.} 2019, \nat, 572, 352,
  \dodoi{10.1038/s41586-019-1389-7}

\bibitem[{{Ravi} {et~al.}(2022){Ravi}, {Law}, {Li}, {Aggarwal}, {Bhardwaj},
  {Burke-Spolaor}, {Connor}, {Lazio}, {Simard}, {Somalwar}, \&
  {Tendulkar}}]{Ravi2022}
{Ravi}, V., {Law}, C.~J., {Li}, D., {et~al.} 2022, \mnras, 513, 982,
  \dodoi{10.1093/mnras/stac465}

\bibitem[{{Ravi} {et~al.}(2023){Ravi}, {Catha}, {Chen}, {Connor}, {Faber},
  {Lamb}, {Hallinan}, {Harnach}, {Hellbourg}, {Hobbs}, {Hodge}, {Hodges},
  {Law}, {Rasmussen}, {Sharma}, {Sherman}, {Shi}, {Simard}, {Squillace},
  {Weinreb}, {Woody}, {Yadlapalli}, {Ahumada}, {Dong}, {Fremling}, {Huang},
  {Karambelkar}, \& {Miller}}]{Ravi2023}
{Ravi}, V., {Catha}, M., {Chen}, G., {et~al.} 2023, \apjl, 949, L3,
  \dodoi{10.3847/2041-8213/acc4b6}

\bibitem[{{Ravi} {et~al.}(2025){Ravi}, {Catha}, {Chen}, {Connor}, {Cordes},
  {Faber}, {Lamb}, {Hallinan}, {Harnach}, {Hellbourg}, {Hobbs}, {Hodge},
  {Hodges}, {Law}, {Rasmussen}, {Sharma}, {Sherman}, {Shi}, {Simard},
  {Somalwar}, {Squillace}, {Weinreb}, {Woody}, {Yadlapalli}, \& {Deep Synoptic
  Array Team}}]{Ravi2025}
---. 2025, \aj, 169, 330, \dodoi{10.3847/1538-3881/adc725}

\bibitem[{{Ryder} {et~al.}(2023){Ryder}, {Bannister}, {Bhandari}, {Deller},
  {Ekers}, {Glowacki}, {Gordon}, {Gourdji}, {James}, {Kilpatrick}, {Lu},
  {Marnoch}, {Moss}, {Prochaska}, {Qiu}, {Sadler}, {Simha}, {Sammons}, {Scott},
  {Tejos}, \& {Shannon}}]{Ryder2023}
{Ryder}, S.~D., {Bannister}, K.~W., {Bhandari}, S., {et~al.} 2023, Science,
  382, 294, \dodoi{10.1126/science.adf2678}

\bibitem[{{Sammons} {et~al.}(2023){Sammons}, {Deller}, {Glowacki}, {Gourdji},
  {James}, {Prochaska}, {Qiu}, {Scott}, {Shannon}, \& {Trott}}]{Sammons2023}
{Sammons}, M.~W., {Deller}, A.~T., {Glowacki}, M., {et~al.} 2023, \mnras, 525,
  5653, \dodoi{10.1093/mnras/stad2631}

\bibitem[{{Shannon} {et~al.}(2024){Shannon}, {Bannister}, {Bera}, {Bhandari},
  {Day}, {Deller}, {Dial}, {Dobie}, {Ekers}, {Fong}, {Glowacki}, {Gordon},
  {Gourdji}, {Jaini}, {James}, {Kumar}, {Mahony}, {Marnoch}, {Muller},
  {Prochaska}, {Qiu}, {Ryder}, {Sadler}, {Scott}, {Tejos}, {Uttarkar}, \&
  {Wang}}]{Shannon2024}
{Shannon}, R.~M., {Bannister}, K.~W., {Bera}, A., {et~al.} 2024, arXiv
  e-prints, arXiv:2408.02083, \dodoi{10.48550/arXiv.2408.02083}

\bibitem[{{Sharma} {et~al.}(2024){Sharma}, {Ravi}, {Connor}, {Law}, {Ocker},
  {Sherman}, {Kosogorov}, {Faber}, {Hallinan}, {Harnach}, {Hellbourg}, {Hobbs},
  {Hodge}, {Hodges}, {Lamb}, {Rasmussen}, {Somalwar}, {Weinreb}, {Woody},
  {Leja}, {Anand}, {Das}, {Qin}, {Rose}, {Dong}, {Miller}, \&
  {Yao}}]{Sharma2024}
{Sharma}, K., {Ravi}, V., {Connor}, L., {et~al.} 2024, \nat, 635, 61,
  \dodoi{10.1038/s41586-024-08074-9}

\bibitem[{{Shull} {et~al.}(2012){Shull}, {Smith}, \& {Danforth}}]{Shull2012}
{Shull}, J.~M., {Smith}, B.~D., \& {Danforth}, C.~W. 2012, \apj, 759, 23,
  \dodoi{10.1088/0004-637X/759/1/23}

\bibitem[{{Skrutskie} {et~al.}(2006){Skrutskie}, {Cutri}, {Stiening},
  {Weinberg}, {Schneider}, {Carpenter}, {Beichman}, {Capps}, {Chester},
  {Elias}, {Huchra}, {Liebert}, {Lonsdale}, {Monet}, {Price}, {Seitzer},
  {Jarrett}, {Kirkpatrick}, {Gizis}, {Howard}, {Evans}, {Fowler}, {Fullmer},
  {Hurt}, {Light}, {Kopan}, {Marsh}, {McCallon}, {Tam}, {Van Dyk}, \&
  {Wheelock}}]{2MASS}
{Skrutskie}, M.~F., {Cutri}, R.~M., {Stiening}, R., {et~al.} 2006, \aj, 131,
  1163, \dodoi{10.1086/498708}

\bibitem[{{Tacconi} {et~al.}(2018){Tacconi}, {Genzel}, {Saintonge}, {Combes},
  {Garc{\'\i}a-Burillo}, {Neri}, {Bolatto}, {Contini}, {F{\"o}rster Schreiber},
  {Lilly}, {Lutz}, {Wuyts}, {Accurso}, {Boissier}, {Boone}, {Bouch{\'e}},
  {Bournaud}, {Burkert}, {Carollo}, {Cooper}, {Cox}, {Feruglio}, {Freundlich},
  {Herrera-Camus}, {Juneau}, {Lippa}, {Naab}, {Renzini}, {Salome}, {Sternberg},
  {Tadaki}, {{\"U}bler}, {Walter}, {Weiner}, \& {Weiss}}]{Tacconi2018}
{Tacconi}, L.~J., {Genzel}, R., {Saintonge}, A., {et~al.} 2018, \apj, 853, 179,
  \dodoi{10.3847/1538-4357/aaa4b4}

\bibitem[{Tendulkar {et~al.}(2017)Tendulkar, Bassa, Cordes, Bower, Law,
  Chatterjee, Adams, Bogdanov, Burke-Spolaor, Butler, Demorest, Hessels, Kaspi,
  Lazio, Maddox, Marcote, McLaughlin, Paragi, Ransom, Scholz, Seymour, Spitler,
  van Langevelde, \& Wharton}]{Tendulkar2017}
Tendulkar, S.~P., Bassa, C.~G., Cordes, J.~M., {et~al.} 2017, The Astrophysical
  Journal, 834, L7, \dodoi{10.3847/2041-8213/834/2/l7}

\bibitem[{{Tully} \& {Pierce}(2000)}]{Tully2000}
{Tully}, R.~B., \& {Pierce}, M.~J. 2000, \apj, 533, 744, \dodoi{10.1086/308700}

\bibitem[{{Wada} \& {Norman}(2007)}]{Wada2007}
{Wada}, K., \& {Norman}, C.~A. 2007, \apj, 660, 276, \dodoi{10.1086/513002}

\bibitem[{{Walker} {et~al.}(2024){Walker}, {Spitler}, {Ma}, {Cheng}, {Artale},
  \& {Hummels}}]{Walker2024}
{Walker}, C. R.~H., {Spitler}, L.~G., {Ma}, Y.-Z., {et~al.} 2024, \aap, 683,
  A71, \dodoi{10.1051/0004-6361/202347139}

\bibitem[{{Wang} {et~al.}(2022){Wang}, {Zhang}, {Dai}, \& {Cheng}}]{Wang2022}
{Wang}, F.~Y., {Zhang}, G.~Q., {Dai}, Z.~G., \& {Cheng}, K.~S. 2022, Nature
  Communications, 13, 4382, \dodoi{10.1038/s41467-022-31923-y}

\bibitem[{{Wei} \& {Melia}(2023)}]{Wei2023}
{Wei}, J.-J., \& {Melia}, F. 2023, \apj, 955, 101,
  \dodoi{10.3847/1538-4357/acefb8}

\bibitem[{{Wu} \& {Wang}(2024)}]{Wu2024}
{Wu}, Q., \& {Wang}, F.-Y. 2024, Chinese Physics Letters, 41, 119801,
  \dodoi{10.1088/0256-307X/41/11/119801}

\bibitem[{{Wu} {et~al.}(2022){Wu}, {Zhang}, \& {Wang}}]{Wu2022}
{Wu}, Q., {Zhang}, G.-Q., \& {Wang}, F.-Y. 2022, \mnras, 515, L1,
  \dodoi{10.1093/mnrasl/slac022}

\bibitem[{{Xiao} {et~al.}(2021){Xiao}, {Wang}, \& {Dai}}]{Xiao2021}
{Xiao}, D., {Wang}, F., \& {Dai}, Z. 2021, Science China Physics, Mechanics,
  and Astronomy, 64, 249501, \dodoi{10.1007/s11433-020-1661-7}

\bibitem[{{Xu} \& {Han}(2015)}]{XuHan2015}
{Xu}, J., \& {Han}, J.~L. 2015, Research in Astronomy and Astrophysics, 15,
  1629, \dodoi{10.1088/1674-4527/15/10/002}

\bibitem[{{Yang} {et~al.}(2022){Yang}, {Wu}, \& {Wang}}]{YangKB2022}
{Yang}, K.~B., {Wu}, Q., \& {Wang}, F.~Y. 2022, \apjl, 940, L29,
  \dodoi{10.3847/2041-8213/aca145}

\bibitem[{{Yang} \& {Dai}(2019)}]{Yang2019}
{Yang}, Y.-H., \& {Dai}, Z.-G. 2019, \apj, 885, 149,
  \dodoi{10.3847/1538-4357/ab48dd}

\bibitem[{{Yang} \& {Zhang}(2017)}]{Yang2017}
{Yang}, Y.-P., \& {Zhang}, B. 2017, \apj, 847, 22,
  \dodoi{10.3847/1538-4357/aa8721}

\bibitem[{{Zhang}(2018)}]{90}
{Zhang}, B. 2018, \apjl, 854, L21, \dodoi{10.3847/2041-8213/aaadba}

\bibitem[{{Zhang}(2023)}]{Zhang2023Review}
---. 2023, Reviews of Modern Physics, 95, 035005,
  \dodoi{10.1103/RevModPhys.95.035005}

\bibitem[{{Zhang} \& {Wang}(2019)}]{Zhang2019}
{Zhang}, G.~Q., \& {Wang}, F.~Y. 2019, \mnras, 487, 3672,
  \dodoi{10.1093/mnras/stz1566}

\bibitem[{{Zhang} {et~al.}(2020){Zhang}, {Yu}, {He}, \& {Wang}}]{Zhang2020}
{Zhang}, G.~Q., {Yu}, H., {He}, J.~H., \& {Wang}, F.~Y. 2020, \apj, 900, 170,
  \dodoi{10.3847/1538-4357/abaa4a}

\bibitem[{{Zhang} \& {Zhang}(2022)}]{Zhang2022b}
{Zhang}, R.~C., \& {Zhang}, B. 2022, \apjl, 924, L14,
  \dodoi{10.3847/2041-8213/ac46ad}

\bibitem[{{Zhang} {et~al.}(2025){Zhang}, {Nagamine}, {Oku}, {Lee}, {Fukushima},
  {Tomaru}, {Zhang}, {Medlock}, \& {Nagai}}]{ZhangZ2025}
{Zhang}, Z., {Nagamine}, K., {Oku}, Y., {et~al.} 2025, arXiv e-prints,
  arXiv:2503.12741, \dodoi{10.48550/arXiv.2503.12741}

\bibitem[{{Zhang} {et~al.}(2021){Zhang}, {Yan}, {Li}, {Zhang}, \&
  {Wang}}]{ZhangZJ2021}
{Zhang}, Z.~J., {Yan}, K., {Li}, C.~M., {Zhang}, G.~Q., \& {Wang}, F.~Y. 2021,
  \apj, 906, 49, \dodoi{10.3847/1538-4357/abceb9}

\bibitem[{{Zhang} \& {Zhang}(2025)}]{Zhang2025}
{Zhang}, Z.-L., \& {Zhang}, B. 2025, arXiv e-prints, arXiv:2504.13132,
  \dodoi{10.48550/arXiv.2504.13132}

\bibitem[{{Zhao} \& {Wang}(2021)}]{Zhao2021}
{Zhao}, Z.~Y., \& {Wang}, F.~Y. 2021, \apjl, 923, L17,
  \dodoi{10.3847/2041-8213/ac3f2f}

\bibitem[{{Zhao} {et~al.}(2023){Zhao}, {Zhang}, {Wang}, \& {Dai}}]{Zhao2023}
{Zhao}, Z.~Y., {Zhang}, G.~Q., {Wang}, F.~Y., \& {Dai}, Z.~G. 2023, \apj, 942,
  102, \dodoi{10.3847/1538-4357/aca66b}

\end{thebibliography}

\clearpage

\begin{table}[!ht]
\vspace{30pt}
\centering
\rotatebox[origin=c]{90}{%
\begin{varwidth}{\textheight}
\caption{Properties of FRBs and Host Galaxies}
\label{tab:data1}
\setlength{\tabcolsep}{7pt}
\begin{tabular}{lrrrllllllllll}
\toprule
FRB name & Redshift & DM & $\rm{}DM_{exc}$ & $\tau_{\rm{}exc}$ & $i$ & Offset & Area & $\rm{}\log(M)$ & SFR & $\rm{}\log(sSFR)$ & Ref. & Quality\\
         &     & (pc cm$^{-3}$) & (pc cm$^{-3}$) & (ms) & (deg) & (kpc) & ($\rm{}kpc^{2}$) & (M$_\odot$) & (M$_\odot$ yr$^{-1}$) & (yr$^{-1}$) &  \\
         &    & (1) & (2) & (3) & (4) & (5) & (6) & (7) & (8) & (9) & (10) & (11) \\
\hline
\midrule
20231229A & $0.019$ & $198.50$ & $140.32$ & \ & \ & \ & \ & $9.50 \pm 0.10$ & $0.25 \pm 0.15$ & $-10.11 \pm 0.36$ & 4 & 1 \\
20231230A & $0.0298$ & $131.40$ & $69.82$ & \ & $27.23$ & 8.15 & \ & $9.57 \pm 0.04$ & $0.54 \pm 0.34$ & $-9.84 \pm 0.27$ & 4 & 1,43 \\
20231206A & $0.0659$ & $457.70$ & $398.56$ & \ & $69.84$ & \ & $144.38 \pm 0.45$ & $10.56 \pm 0.07$ & $8.19 \pm 2.90$ & $-9.66 \pm 0.21$ & 0 & 1 \\
20230926A & $0.0553$ & $222.80$ & $170.18$ & \ & $64.12$ & \ & $163.72 \pm 0.93$ & $10.64 \pm 0.05$ & $0.63 \pm 0.18$ & $-10.83 \pm 0.14$ & 0 & 1 \\
20231223C & $0.1059$ & $165.80$ & $117.93$ & \ & $30.94$ & \ & $201.92 \pm 1.24$ & $10.54 \pm 0.06$ & $1.07 \pm 0.43$ & $-10.48 \pm 0.20$ & 4 & 1 \\
20231128A & $0.1079$ & $331.60$ & $306.59$ & \ & $40.43$ & \ & $82.06 \pm 0.07$ & $10.43 \pm 0.09$ & $16.28 \pm 4.24$ & $-9.21 \pm 0.18$ & 0 & 1 \\
20230703A & $0.1184$ & $291.30$ & $264.37$ & \ & $36.57$ & \ & $68.16 \pm 0.07$ & $10.12 \pm 0.09$ & $3.25 \pm 1.20$ & $-9.59 \pm 0.21$ & 0 & 1 \\
20230222B & $0.11$ & $187.80$ & $160.07$ & \ & $22.47$ & \ & $180.55 \pm 1.03$ & $10.37 \pm 0.08$ & $1.81 \pm 0.57$ & $-10.12 \pm 0.19$ & 0 & 1 \\
20231204A & $0.0644$ & $221.00$ & $191.17$ & \ & $26.80$ & \ & $279.63 \pm 3.02$ & $10.69 \pm 0.07$ & $4.32 \pm 2.63$ & $-10.04 \pm 0.29$ & 0 & 1 \\
20230222A & $0.1223$ & $706.10$ & $571.90$ & \ & $35.56$ & \ & $45.98 \pm 0.81$ & $10.16 \pm 0.17$ & $13.16 \pm 8.34$ & $-9.04 \pm 0.41$ & 0 & 1 \\
20231005A & $0.0713$ & $189.40$ & $155.93$ & \ & $71.22$ & \ & $44.94 \pm 0.09$ & $10.11 \pm 0.10$ & $0.67 \pm 1.41$ & $-10.32 \pm 1.54$ & 0 & 1 \\
20230203A & $0.1464$ & $420.10$ & $383.84$ & \ & $62.50$ & \ & $158.26 \pm 0.35$ & $10.63 \pm 0.10$ & $5.24 \pm 3.49$ & $-9.91 \pm 0.31$ & 0 & 1 \\
20231011A & $0.0783$ & $186.30$ & $115.95$ & \ & $75.33$ & \ & $120.90 \pm 2.77$ & $9.50 \pm 0.18$ & $1.20 \pm 0.57$ & $-9.42 \pm 0.39$ & 0 & 1 \\
20231017A & $0.245$ & $344.20$ & $279.65$ & \ & $59.03$ & $6.60$ & $200.33 \pm 18.13$ & $9.99 \pm 0.18$ & $0.00 \pm 0.29$ & $-12.94 \pm 5.76$ & 0 & 1 \\
20231201A & $0.1119$ & $169.40$ & $99.45$ & \ & $37.47$ & \ & $159.35 \pm 13.04$ & $9.44 \pm 0.18$ & $0.75 \pm 0.34$ & $-9.55 \pm 0.30$ & 0 & 1 \\
20231025B & $0.3238$ & $368.70$ & $320.11$ & \ & $28.57$ & \ & $185.11 \pm 1.48$ & $9.74 \pm 0.12$ & $2.06 \pm 0.81$ & $-9.43 \pm 0.24$ & 0 & 1 \\
20231123A & $0.0729$ & $302.10$ & $212.35$ & \ & $34.33$ & \ & $71.32 \pm 1.71$ & $9.56 \pm 0.24$ & $0.83 \pm 0.76$ & $-9.62 \pm 1.00$ & 0 & 1 \\
20230311A & $0.1918$ & $364.30$ & $271.84$ & \ & $27.08$ & $11.00$ & $372.23 \pm 3.04$ & $10.07 \pm 0.12$ & $4.07 \pm 1.87$ & $-9.47 \pm 0.25$ & 0 & 1 \\
20230730A & $0.2115$ & $312.50$ & $227.34$ & \ & \ & \ & \ & \ & \ & \ & \ & 1 \\
20180924B & $0.3214$ & $361.42$ & $319.54$ & $9.28$ & $34.07$ & $3.40$ & $145.95 \pm 0.49$ & $10.25 \pm 0.10$ & $4.31 \pm 1.72$ & $-9.61 \pm 0.24$ & 0 & 2 \\
20190102C & $0.2913$ & $364.50$ & $307.69$ & $0.83$ & $45.66$ & $2.00$ & $166.56 \pm 1.35$ & $9.57 \pm 0.14$ & $1.46 \pm 0.92$ & $-9.41 \pm 0.32$ & 1 & 3 \\
20180301A & $0.3304$ & $536.00$ & $384.28$ & $1.51$ & \ & $10.80$ & \ & \ & \ & \ & \ & 4,5 \\
20191228A & $0.2432$ & $297.50$ & $263.56$ & $136.95$ & \ & $5.70$ & \ & \ & \ & \ & \ & 4,6 \\
20200906A & $0.3688$ & $577.80$ & $544.16$ & \ & $63.00$ & $5.90$ & $260.69 \pm 0.77$ & $10.29 \pm 0.11$ & $8.09 \pm 4.18$ & $-9.36 \pm 0.25$ & 0 & 7 \\
20210117A & $0.214$ & $728.95$ & $693.85$ & $5.28$ & \ & $2.80$ & \ & \ & \ & \ & \ & 8 \\
20181030A & $0.0039$ & $103.50$ & $63.06$ & $1.60$ & \ & $14.00$ & \ & \ & \ & \ & \ & 8 \\
20181220A & $0.0275$ & $209.40$ & $83.40$ & $0.44$ & $81.53$ & \ & $39.82 \pm 0.61$ & $10.15 \pm 0.11$ & $1.81 \pm 0.92$ & $-9.89 \pm 0.34$ & 0 & 9 \\
20190418A & $0.0715$ & $184.50$ & $113.50$ & $0.78$ & $52.81$ & \ & $20.28 \pm 0.43$ & $10.16 \pm 0.09$ & $1.33 \pm 0.87$ & $-10.05 \pm 0.32$ & 0 & 9 \\
20190425A & $0.03122$ & $128.20$ & $79.20$ & $0.38$ & $58.38$ & \ & $1002.76 \pm 6.63$ & $10.21 \pm 0.06$ & $1.56 \pm 0.60$ & $-10.02 \pm 0.20$ & 0 & 9 \\
20181223C & $0.0302$ & $112.50$ & $92.50$ & $0.11$ & $40.95$ & \ & $23.39 \pm 0.04$ & $9.25 \pm 0.06$ & $0.18 \pm 0.04$ & $-10.00 \pm 0.11$ & 0 & 9 \\
20210410D & $0.1415$ & $578.78$ & $522.58$ & $226.90$ & $42.48$ & $2.90$ & $76.83 \pm 1.03$ & $9.37 \pm 0.13$ & $0.14 \pm 0.38$ & $-10.20 \pm 3.72$ & 1 & 10 \\
\bottomrule
\end{tabular}
\end{varwidth}}
\end{table}

\begin{table}[!ht]
\vspace{30pt}
\centering
\rotatebox[origin=c]{90}{%
\begin{varwidth}{\textheight}
\caption{(Continued)}
\label{tab:data2}
\setlength{\tabcolsep}{7pt}
\begin{tabular}{lrrrllllllllll}
\toprule
FRB name & Redshift & DM & $\rm{}DM_{exc}$ & $\tau_{\rm{}exc}$ & $i$ & Offset & Area & $\rm{}\log(M)$ & SFR & $\rm{}\log(sSFR)$ & Ref. & Quality\\
         &     & (pc cm$^{-3}$) & (pc cm$^{-3}$) & (ms) & (deg) & (kpc) & ($\rm{}kpc^{2}$) & (M$_\odot$) & (M$_\odot$ yr$^{-1}$) & (yr$^{-1}$) &  \\
          &    & (1) & (2) & (3) & (4) & (5) & (6) & (7) & (8) & (9) & (10) & (11) \\
\hline
\midrule
20210603A & $0.1772$ & $500.15$ & $460.15$ & $0.16$ & $64.84$ & $7.20$ & $195.96 \pm 0.64$ & $10.97 \pm 0.10$ & $0.00 \pm 0.22$ & $-15.28 \pm 6.84$ & 0 & 11 \\
20190608B & $0.1178$ & $339.50$ & $301.44$ & $66.76$ & $36.81$ & $6.60$ & $78.03 \pm 0.13$ & $10.21 \pm 0.10$ & $6.37 \pm 2.27$ & $-9.41 \pm 0.25$ & 0 & 12,3 \\
20230930A & $0.0925$ & $456.00$ & $422.27$ & \ & $30.60$ & \ & \ & \ & \ & \ & \ & 13 \\
20230506C & $0.3896$ & $761.00$ & $740.52$ & \ & \ & \ & \ & \ & \ & \ & \ & 13 \\
20240213A & $0.1185$ & $357.40$ & $317.30$ & \ & $70.12$ & \ & $93.64 \pm 0.15$ & $10.13 \pm 0.11$ & $4.94 \pm 2.65$ & $-9.44 \pm 0.29$ & 0 & 14 \\
20240215A & $0.21$ & $549.50$ & $501.50$ & \ & $40.97$ & \ & $121.18 \pm 0.34$ & $9.34 \pm 0.10$ & $2.60 \pm 1.39$ & $-8.91 \pm 0.28$ & 1 & 14 \\
20240229A & $0.287$ & $491.15$ & $453.25$ & \ & $23.77$ & \ & $196.57 \pm 1.49$ & $10.08 \pm 0.20$ & $1.80 \pm 2.08$ & $-9.79 \pm 1.55$ & 0 & 14 \\
20231220A & $0.3355$ & $491.20$ & $441.30$ & \ & $47.87$ & \ & $261.75 \pm 1.05$ & $10.35 \pm 0.10$ & $5.88 \pm 3.06$ & $-9.57 \pm 0.26$ & 0 & 14 \\
20240119A & $0.376$ & $483.10$ & $445.20$ & \ & $31.73$ & \ & $444.51 \pm 5.91$ & $9.56 \pm 0.14$ & $2.51 \pm 1.54$ & $-9.14 \pm 0.32$ & 0 & 14 \\
20220831A & $0.262$ & $1146.25$ & $1019.55$ & \ & \ & \ & \ & \ & \ & \ & \ & 14 \\
20230814A & $0.553$ & $696.40$ & $591.50$ & \ & \ & \ & \ & \ & \ & \ & \ & 14 \\
20240123A & $0.968$ & $1462.00$ & $1371.70$ & \ & \ & \ & \ & \ & \ & \ & \ & 14 \\
20230521A & $1.354$ & $1342.90$ & $1204.06$ & \ & \ & \ & \ & \ & \ & \ & \ & 14 \\
20210405I & $0.066$ & $565.17$ & $565.17$ & $70.50$ & \ & $10.65$ & \ & \ & \ & \ & \ & 15 \\
20220105A & $0.2784$ & $580.00$ & $557.96$ & \ & $56.26$ & $8.45$ & $100.73 \pm 0.87$ & $9.95 \pm 0.11$ & $0.50 \pm 0.74$ & $-10.23 \pm 1.10$ & 0 & 16 \\
20211203C & $0.3437$ & $635.00$ & $571.27$ & \ & $13.37$ & $5.04$ & $96.84 \pm 0.29$ & $10.04 \pm 0.13$ & $5.67 \pm 1.39$ & $-9.27 \pm 0.20$ & 3 & 16 \\
20200430A & $0.16$ & $380.25$ & $353.24$ & $43.20$ & $43.88$ & $3.00$ & $67.32 \pm 0.64$ & $9.19 \pm 0.13$ & $0.39 \pm 0.38$ & $-9.64 \pm 0.59$ & 4 & 17 \\
20191001A & $0.234$ & $507.90$ & $463.19$ & $11.74$ & $53.47$ & $11.00$ & $185.68 \pm 2.18$ & $10.69 \pm 0.16$ & $15.11 \pm 11.17$ & $-9.51 \pm 0.31$ & 1 & 17 \\
20190714A & $0.2365$ & $504.13$ & $466.15$ & $17.20$ & $51.10$ & $1.90$ & $99.48 \pm 6.43$ & $10.03 \pm 0.11$ & $3.68 \pm 1.43$ & $-9.45 \pm 0.25$ & 0 & 17 \\
20190611B & $0.378$ & $321.40$ & $264.19$ & $3.64$ & \ & $11.00$ & \ & \ & \ & \ & \ & 17 \\
20190711A & $0.522$ & $593.10$ & $537.18$ & $24.70$ & \ & $3.20$ & \ & \ & \ & \ & \ & 17 \\
20200223B & $0.0602$ & $201.80$ & $156.20$ & $2.70$ & \ & \ & \ & \ & \ & \ & \ & 18 \\
20191106C & $0.10775$ & $332.20$ & $307.20$ & $10.00$ & \ & \ & \ & \ & \ & \ & \ & 18\ \\
20190110C & $0.12244$ & $221.60$ & $184.50$ & $1.50$ & \ & \ & \ & \ & \ & \ & \ & 18 \\
20190614D & $0.63$ & $959.20$ & $875.67$ & \ & \ & \ & \ & \ & \ & \ & \ & 19 \\
20220207C & $0.04304$ & $262.38$ & $183.08$ & \ & $79.94$ & $2.21$ & $50.11 \pm 1.29$ & $10.39 \pm 0.21$ & $0.01 \pm 1.08$ & $-12.42 \pm 5.90$ & 0 & 20 \\
20220920A & $0.15824$ & $314.99$ & $274.69$ & \ & $26.47$ & $7.32$ & $203.18 \pm 0.24$ & $10.24 \pm 0.11$ & $4.67 \pm 1.88$ & $-9.57 \pm 0.25$ & 0 & 20 \\
20220825A & $0.2414$ & $651.24$ & $571.54$ & \ & $45.15$ & $2.25$ & $222.44 \pm 23.79$ & $10.23 \pm 0.17$ & $5.45 \pm 2.82$ & $-9.49 \pm 0.35$ & 0 & 20 \\
20220307B & $0.24812$ & $499.27$ & $363.57$ & \ & $43.37$ & $7.09$ & $206.12 \pm 21.61$ & $10.73 \pm 4.14$ & $21.63 \pm 257.21$ & $-9.50 \pm 1.34$ & 1 & 20 \\
20221012A & $0.28467$ & $441.08$ & $386.68$ & \ & $44.81$ & $14.14$ & $267.17 \pm 0.38$ & $11.17 \pm 0.08$ & $0.00 \pm 0.02$ & $-16.89 \pm 6.45$ & 0 & 20 \\
20220310F & $0.47796$ & $462.24$ & $416.84$ & \ & $36.50$ & $4.39$ & $454.11 \pm 3.43$ & $10.24 \pm 0.10$ & $5.20 \pm 1.93$ & $-9.53 \pm 0.25$ & 0 & 20 \\
\bottomrule
\end{tabular}
\end{varwidth}}
\end{table}

\begin{table}[!ht]
\vspace{30pt}
\centering
\rotatebox[origin=c]{90}{%
\begin{varwidth}{\textheight}
\caption{(Continued)}
\label{tab:data3}
\setlength{\tabcolsep}{7pt}
\begin{tabular}{lrrrllllllllll}
\toprule
FRB name & Redshift & DM & $\rm{}DM_{exc}$ & $\tau_{\rm{}exc}$ & $i$ & Offset & Area & $\rm{}\log(M)$ & SFR & $\rm{}\log(sSFR)$ & Ref. & Quality\\
         &     & (pc cm$^{-3}$) & (pc cm$^{-3}$) & (ms) & (deg) & (kpc) & ($\rm{}kpc^{2}$) & (M$_\odot$) & (M$_\odot$ yr$^{-1}$) & (yr$^{-1}$) &  \\
         &    & (1) & (2) & (3) & (4) & (5) & (6) & (7) & (8) & (9) & (10) & (11) \\
\hline
\midrule
20220506D & $0.30039$ & $396.97$ & $307.87$ & \ & \ & $2.08$ & \ & \ & \ & \ & \ & 20 \\
20220418A & $0.622$ & $623.25$ & $585.65$ & \ & \ & $15.15$ & \ & \ & \ & \ & \ & 20 \\
20220509G & $0.0894$ & $269.53$ & $214.33$ & $2.37$ & $59.39$ & $3.80$ & $41.57 \pm 0.96$ & $10.64 \pm 0.08$ & $1.75 \pm 0.90$ & $-10.38 \pm 0.22$ & 3 & 20 \\
20220914A & $0.1139$ & $631.28$ & $576.08$ & $2.37$ & $45.35$ & $9.87$ & $39.10 \pm 2.41$ & $9.45 \pm 0.12$ & $0.76 \pm 0.31$ & $-9.60 \pm 0.24$ & 0 & 20 \\
20171020A & $0.00867$ & $114.10$ & $77.46$ & \ & $68.75$ & \ & $11.27 \pm 0.06$ & $8.44 \pm 0.21$ & $0.30 \pm 0.25$ & $-8.96 \pm 0.44$ & 4 & 21 \\
20180916B & $0.0337$ & $348.76$ & $149.80$ & $0.88$ & \ & $5.50$ & \ & \ & \ & \ & \ & 22 \\
20190303A & $0.064$ & $222.40$ & $193.09$ & $2.96$ & \ & \ & \ & \ & \ & \ & \ & \ 23 \\
20180814A & $0.06835$ & $189.40$ & $106.30$ & $3.09$ & \ & \ & \ & \ & \ & \ & \ & 23 \\
20190520B & $0.241$ & $1210.30$ & $1152.11$ & $188.00$ & $17.46$ & $1.30$ & $60.58 \pm 2.28$ & \ & \ & \ & \ & 24 \\
20181112A & $0.4755$ & $589.27$ & $546.86$ & $0.45$ & \ & $1.70$ & \ & \ & \ & \ & \ & 25 \\
20201123A & $0.0507$ & $433.55$ & $180.47$ & \ & \ & \ & \ & \ & \ & \ & \ & 26 \\
20220717A & $0.36295$ & $637.00$ & $519.00$ & $63.27$ & \ & \ & $141.50 \pm 17.56$ & $10.84 \pm 0.60$ & $3.23 \pm 11.08$ & $-10.13 \pm 4.01$ & 1 & 27 \\
20190523A & $0.66$ & $760.80$ & $723.93$ & $10.80$ & \ & $27.00$ & \ & $10.76 \pm 0.20$ & $30.84 \pm 48.40$ & $-9.32 \pm 0.53$ & 1 & 28 \\
20201124A & $0.098$ & $413.52$ & $290.36$ & $8.05$ & \ & $1.32$ & \ & \ & \ & \ & \ & 29 \\
20220319D & $0.01123$ & $110.95$ & $110.95$ & \ & $24.66$ & $2.14$ & $12.29 \pm 0.65$ & $9.91 \pm 0.06$ & $0.21 \pm 0.10$ & $-10.58 \pm 0.25$ & 4 & 30 \\
20220912A & $0.0771$ & $219.46$ & $99.03$ & $2.94$ & $42.78$ & \ & $22.74 \pm 1.30$ & $9.60 \pm 0.19$ & $0.92 \pm 0.56$ & $-9.63 \pm 0.41$ & 0 & 31 \\
20220610A & $1.016$ & $1457.62$ & $1425.25$ & $6.84$ & \ & \ & \ & \ & \ & \ & \ & 32 \\
20230708A & $0.105$ & $411.51$ & $361.51$ & $0.94$ & \ & 1.10 & \ & $7.78 \pm 0.14$ & $0.04 \pm 0.03$ & $-9.14 \pm 0.38$ & 1 & 33,34,43 \\
20230718A & $0.035$ & $477.00$ & $81.00$ & \ & \ & \ & \ & \ & \ & \ & \ & 33,35 \\
20210320C & $0.2797$ & $384.80$ & $342.60$ & $0.88$ & $27.62$ & $3.22$ & $272.33 \pm 20.48$ & $10.22 \pm 0.14$ & $7.13 \pm 3.10$ & $-9.36 \pm 0.26$ & 0 & 33,6 \\
20240210A & $0.023686$ & $283.73$ & $252.73$ & \ & $51.95$ & 4.38 & $72.54 \pm 0.38$ & $9.71 \pm 0.05$ & $0.81 \pm 0.13$ & $-9.79 \pm 0.10$ & 0 & 33,43 \\
20240201A & $0.042729$ & $374.50$ & $336.50$ & \ & $82.45$ & 4.38 & $29.17 \pm 0.04$ & $9.99 \pm 0.07$ & $1.18 \pm 0.83$ & $-9.92 \pm 0.42$ & 0 & 33,43 \\
20240310A & $0.127$ & $601.80$ & $565.80$ & \ & $56.27$ & 2.18 & $30.76 \pm 0.09$ & $9.14 \pm 0.12$ & $0.38 \pm 0.18$ & $-9.56 \pm 0.28$ & 0 & 33,43 \\
20230526A & $0.157$ & $361.40$ & $311.40$ & \ & $20.63$ & 1.76 & $30.28 \pm 0.21$ & $9.44 \pm 0.13$ & $0.59 \pm 0.62$ & $-9.67 \pm 0.71$ & 1 & 33,43 \\
20220725A & $0.1926$ & $290.40$ & $259.40$ & \ & \ & 2.03 & \ & $10.70 \pm 0.10$ & $10.73 \pm 4.13$ & $-9.67 \pm 0.22$ & 0 & 30,43 \\
20221106A & $0.2044$ & $343.80$ & $308.80$ & \ & $39.69$ & 4.52 & $197.09 \pm 0.68$ & $10.78 \pm 0.10$ & $2.37 \pm 2.20$ & $-10.39 \pm 0.51$ & 0 & 33,43 \\
20230902A & $0.3619$ & $440.10$ & $406.10$ & \ & $34.67$ & 4.44 & $108.45 \pm 0.62$ & $9.58 \pm 0.13$ & $1.77 \pm 1.16$ & $-9.34 \pm 0.34$ & 1 & 33,43 \\
20231226A & $0.1569$ & $329.90$ & $184.90$ & \ & $35.63$ & 4.84 & $216.13 \pm 0.73$ & $9.71 \pm 0.08$ & $2.43 \pm 0.74$ & $-9.33 \pm 0.19$ & 0 & 33,43 \\
20220501C & $0.381$ & $449.50$ & $418.50$ & \ & \ & \ & \ & \ & \ & \ & \ & 33 \\
20220918A & $0.491$ & $656.80$ & $615.80$ & \ & \ & 5.83 & \ & \ & \ & \ & \ & 33,43 \\
20221219A & $0.553$ & $706.70$ & $662.32$ & $569.13$ & $46.25$ & $5.92$ & $568.09 \pm 5.90$ & $10.53 \pm 0.16$ & $1.96 \pm 2.49$ & $-10.22 \pm 0.52$ & 0 & 36,37 \\
\bottomrule
\end{tabular}
\end{varwidth}}
\end{table}

\begin{table}[!ht]
\vspace{30pt}
\centering
\rotatebox[origin=c]{90}{%
\begin{varwidth}{\textheight}
\caption{(Continued)}
\label{tab:data4}
\setlength{\tabcolsep}{7pt}
\begin{tabular}{lrrrllllllllll}
\toprule
FRB name & Redshift & DM & $\rm{}DM_{exc}$ & $\tau_{\rm{}exc}$ & $i$ & Offset & Area & $\rm{}\log(M)$ & SFR & $\rm{}\log(sSFR)$ & Ref. & Quality\\
         &     & (pc cm$^{-3}$) & (pc cm$^{-3}$) & (ms) & (deg) & (kpc) & ($\rm{}kpc^{2}$) & (M$_\odot$) & (M$_\odot$ yr$^{-1}$) & (yr$^{-1}$) &  \\
         &    & (1) & (2) & (3) & (4) & (5) & (6) & (7) & (8) & (9) & (10) & (11) \\
\hline
\midrule
20231120A & $0.0368$ & $438.90$ & $395.10$ & \ & $71.74$ & $0.87$ & $45.13 \pm 0.15$ & $10.58 \pm 0.06$ & $0.00 \pm 0.00$ & $-18.98 \pm 9.42$ & 0 & 36 \\
20230124A & $0.0939$ & $590.60$ & $551.99$ & \ & $10.17$ & $1.14$ & $43.23 \pm 0.09$ & $9.69 \pm 0.11$ & $0.91 \pm 0.48$ & $-9.72 \pm 0.26$ & 0 & 36 \\
20230628A & $0.127$ & $345.15$ & $306.05$ & \ & $30.70$ & \ & $77.25 \pm 0.35$ & $9.43 \pm 0.13$ & $1.40 \pm 0.55$ & $-9.29 \pm 0.27$ & 0 & 36 \\
20221101B & $0.2395$ & $490.70$ & $359.45$ & \ & \ & $10.16$ & \ & $11.25 \pm 0.15$ & $0.01 \pm 3.17$ & $-13.40 \pm 5.56$ & 0 & 36 \\
20231123B & $0.2621$ & $396.70$ & $356.50$ & \ & $65.18$ & $14.40$ & $574.05 \pm 1.18$ & $10.99 \pm 0.12$ & $9.15 \pm 6.64$ & $-10.03 \pm 0.37$ & 0 & 36 \\
20230307A & $0.2706$ & $608.90$ & $571.32$ & \ & $31.95$ & $7.93$ & $538.26 \pm 1.78$ & $10.59 \pm 0.09$ & $7.07 \pm 4.20$ & $-9.71 \pm 0.25$ & 0 & 36 \\
20221116A & $0.2764$ & $640.60$ & $508.34$ & \ & \ & $18.41$ & \ & $10.92 \pm 0.21$ & $0.95 \pm 8.36$ & $-11.03 \pm 4.11$ & 0 & 36 \\
20230626A & $0.327$ & $451.20$ & $411.95$ & \ & $27.51$ & $3.71$ & $83.52 \pm 0.30$ & $10.62 \pm 0.13$ & $0.70 \pm 2.44$ & $-10.78 \pm 2.86$ & 0 & 36 \\
20220330D & $0.3714$ & $468.10$ & $429.56$ & \ & $25.44$ & $16.72$ & $113.15 \pm 1.87$ & $10.40 \pm 0.10$ & $0.00 \pm 0.00$ & $-17.82 \pm 7.14$ & 0 & 36 \\
20230216A & $0.531$ & $828.00$ & $789.57$ & \ & $30.87$ & $23.10$ & $159.78 \pm 1.93$ & $9.78 \pm 0.13$ & $28.06 \pm 10.66$ & $-8.34 \pm 0.26$ & 0 & 36 \\
20221029A & $0.975$ & $1391.05$ & $1347.21$ & \ & $47.00$ & $26.97$ & $405.65 \pm 20.42$ & $10.33 \pm 0.36$ & $22.21 \pm 40.89$ & $-9.10 \pm 0.36$ & 2 & 36 \\
20221113A & $0.2505$ & $411.40$ & $319.67$ & \ & \ & $3.62$ & \ & \ & \ & \ & \ & 36 \\
20230501A & $0.3015$ & $532.50$ & $406.80$ & \ & \ & $3.74$ & \ & \ & \ & \ & \ & 36 \\
20220208A & $0.351$ & $437.00$ & $335.44$ & \ & \ & $25.88$ & \ & \ & \ & \ & \ & 36 \\
20220726A & $0.3619$ & $686.55$ & $597.03$ & \ & \ & $8.33$ & \ & \ & \ & \ & \ & 36 \\
20220204A & $0.4012$ & $612.20$ & $561.50$ & \ & \ & $12.07$ & \ & \ & \ & \ & \ & 36 \\
20230712A & $0.4525$ & $586.96$ & $547.76$ & \ & \ & $12.24$ & \ & \ & \ & \ & \ & 36 \\
20221027A & $0.5422$ & $452.50$ & $405.19$ & \ & $68.64$ & $8.80$ & $279.68 \pm 5.98$ & \ & \ & \ & \ & 36 \\
20240114A & $0.1306$ & $527.70$ & $478.03$ & $1.89$ & $30.57$ & \ & $30.78 \pm 0.63$ & $8.36 \pm 0.14$ & $0.10 \pm 0.05$ & $-9.34 \pm 0.28$ & 1 & 37,38 \\
20211212A & $0.0707$ & $206.00$ & $178.90$ & \ & $33.26$ & $2.68$ & $118.94 \pm 0.29$ & $10.22 \pm 0.07$ & $3.64 \pm 1.28$ & $-9.66 \pm 0.21$ & 0 & 39 \\
20210807D & $0.1293$ & $251.30$ & $130.13$ & \ & $44.50$ & $11.89$ & $1024.79 \pm 64.95$ & $10.41 \pm 0.16$ & $11.33 \pm 4.15$ & $-9.38 \pm 0.26$ & 0 & 39 \\
20211127I & $0.0469$ & $234.83$ & $192.33$ & \ & $13.99$ & $1.59$ & $963.25 \pm 23.20$ & $10.03 \pm 0.02$ & $1.33 \pm 0.27$ & $-9.91 \pm 0.09$ & 4 & 39 \\
20220529A & $0.1839$ & $246.00$ & $206.05$ & \ & \ & \ & \ & \ & \ & \ & \ & 40 \\
20121102A & $0.19273$ & $557.00$ & $368.58$ & $11.40$ & \ & $0.60$ & \ & \ & \ & \ & \ & 41,42 \\
\bottomrule
\end{tabular}
\vspace{10pt}  
\begin{minipage}{0.8\textwidth}  
    \textit{Note:} 
    (1) The DM is derived from the first detected burst for repeaters. 
    (2) $\rm{}DM_{exc} = \rm{}DM_{obs} - \rm{}DM_{MW}$, where $\rm{}DM_{MW}$ is estimated using the NE2001 model\citep{ne2001}. 
    (3) The scattering time $\tau_{\rm{}exc}$ is scaled to 600 MHz, assuming $\tau_{\rm{}exc} \propto \nu^{-4}$. 
    (4) The inclination angle is derived from the Sérsic model. Details of the fitting methods, see Section \ref{subsec:fitting methods}.
    (5) Offset is defined as the distance from locations of FRBs to the optical center of the host galaxies.
    (6) The galaxy area is estimated using the lengths of the major axis ($a$) and minor axis ($b$) obtained by fitting the Sérsic model to the assumed elliptical 2D projection shape of the galaxy. 
    (7) The stellar mass of the host galaxy is given on a logarithmic scale.
    (8) We provide only upper limits for galaxies with SFR $< 0.01$ M$_\odot$ yr$^{-1}$. 
    (9) The specific star formation rate (sSFR) is defined as $ \rm{sSFR =SFR} / M $. The error is calculated using the error propagation formula.
    (10) References: 1 \cite{Amiri2025}. 2 \cite{Bannister2019}. 3 \cite{Bhandari2020}. 4 \cite{Bhandari2020}. 5 \cite{Kumar2023}. 6 \cite{Sammons2023}. 7 \cite{Bhandari2022}. 8 \cite{Bhardwaj2021}. 9 \cite{Bhardwaj2024} 10 \cite{Caleb2023}. 11 \cite{Cassanelli2024}. 12 \cite{Chittidi2021}. 13 \cite{Anna-Thomas2025}. 14 \cite{Connor2024}. 15 \cite{Driessen2024}. 16 \cite{Gordon2023}. 17 \cite{Heintz2020}. 18 \cite{Ibik2024}. 19 \cite{law2020}. 20 \cite{Law2024}. 21 \cite{Mahony2018}. 22 \cite{Marcote2020}. 23 \cite{Michilli2023}. 24 \cite{Niu2022}. 25 \cite{Prochaska2019}. 26 \cite{Rajwade2022}. 27 \cite{Rajwade2024}. 28 \cite{Ravi2019}. 29 \cite{Ravi2022}. 30 \cite{Ravi2025}. 31 \cite{Ravi2023}. 32 \cite{Ryder2023}. 33 \cite{Shannon2024}. 34 \cite{Dial2025}. 35 \cite{Glowacki2024}. 36 \cite{Sharma2024}. 37 \cite{Chen2025}. 38 \cite{Panda2024}. 39 Deller et al. (in prep.). 40 \cite{LiY2025}. 41 \cite{Tendulkar2017}. 42 \cite{Chatterjee2017}. 43 \cite{Gordon2025}.
    (11) The SED fitting results are visually evaluated and classified into five quality levels: 0, 1, 2, 3, and 4. In this work, only FRBs with Quality levels 0 ,1 and 3 are included in the analysis.
\end{minipage}
\end{varwidth}}
\end{table}

\end{document}